\documentclass[iop]{emulateapj}
\usepackage{amsmath}
\usepackage{graphicx}
\usepackage[breaklinks,colorlinks,
   urlcolor=blue, citecolor=blue, linkcolor=blue]{hyperref}
\usepackage{rotating}
\usepackage{epstopdf}
\usepackage{xcolor}
\usepackage[T1]{fontenc}
\usepackage[english]{babel}
\usepackage[utf8]{inputenc}

\slugcomment{Accepted for publication in ApJ May 21, 2019}

\shorttitle{Spotted stars in the OGLE Galactic bulge data}
\shortauthors{Iwanek et al.}

\begin{document}

\title{12\;660 spotted stars toward the OGLE Galactic bulge fields}

\author{P. Iwanek\altaffilmark{1$^\dagger$}, I. Soszy\' nski\altaffilmark{1}, 
J. Skowron\altaffilmark{1}, A. Udalski\altaffilmark{1}, K. St{\k e}pie\' n\altaffilmark{1}, 
S. Koz\l owski\altaffilmark{1}, P. Mr\' oz\altaffilmark{1}, 
\newline R. Poleski\altaffilmark{2,1}, D. Skowron\altaffilmark{1}, M. K. Szyma\' nski\altaffilmark{1}, 
P. Pietrukowicz\altaffilmark{1}, K. Ulaczyk\altaffilmark{3,1}, 
\L. Wyrzykowski\altaffilmark{1},  \newline K. Kruszy\' nska\altaffilmark{1}
and K. Rybicki\altaffilmark{1}}

\affil{$^1$ Warsaw University Observatory, Al. Ujazdowskie 4, 00-478 Warsaw, 
Poland}
\affil{$^2$ Department of Astronomy, Ohio State University, 140 W. 18th Ave., 
Columbus, OH 43210, USA}
\affil{$^3$ Department of Physics, University of Warwick, Coventry CV4 7AL, UK}
\affil{$^\dagger$ Corresponding author: piwanek@astrouw.edu.pl}

\begin{abstract}

We present the discovery and statistical analysis of $12\;660$ spotted variable stars
toward and inside the Galactic bulge from over two-decade-long 
\textit{Optical Gravitational Lensing Experiment} (OGLE) data.
We devise a new method of dereddening of individual stars toward the Galactic bulge where
strong and highly nonuniform extinction is present. 
In effect, $11\;812$ stars were classified as giants and $848$ as dwarfs. Well defined correlations between the luminosity, variability amplitude and rotation period were found for the giants. 
Rapidly rotating dwarfs with periods $P \leq 2$ d show {\textit I-}band amplitudes lower than 0.2~mag which is substantially less than the amplitudes of up to 0.8~mag observed in giants and slowly
rotating dwarfs. We also notice that amplitudes of stars brighter than $I_0 \approx 16$ mag do not exceed 0.3-0.4 mag. We divide the stars into three groups characterized by correlation between light and color variations. The positive correlation is characteristic for stars that are cooler when fainter, which results
from the variable coverage of the stellar surface with spots similar to the sunspots.
The variability of stars that are cooler when brighter (negative correlation) can be characterized by
chemical spots with overabundance of heavy elements inside and variable {\it line-blanketing} effect,
which is observed in chemically peculiar stars. The null correlation may results from very high 
level of the magnetic activity with rapidly variable magnetic fields. 
This division is readily visible on the color--magnitude diagram (CMD), 
which suggests that it may depend on the radius of the stars. We detect 79 flaring
objects and discuss briefly their properties. Among others, we find that relative brightening during flares is correlated with brightness amplitude.

\end{abstract}


\section{Introduction}

Studies of starspots started hundreds years ago with the observations of 
the Sun with naked eye. The first thorough examination of the Sun's spots 
migrations was done by \citet{Schwabe44, Schwabe45} who found out that 
heliographic latitudes of spots vary with an 11-year cycle. Nowadays Schwabe's 
discovery is known as the solar activity cycle. Direct observations of spots
on other stars are not possible, however the influence of active regions on
observational stellar parameters ({\em e.g.} brightness variability, colors, spectral features etc.)
has been observed for decades. Unfortunately, changes in stellar parameters caused
by spots were not properly explained for many years. As an illustration, \citet{Hoffmeister15} was the first who observed light variations on the RS~CVn binary, while \citet{Ross26} discovered the first
BY Dra-type variable -- HH~And. The first person who began to consider light curves
distortions as caused by spots was \citet{Kron47, Kron50}. As another example in
a good tone is to mention the work done by \citet{Maury97}, who classified star
$\alpha^2$~CVn as A0p, where ``p'' marked some peculiarity in spectral lines.
Less than two decades later, \citet{Guthnick14}, using a photoelectric photometer,
discovered that this star is variable with a period of $5.54$ d, and
with the brightness amplitude of $0.05$ mag. The variability is
explained as a result of nonuniform distribution of elements on
the stellar surface (chemical spots), associated with the magnetic
field. This object became a prototype of stars with a peculiar
chemical composition, called Ap, or more generally CP (Chemically Peculiar).
The variability of magnetic, chemically peculiar stars, based on 16 objects, was described
by \citet{Stepien68}, while classification of stars with this type of spots was
systematized by \citet{Preston74}.

The following years of research showed that cool spots are present in
various types of stars.
Many stars produce significantly bigger spots with longer lifetimes 
than their solar counterparts. Recent research shows that cool spots can possibly exist 
on hot stars ({\em e.g.} \citealt{Balona15, Balona17}) although other
explanations of their observations have also been offered. Thanks to the modern 
methods of analysis, availability of space telescopes and long time-series observations it
has been possible to measure magnetic activity cycles in many stars ({\em e.g.} \citealt{Olah02, 
Messina02, Olah09, Olah16, Reinhold17, Montet17}). Nowadays many studies of stars' magnetic fields are also based on the spectropolarimetric observations, Doppler Imaging and Zeeman-Doppler Imaging ({\em e.g.} \citealt{Roettenbacher16, Wade16, Kochukhov19a, Kochukhov19b, Strassmeier19, Hubrig19}).

\begin{figure*}[h!]
\begin{center}
\includegraphics[scale=0.13]{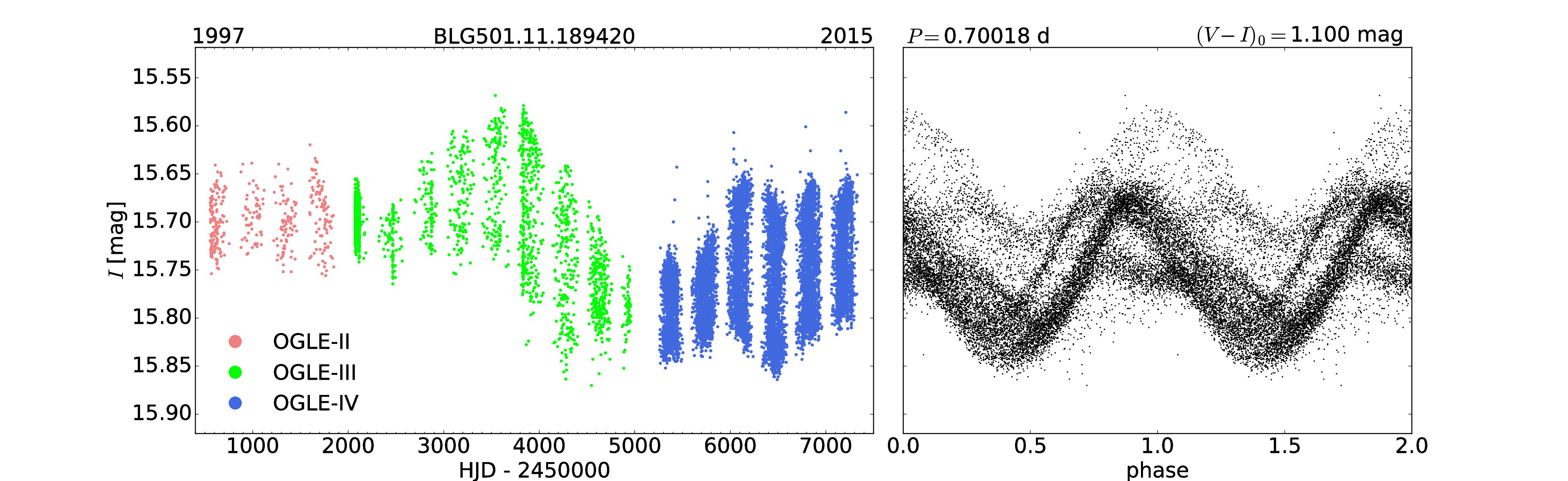}
\includegraphics[scale=0.13]{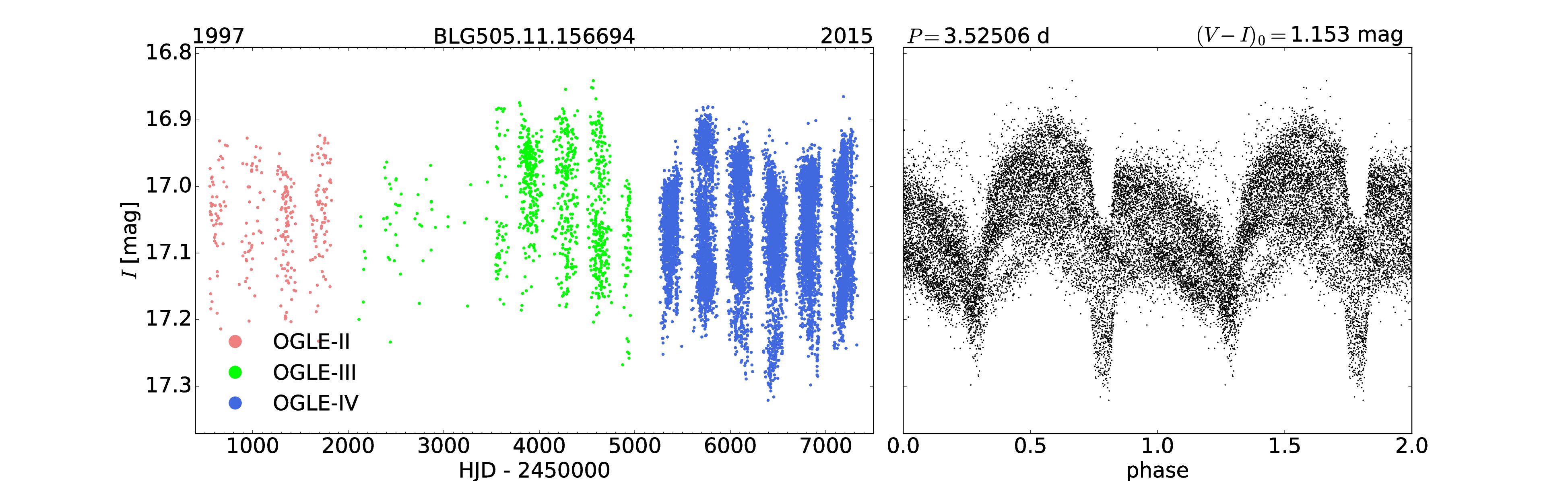}
\includegraphics[scale=0.13]{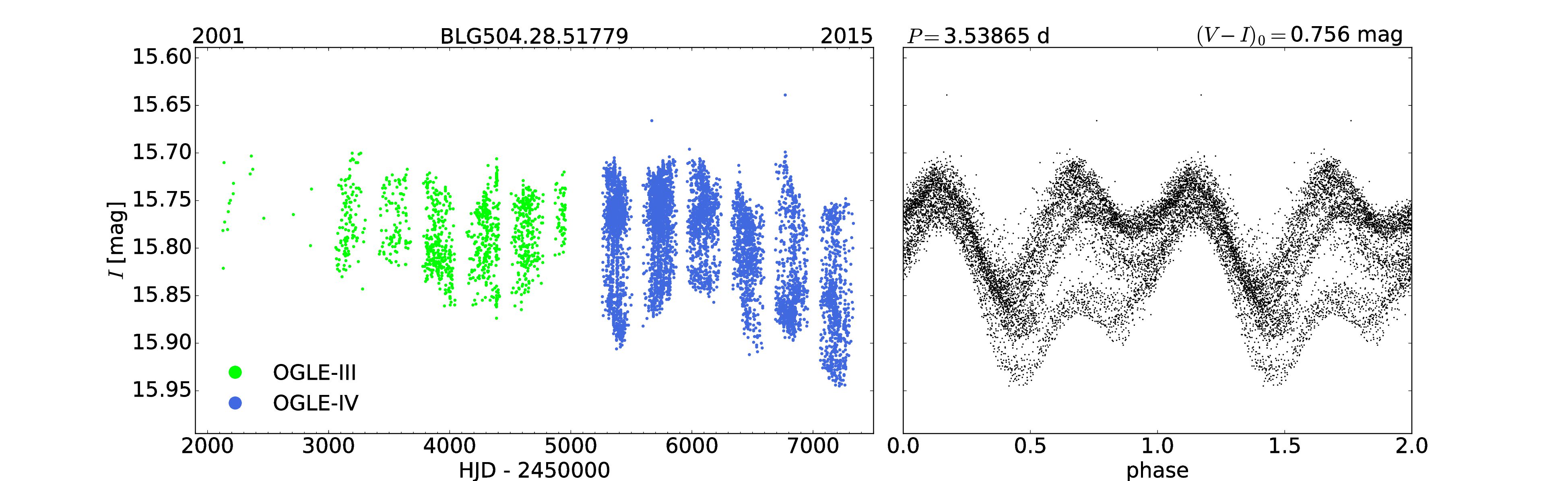}
\includegraphics[scale=0.13]{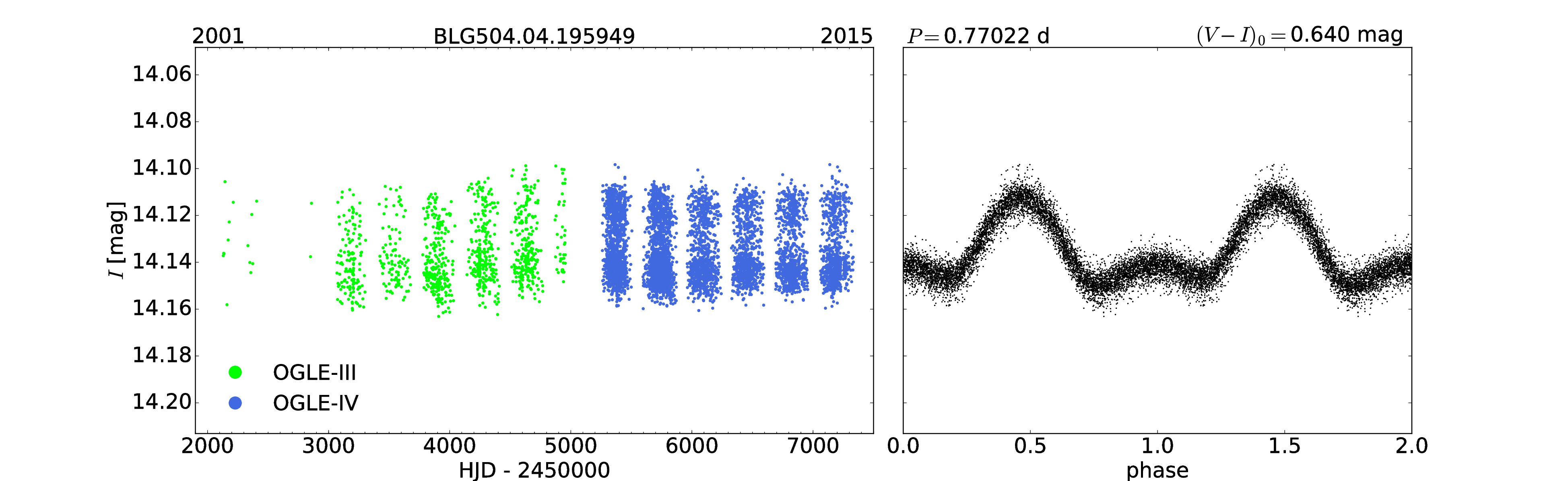}
\caption{Four examples of the spotted stars' light curves from our collection. 
In the top row we can see a star with one dominant dark cool spot and no clear 
indication of binarity. In the second row, we can see a typical RS CVn-type 
eclipsing system. In the third row, we present a star with two dominant dark cool spots.
The bottom row presents star with chemically peculiar spots. Left hand-side panels show 
unfolded light curves, while the right hand-side panels show phase-folded light curves with rotation 
periods $P$~(in the case of second plot we show the orbital period). The rotation period $P$ 
and color index $(V-I)_0$ corrected for the interstellar extinction are 
indicated above the plots. The dates at the top of the non-phased plot mark the 
year when the observations started, and the year of the last used observations.
Individual phases of the OGLE project (OGLE-II, OGLE-III, and current 
OGLE-IV) are marked with different colors.}
\label{fig1}
\end{center}
\end{figure*}

\newpage
Examination of stellar spots and stellar activity is closely related to the 
study of mechanisms responsible for the generation of magnetic field in stars. 
The~first model of magnetohydrodynamic dynamo was provided by 
\citet{Parker55}. In this model, active regions of stars (including spots) arise 
as a~result of ejecting a magnetic flux from deep layers of the convective 
zone onto the stars' surface. Contemporary research has shown that rotation is a 
key in the generation of magnetic field of stars, and thus in the generation of 
phenomenon of stellar activity \citep{Kraft67, Skumanich72, Durney78}. The examination of 
quasi-periodic brightness variations due to the magnetically active regions on 
stellar surfaces is closely related to the study of their rotation ({\em e.g.} 
\citealt{Irwin09}). Stars with clear manifestations of the magnetic 
field activity are an excellent laboratory for testing dynamo models and a mine of 
knowledge about differential rotation ({\em e.g.} \citealt{Guerrero13, Guerrero16a, 
Guerrero16b, Guerrero18}). One of the largest samples of stellar rotation periods 
to date was published by \citet{McQuillan14} from the {\it{Kepler}} space mission 
data and it consists of 34\;030 records.

Another studies of the stellar activity showed a strong correlation 
between rotation periods and X-ray emission \citep{Pallavicini81}, H$\alpha$ 
emission \citep{Mekkaden85}, UV emission \citep{Simon87}, radio emission 
\citep{Drake89}, and frequency of flares \citep{Maehara12}.

The era of large-scale sky surveys initiated massive studies of stellar 
variability, including the variability caused by manifestation of the magnetic 
field. An extensive statistical analysis of spotted giants and subgiants,
based on MACHO data (MAssive Compact Halo Objects, \citealt{Alcock95}) was
obtained by \citet{Drake06}. \citet{Kiraga12} and \citet{Kiraga13} analyzed
over 2000 spotted stars based on the ASAS survey \citep{Pojmanski94, Pojmanski02}.
Recently \citet{Lanzafame18a, Lanzafame18b} examined almost $150\;000$ late-type 
spotted dwarfs candidates found in the {\textit {Gaia Data Release 2}} (Gaia 
DR2; \citealt{Gaia16, Gaia18}).

Here we present a detection and statistical analysis of $12\;660$ stars inside and toward
the Galactic bulge found in the data collected by the OGLE project 
({\textit{Optical Gravitational Lensing Experiment}}; \citealt{Udalski15}).
As the Galactic bulge is one of the most challenging regions 
to explore in the Milky Way, because of the large and highly
nonuniform interstellar extinction toward this
direction, we propose here a new approach for the dereddening procedure.
In the first part of this work, we try to verify correlations found by \citet{Drake06}
and \citet{Lanzafame18a, Lanzafame18b}. 
The second part of the paper is devoted to the two linear correlations between
changes in brightness and changes in color that we discover for spotted stars.
In the third part of this work, we conduct a basic analysis of stars with flares, which we find
among all the spotted stars from our sample.

\section{Observations}

Long-term, high-cadence, multi-color, precise photometric observations play a key role in the analysis of 
stellar magnetic activity. For this reason, we use observations 
collected during three subsequent phases of the OGLE project:
OGLE-II (1997--2000), OGLE-III (2001--2009), and OGLE-IV (2010--2015).
Each phase marks the improvement in CCD camera size. The fourth phase of the OGLE
project is still ongoing. However, for the purposes of this analysis, 
we use data collected by the end of the 2015. All data were obtained with the 
1.3-meter Warsaw Telescope located at Las Campanas Observatory, Chile, which is 
operated by the Carnegie Institution for Science. During the OGLE-II phase about 
30 million stars in the area of 11 square degrees in the central parts of the 
Milky Way were permanently monitored. In the next two phases, the sky coverage 
was extended, and now, during the OGLE-IV phase, we regularly observe more than 
400 million stars in the area of 182 square degrees of the Galactic bulge.

During the ongoing OGLE-IV phase, a 32-detector mosaic CCD camera covering an 
area of about 1.4 square degrees in the sky is used. Observations are carried 
out in two standard \textit{I}- and \textit{V}-band filters of the 
Johnson-Cousins photometric system. Most of our observations are taken in the 
\textit{I}-band with an exposure time of 100~s. The number of collected data 
points greatly varies between individual fields -- for the most sampled fields 
we obtained over 13\,000 data points and for the least sampled $\sim$100 data 
points. In the \textit{V}-band, we have secured from several up to over 100 epochs 
per star with an integration time of 150~s.

All images collected by OGLE are reduced with the standard Difference Image 
Analysis (DIA) technique \citep{Alard98, Wozniak00}. For more details about the 
OGLE instrumentation, data reductions, calibrations see \citet{Udalski15}. 
Details about the sky coverage and observing cadence can be also found on the OGLE 
website\footnote{http://ogle.astrouw.edu.pl/}.

\section{Selection and data sample}

The majority of spotted stars in our collection have been found as a by-product 
of the massive search for eclipsing binary systems in the Galactic bulge 
\citep{Soszynski16}, but many more have been found during searching for pulsating stars in the Milky Way
(\citealt{Udalski18}, Soszy\' nski et al., in prep.). Periodic light curves with a characteristic long-term 
modulation of amplitudes and mean magnitudes were flagged as candidates for 
spotted variables. Typical long-term modulation of stellar brightness of spotted variables
has time-scale of order of several years, and it is directly related to the magnetic activity cycles, 
similar to the Sun. In addition, strictly periodic stars (with no significant changes of the mean brightness) were
also included (very likely candidates for CP stars). Each light curve in our sample has been visually inspected 
at least once, and usually a few times by more than one experienced 
astronomer. The light variability caused by stellar spots may be confused with
semiregular variations due to radial and non-radial pulsations in red
giant stars (long-period variables, e.g. \citealt{Soszynski13}).
However, in the most cases we were able to distinguish between both types of
variable stars using their time-series photometry. As shown in
Figure 1 of \citet{Soszynski13}, long-period variables change their
amplitudes from cycle to cycle, while spotted variables exhibit much
slower variability of the amplitudes. In most cases, the well-covered,
long-term OGLE light curves allowed us to unambiguously separate spotted
stars from pulsating variables. In total, we found over 20\,000 candidates for spotted stars toward 
the Galactic bulge in the OGLE-IV database. From all of the spotted stars found 
in the OGLE-IV data, we have chosen 12\,660 objects which were also observed 
during the OGLE-III phase because the interstellar extinction 
maps are available for these stars \citep{Nataf13}. Stars observed
only during the fourth phase of the OGLE project were not taken into account
as the interstellar extinction is not yet available for them.
The final sample contains stars with clear magnetic field manifestations,
including classical dark spots, peculiar chemical spots or even flares.
In Figure \ref{fig1}, we present examples of four light curves of spotted stars from our collection.

\begin{figure}[h!]
\begin{center}
\includegraphics[scale=0.16]{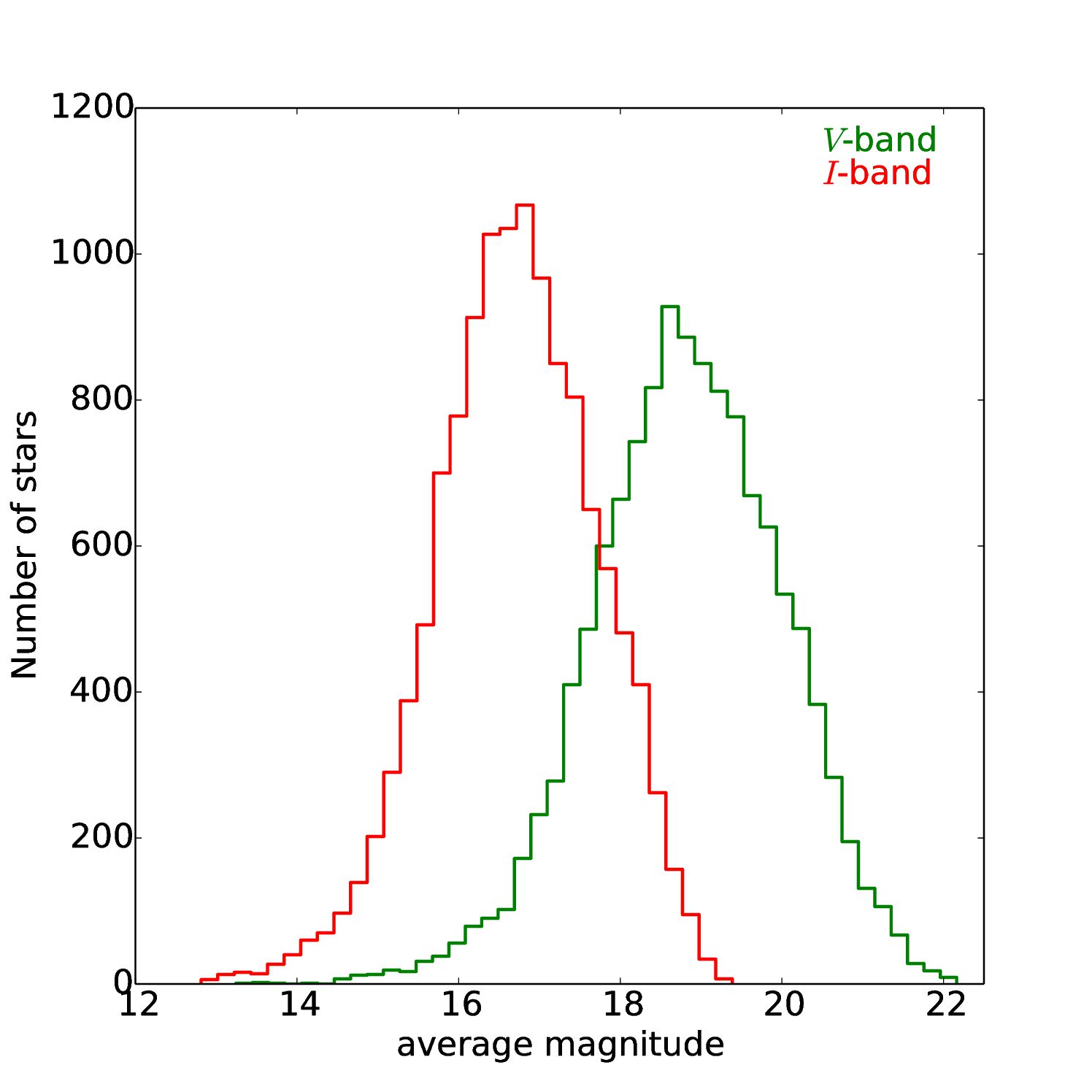}
\caption{Histogram of average brightness in \textit{I-} and \textit{V-}band for spotted stars.}
\label{fig3}
\end{center}
\end{figure}

In Figure \ref{fig3}, we present a histogram of the apparent mean
brightness of the spotted stars in 
the {\it I}- and {\it V}-band filters. It is important to note that
beyond the maximum number of stars at $16.7$ mag in the
\textit{I}-band the number of objects decreases rapidly down to a
limiting value of about 
$19.4$ mag, where no more stars with spots were detected. In the case of 
eclipsing binaries observed toward the Galactic bulge, the most common 
brightness value is about $18.4$ mag \citep{Soszynski16}.
Moreover, the OGLE photometry is complete and very accurate up to $18.7$-$19.5$ mag
in the {\textit{I}} passband (depending on the crowding; see Fig. 17 in \citealt{Udalski15}). 
This may suggests that stars 
with strong magnetic activity are statistically brighter than
stars that do not show such activity. However, we 
cannot rule out a possibility that this may be a selection effect as we did not calculate 
the detection efficiency neither for eclipsing nor for spotted stars. For the central regions of 
the Milky Way bulge the limit of brightness measurement of the Warsaw telescope is 
$I \approx 21$~mag \citep{Udalski15}.

\newpage
\section{Period searching}

The methodology of looking for periodicity in
time-series data is very complex and not trivial.
There exists no fully automated method
that would give periods at $100\%$ level of certainty. Therefore, our method
of searching for periods is semi-automatic. This
means, that we first calculate a periodogram for
each star, from which we pick the most significant
period based on the signal-to-noise ratio. 
In the second step, each light curve folded with that period
has been carefully verified visually. If we decide
that the found period is incorrect, it is corrected 
manually. This operation is repeated iteratively until
a satisfying period is found. In the case when we
are not able to find the reliable period, we exclude
the star from further analysis.

In our analysis we use \textsc{Fnpeaks} code\footnote{http://helas.astro.uni.wroc.pl/deliverables.php?active=fnpeaks}
based on the standard Discrete Fourier Transform modified 
for unevenly spaced data \citep{Kurtz85}.
On the other hand, one of the most widely used period 
searching method in astronomy is Fourier-like 
least-squares spectral analysis proposed by
\citet{Lomb76} and \citet{Scargle82}. Recently, this
method was discussed in great detail by \citet{VanderPlas18}. 
Here we compare the results obtained from both methods.

Before computing the periodograms, we remove the long-term trends
from the data using splines. Afterwards, for each star we compute high-resolution power spectra using both methods.
We searched a frequency space from $0$ to $10$~d$^{-1}$, with a step of $0.000001$~d$^{-1}$. To compare the results from both methods we introduce three period symbols: $P_{\mathrm{FNN}}$ is the period obtained using \textsc{Fnpeaks} code, without visual inspection of light curves, $P_{\mathrm{FN}}$ is the period obtained using \textsc{Fnpeaks} code but adopted after visual verification and $P_{\mathrm{LS}}$ means period obtained with Lomb-Scargle algorithm.

\begin{figure}
\begin{center}
\includegraphics[scale=0.24]{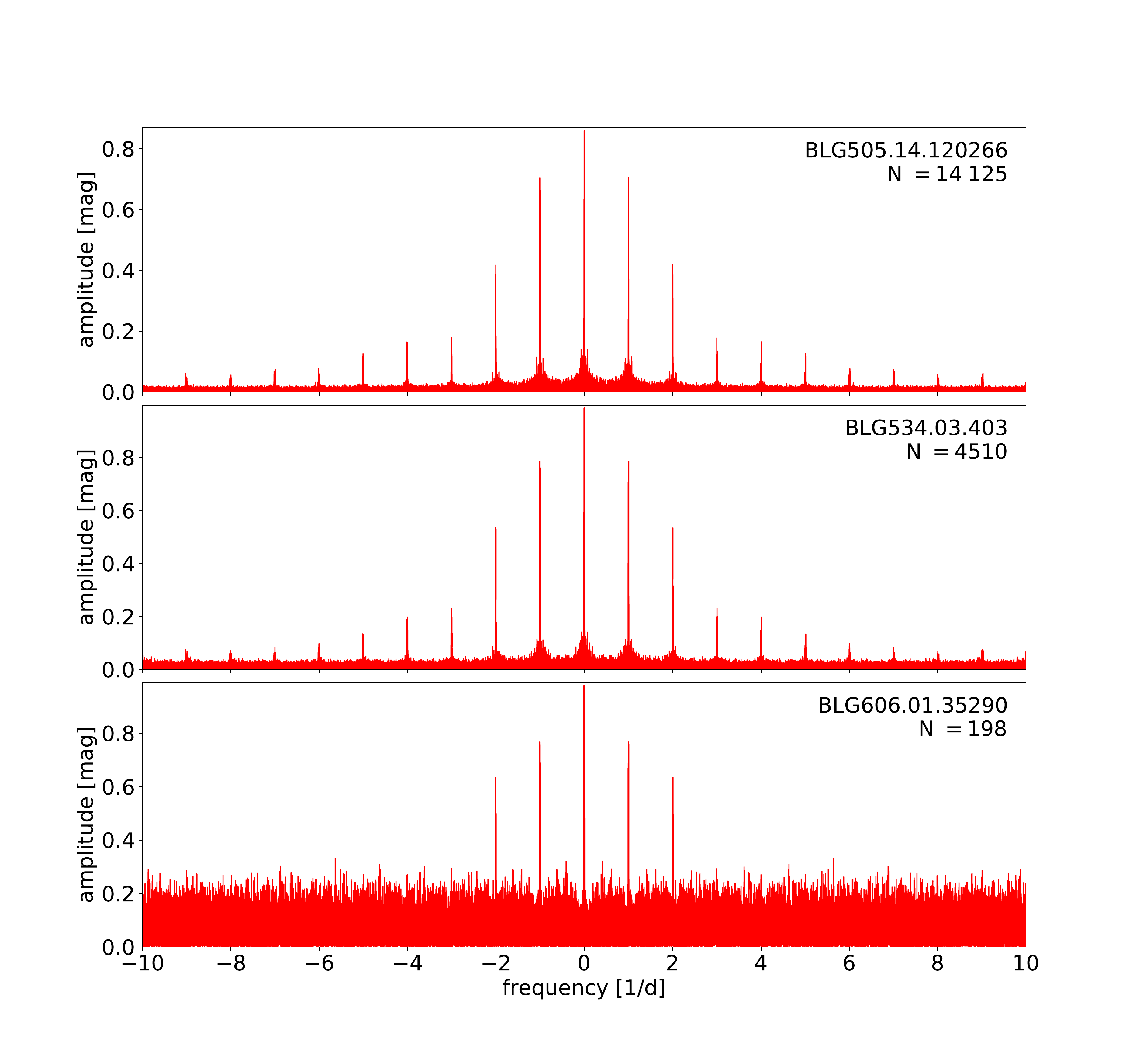}
\caption{Spectral window function for typical OGLE cadences: 
the most frequently observed field (top panel), the moderately 
covered field (middle panel) and the poorly
covered field of our sample (bottom panel). 
By N we denote the number of observations for a given star.}
\label{window-function}
\end{center}
\end{figure}

\begin{figure}
\begin{center}
\includegraphics[scale=0.11]{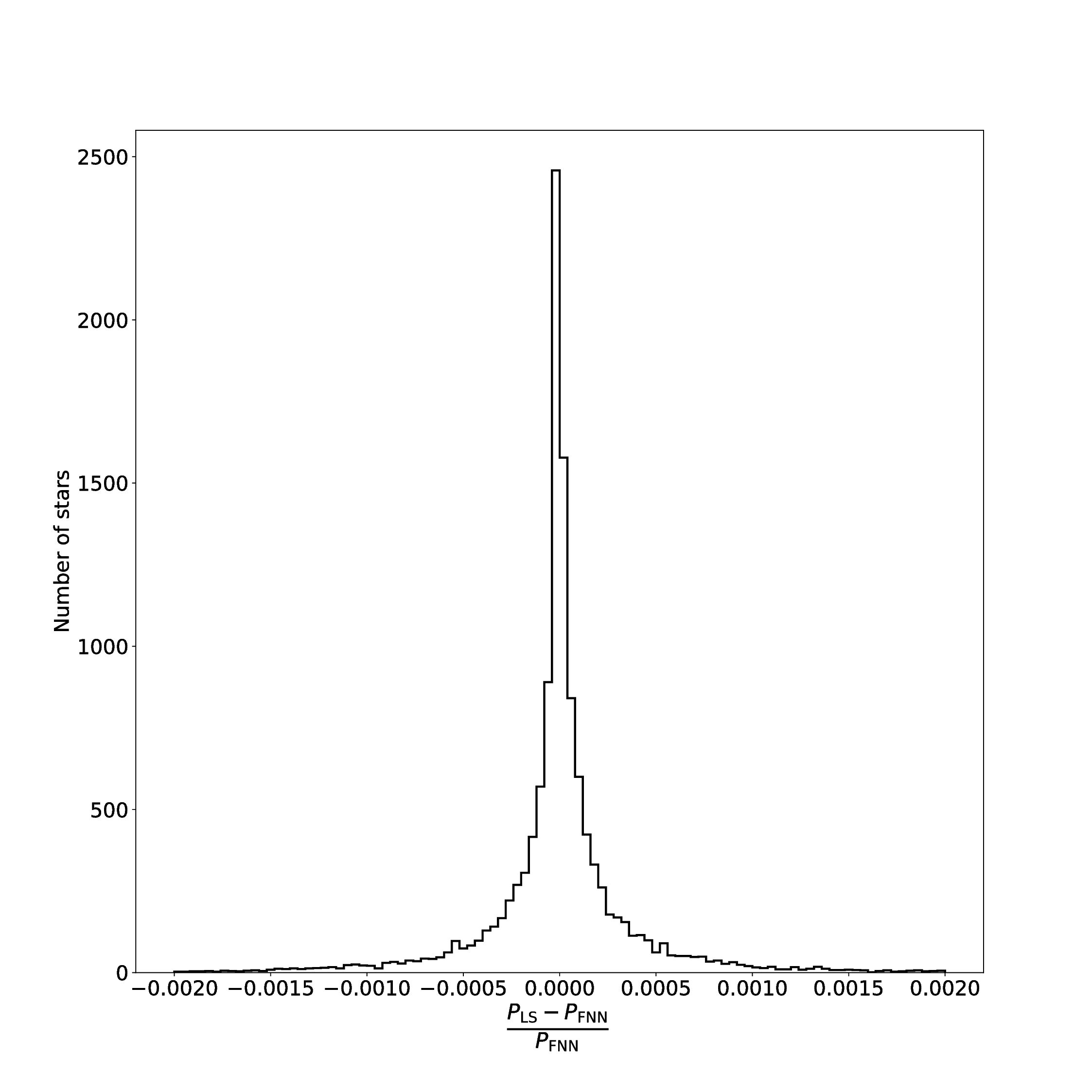}
\caption{Histogram of the relative differences between the
periods obtained using different methods. $P_{\mathrm{LS}}$ means the period found 
using Lomb-Scargle algorithm, while $P_{\mathrm{FNN}}$ denotes the period found 
with \textsc{Fnpeaks} code without visual inspection.}
\label{periods-comparison-hist}
\end{center}
\end{figure}

\begin{figure}[h!]
\includegraphics[scale=0.12]{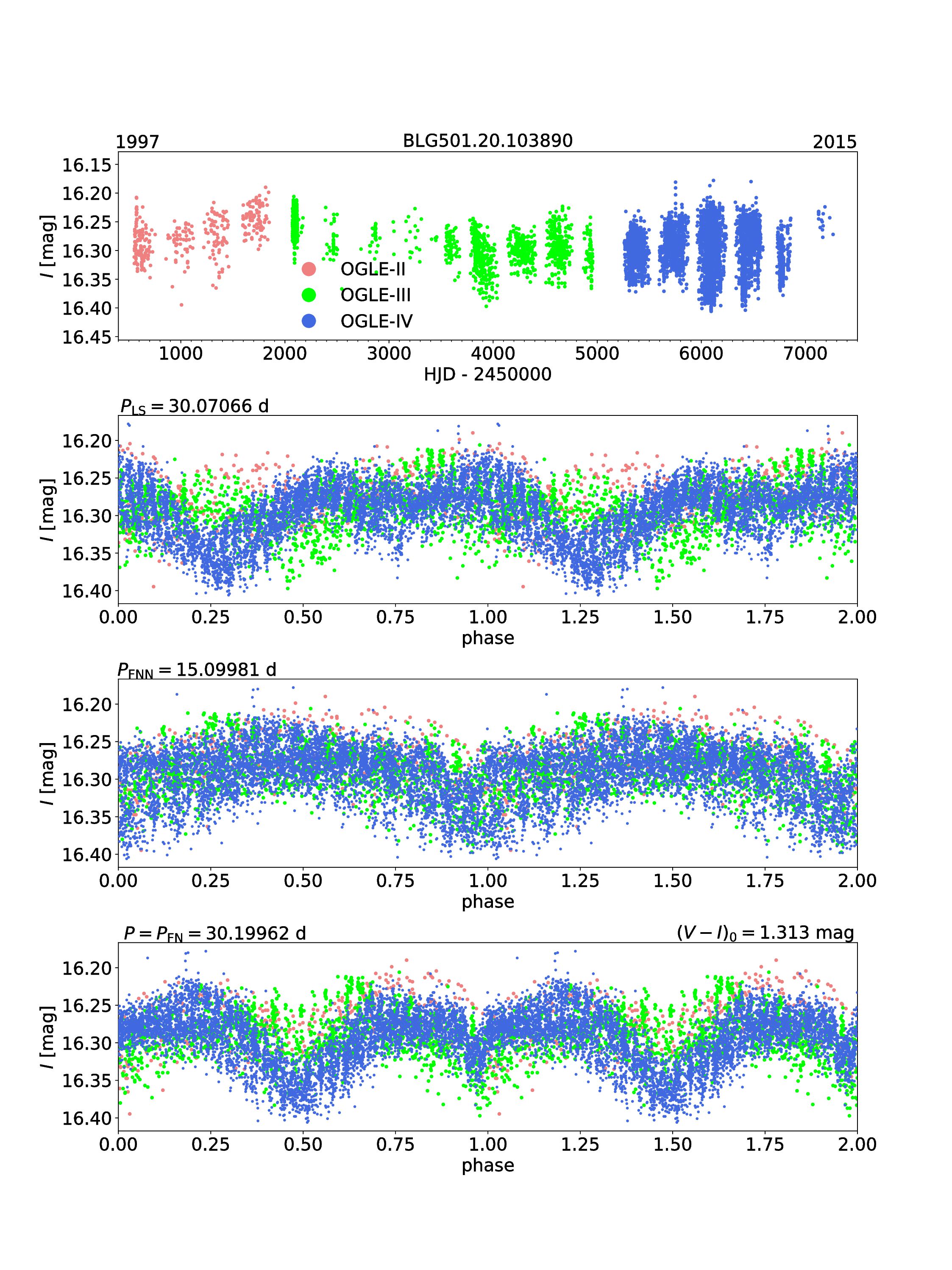}
\caption{An example of a star, for which both period
searching methods failed. The first panel presents the
unfolded light curve. The next three panels
show phase-folded light curves with the found rotation periods 
$P_{\mathrm{LS}}$ (second panel), $P_{\mathrm{FNN}}$ (third panel) and $P_{\mathrm{FN}}$
(fourth panel). The meaning of point colors and dates above the
plot is the same as in Figure \ref{fig1}. Additionally, each
phase-folded light curve is plotted with different colors
related to different phases of the OGLE project. Above the fourth panel we
also provide the color index $(V-I)_0$ corrected for the interstellar extinction.}
\label{periods-comparison-harmonic}
\end{figure}

\begin{figure*}[h]
\begin{center}
\includegraphics[scale=0.105]{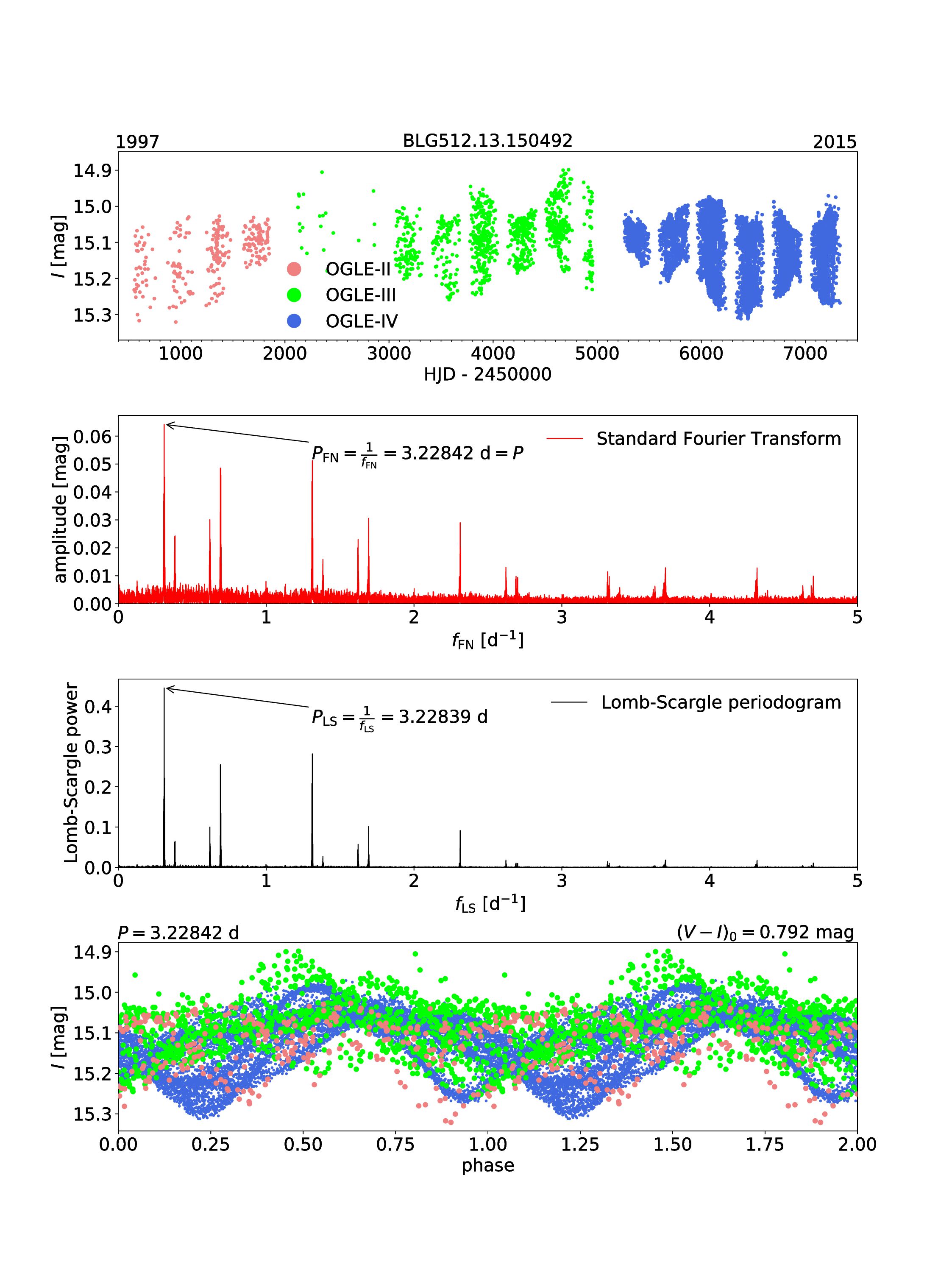}
\includegraphics[scale=0.105]{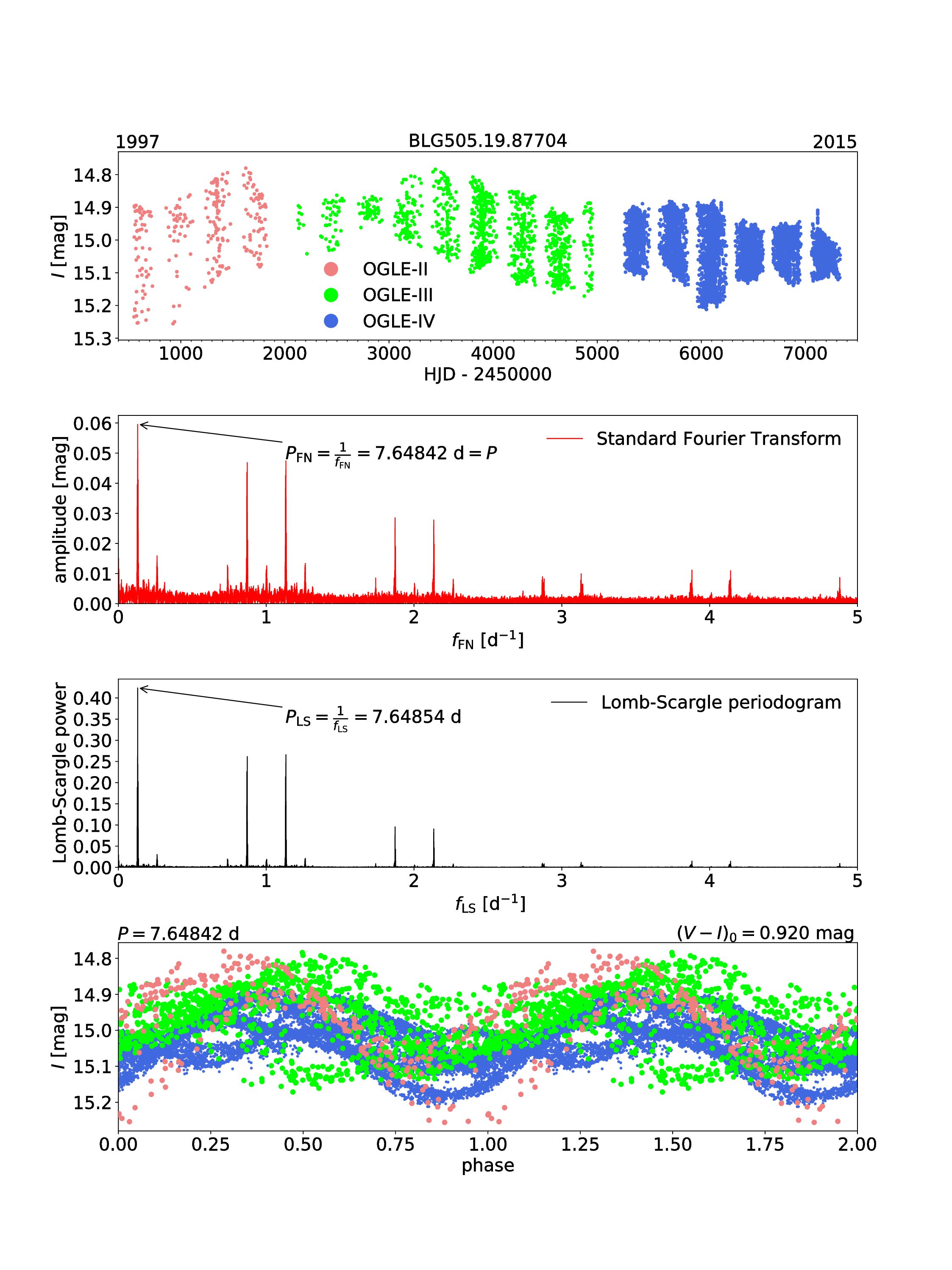}
\includegraphics[scale=0.105]{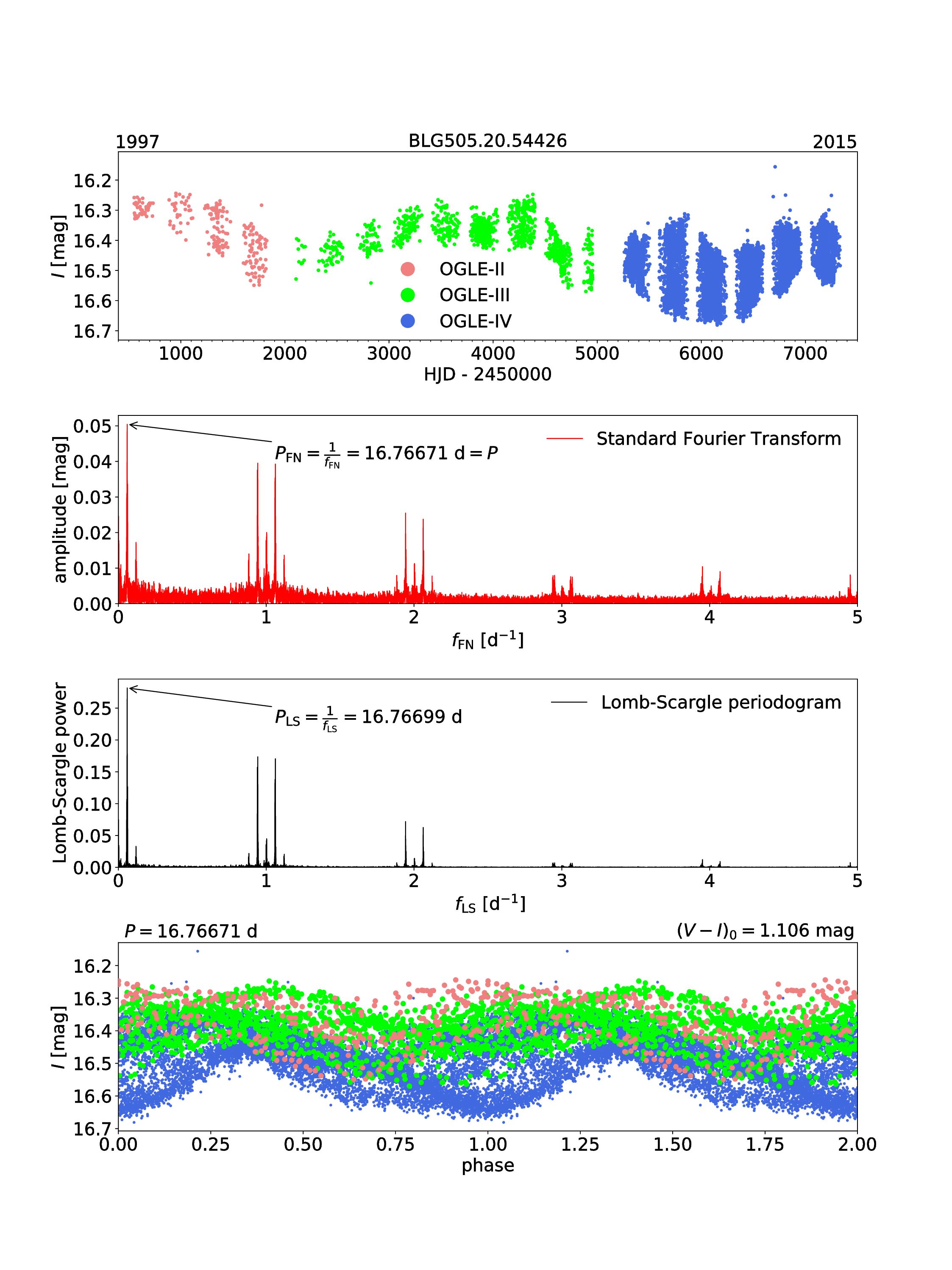}
\includegraphics[scale=0.105]{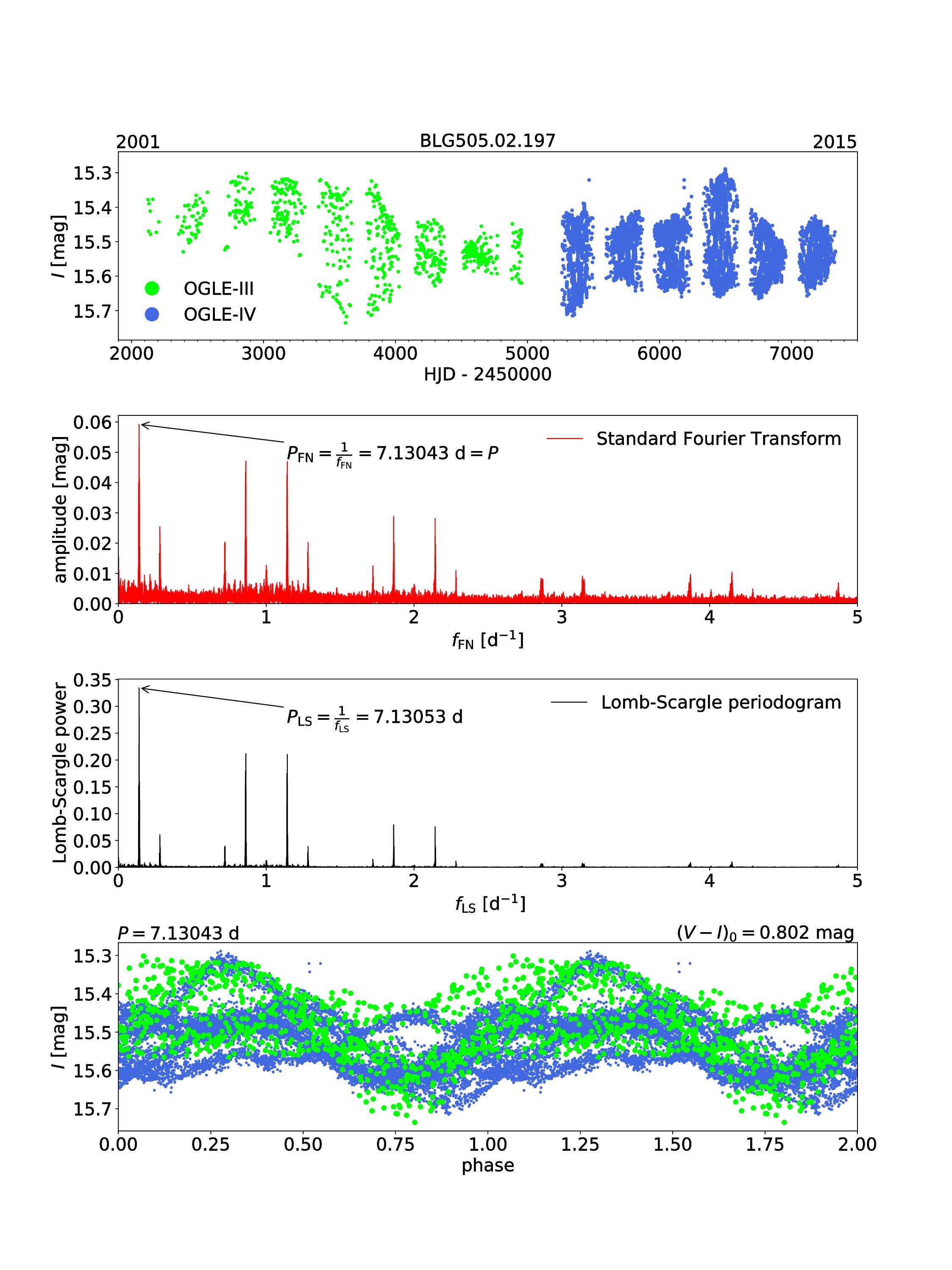}
\caption{Examples of four stars, which show a clearly visible, long-term (from
months to years), quasi-periodic variability due to the evolution and
migration of spots, and activity cycles. The first and last panels on
each plot present time-domain and phase-folded light curves,
respectively. The colors of points and dates above the panels have the
same meaning as in Figure \ref{fig1} and Figure \ref{periods-comparison-harmonic}. The second and third panels
on each plot present typical periodogram patterns, which result from the
stellar rotation combined with the more complex, quasi-periodic, long-term
behavior. The periodograms on the second panels are obtained using
\textsc{Fnpeaks} code, while the third panels show periodograms obtained using
Lomb-Scargle algorithm. On each periodogram we marked the peak
with the highest signal-to-noise ratio by an arrow, with appropriates 
period values following from the peak frequencies. Above the
fourth panels we also provide color indices $(V-I)_0$ corrected for the interstellar extinction.}
\label{periodograms}
\end{center}
\end{figure*}

The OGLE Galactic bulge data (as other
ground-based observations) are contaminated by
two main aliases, $1$-day alias and $1$-year
alias. In addition, the fields we observe have
different cadences and coverages. The spectral
window functions, and therefore typical aliasing
structures arising from the OGLE cadences, are
presented in Figure \ref{window-function}, separately for the most
frequently observed field with the largest number of epochs 
(BLG505; top panel in Figure \ref{window-function}),
moderately covered field (BLG534; middle panel in
Figure \ref{window-function}, and poorly covered field (BLG606;
bottom panel in Figure \ref{window-function}). 
It is clearly seen, that the choice
of the period searching method may have a great importance 
for poorly covered light curves with a
low cadence, because the noise level for them
substantially increases. In such cases the Lomb-Scargle 
algorithm gains, because
it reduces noise, as the noise is chi-square
distributed under the null hypothesis \citep{VanderPlas18}. 
However, it must be noted that our least
covered light curves have less than $200$ epochs
(only $6$ objects out of $12\;660$), while the vast majority 
of spotted stars have more than $500$ observations ($11\;790$ objects). 
More than a quarter of the objects have over $4000$ epochs, and nearly $2000$
of them have a light curve coverage at the level
of more than $10\;000$ epochs.

In Figure \ref{periods-comparison-hist}, we present a comparison of periods 
calculated using both applied methods. We find that for $148$ stars the relative difference 
between found periods are higher than $1\%$. 
This difference is related to the fact that one
of the methods reports $1/2$x or $2$x harmonics of
real period, or one method reports the real period
while the other reports a $1$-day alias. One example
of this behavior can be seen in the light curves
of the star BLG501.20.102890 presented in~Figure~\ref{periods-comparison-harmonic}. 
Please note, that in this case both methods give
wrong periods. Thanks to the careful visual examination
of the light curve, and manual correction ($P_{\mathrm{FNN}}$ period), 
it can be clearly seen that the star is a spotted eclipsing binary. On the other hand, in such binaries we
expect to find at least two periods -- one orbital, resulting 
from eclipses (if inclination is favorable), and the other, resulting from the rotation of the unevenly
spotted primary component. We also expect that these two periods may be 
synchronized, or almost synchronized. Indeed, after substracting 
from the light curve the orbital period, it is possible to find a rotation period of the
spotted primary star. This period was found by
Lomb-Scargle algorithm and is presented in the
second panel of Figure \ref{periods-comparison-harmonic}. 
Looking at the light curve in time-domain, one can see that 
a long-term modulation of the mean brightness
and amplitude is visible due to activity cycle, which
produces additional peaks in low frequencies. All
these cases show that a visual inspection of the
light curve is an indispensable method for examining in 
great detail the nature of a star, and
all fully-automatic methods can fail for stars with
such a complicated, fast-evolving variability.

For other stars from our sample the relative difference between the periods 
found from different methods is much smaller than $1\%$, with
the average value at the level of accuracy of period searching methods. 
In general, there are $751$ object in our collection for which both methods give 
different periods by a factor of $1/2$ or $2$.

The amplitudes of brightness and the mean brightness of
spotted stars vary rapidly in time due to the spots
evolution and activity cycles. Typical time-scales
of such changes are from months to several years. In Figure \ref{periodograms},
we present four stars from our collection, with
such a variability. Beside the light curves (in
time-domain and phase-folded) we also
show in Figure \ref{periodograms} typical power spectra resulting from 
stellar rotation combined with the more complex quasi-periodic,
long-term brightness variations.

Both methods of period searching have advantages and disadvantages. 
However, thanks to the fact that our final
method is not fully automated, there is a really
small chance that some periods reported by us
are not real. Of course, we cannot guarantee
that each of the found period is perfectly correct.
Nonetheless, we are convinced that the number of the misidentified periods is practically negligible and
does not affect at all our statistical studies of spotted stars.
In the further analysis we use periods found using \textsc{Fnpeaks} code $P_{\mathrm{FN}}$ and we name them $P$ throughout the paper.

\begin{figure}[h!]
\begin{center}
\includegraphics[scale=0.15]{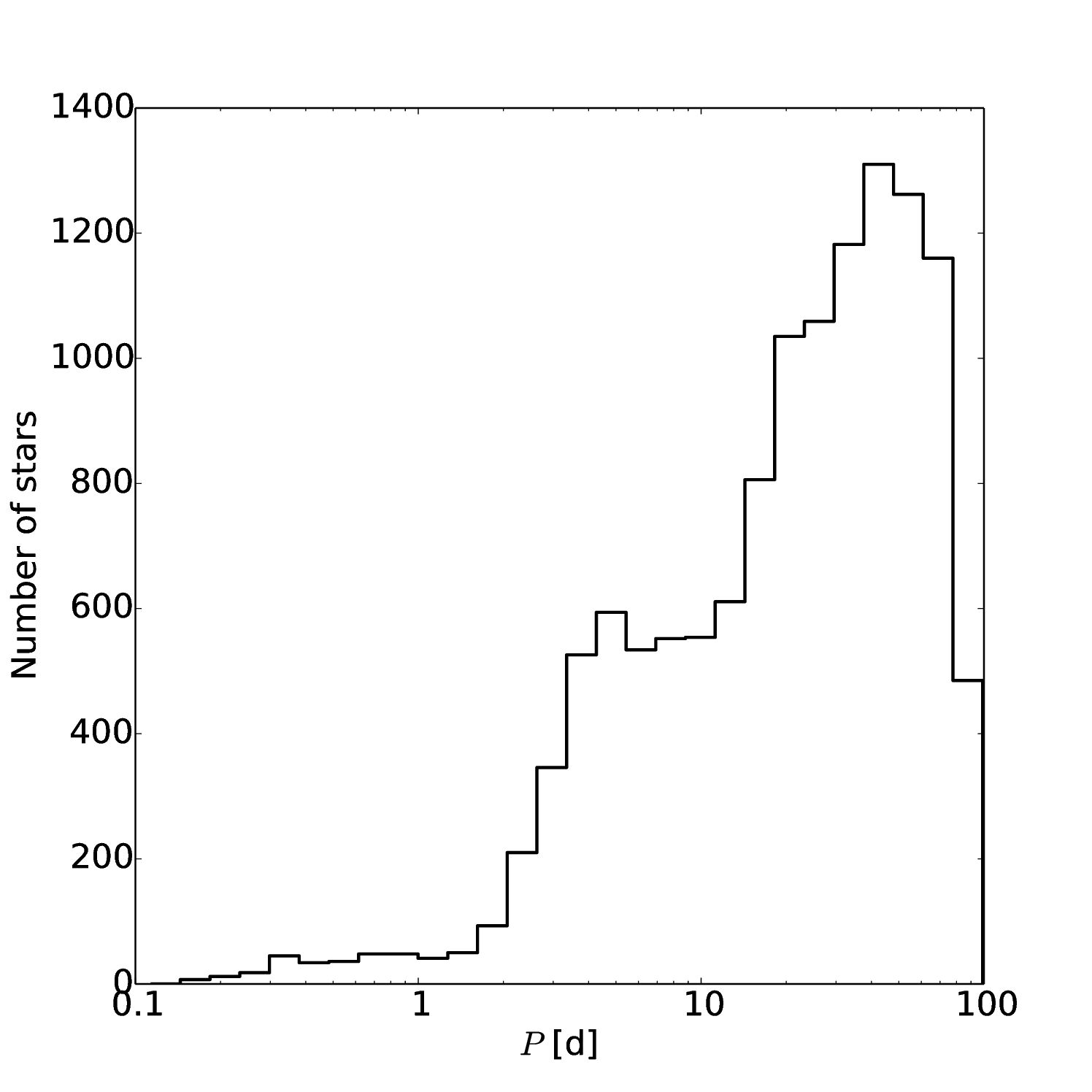}
\caption{Histogram of rotation periods of spotted stars.}
\label{fig2}
\end{center}
\end{figure}

In Figure \ref{fig2}, we present a histogram of rotation periods 
for the spotted variables from our collection. The shortest rotation period in our sample
is 0.113 d and the longest one is 98.951 d. Such a
large range of periods suggests that the presence
of magnetic activity does not depend on the rotation velocity. 
However, the strength and form of the activity caused by the magnetic field may
depend on how fast the star rotates.

\newpage
\section{Dereddening procedure for spotted stars} 
\label{chap:color_calculations}

The Galactic bulge is one of the most challenging regions to explore in the Milky 
Way, as the light of stars observed in this direction is
substantially and unevenly reddened due to 
irregular interstellar clouds of dust. Extinction and reddening can significantly
change brightnesses and colors of stars on a small angular scale (see {\em e.g.} \citealt{Udalski02, Szymanski11}). 
The OGLE observations are carried out mostly in the \textit{I}-band. Observations in the \textit{V}-band are much 
less frequent, but when we observe in the \textit{V}-band, we usually obtain a corresponding
\textit{I-}band measurement just before or after frame in \textit{V} bandpass. 
The majority of observations were made during the OGLE-IV phase,
so for the color and mean brightness calculations we use the data just from this phase. 
The brightness of many spotted stars is rapidly changing, so we choose pairs of brightness 
measurements in both filters taken within 0.1~day and we calculate the observed color indices $(V-I)$.
Typically, there are a few dozen of $(V-I)$ measurements per star, and this number
corresponds to the number of measurements in the {\textit{V-}}band. 
For every star, we calculate the median value of $(V-I)$ index and we reject all points 
deviating more than $\pm 3\sigma$ from the median. After removing outlying measurements
we compute the $(V-I)$ median again.

For the majority of stars the extinction-free brightness $I_0$ and color index $(V-I)_0$ 
can be calculated with extinction $A_I$ and reddening $E(V-I)$ estimated from the colors of
red clump stars by \citet{Nataf13}:

\begin{equation}
I_0 = I - A_I \newline,
\label{eqn:I0}
\end{equation}

\begin{equation}
(V-I)_0 = (V-I) - E(V-I).
\label{eqn:VI0}
\end{equation}

\begin{figure*}
\begin{center}
\includegraphics[scale=0.33]{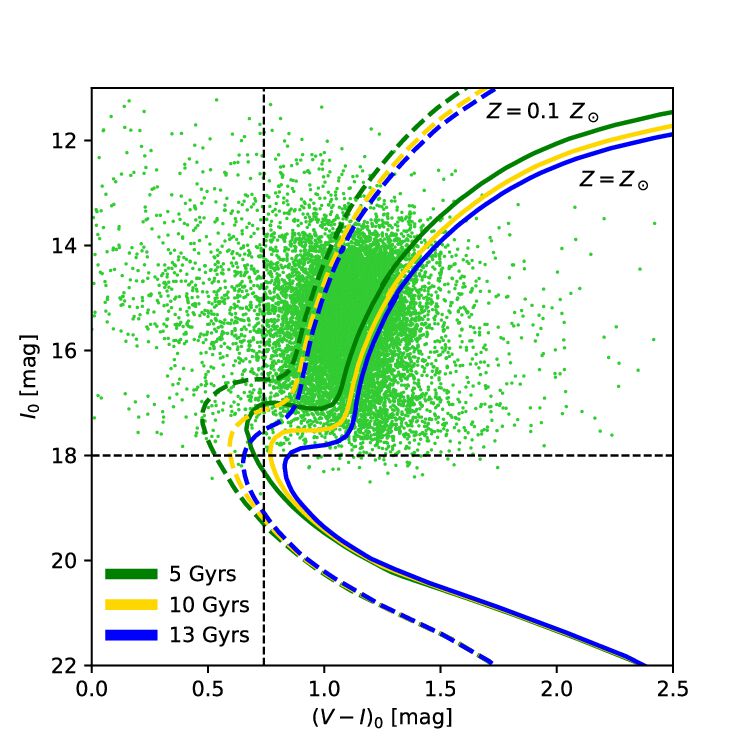}
\includegraphics[scale=0.33]{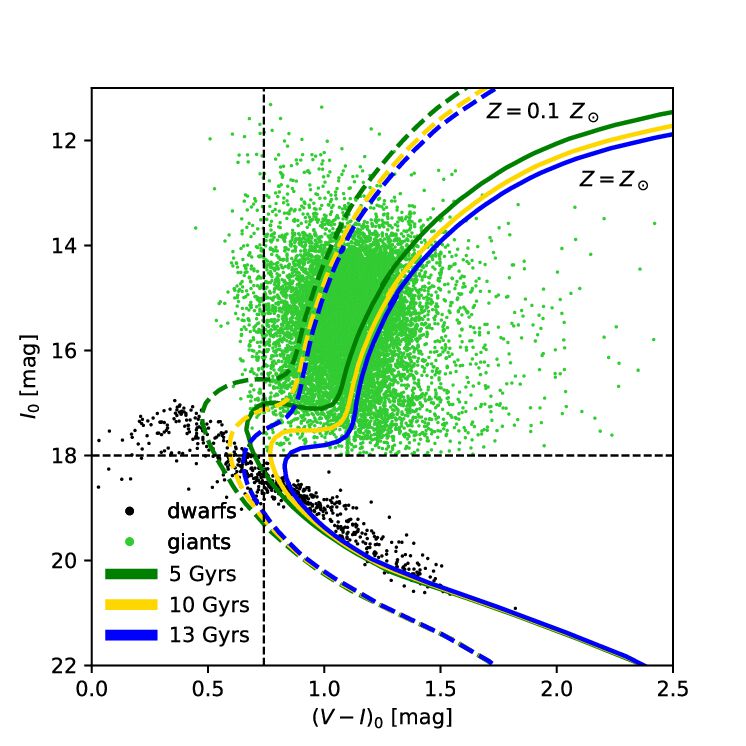}
\caption{Dereddened color--magnitude diagrams (CMD) for the analyzed 
spotted stars in our sample, marked with green dots. In both panels, we plot 
PARSEC isochrones \citep{Bressan12} for solar metallicity ($Z = Z_\odot = 
0.0196$; \citealt{Steiger16}; marked with solid lines), and for $10\%$ of the 
solar metallicity (marked with dashed lines). With colors we marked three 
different ages of stars.
{\textit{Left~panel:}} Dereddened CMD using $A_I$ and $E(V-I)$ calculated by \citet{Nataf13}. 
By dashed black lines we marked a region dominated by the foreground objects for which reddening is
overestimated (top-left). The vertical dashed line is set at $(V-I)_0 = 0.74$ mag, what is the center of RGB region 
shifted by $2\sigma$ toward bluer colors (for RGB $(V-I)_0 = 1.10 \pm 0.18$ mag). The horizontal dashed 
line presents the lowest brightness of giants located in the Galactic bugle, 
${M_{I}} = 3.5$ mag (distance modulus $\mu_0 = 14.5$ mag). 
{\textit{Right panel:}} Dereddened CMD using our new
method. Dashed vertical and horizontal lines have the same meaning as
in the left 
panel. Green dots correspond to giants and black dots to dwarfs.}
\label{fig4}
\end{center}
\end{figure*}

\noindent However, if we assume that the light from all stars experience total 
reddening and extinction observed toward the red clump stars in the 
bulge, we would certainly overestimate the dust influence on a fraction of the 
stars lying closer to us, in the Galactic disk. The results of dereddening using the interstellar extinction map 
made by \citet{Nataf13}, are presented in the left panel of Figure \ref{fig4}.

\begin{figure*}
\begin{center}
\includegraphics[scale=0.30]{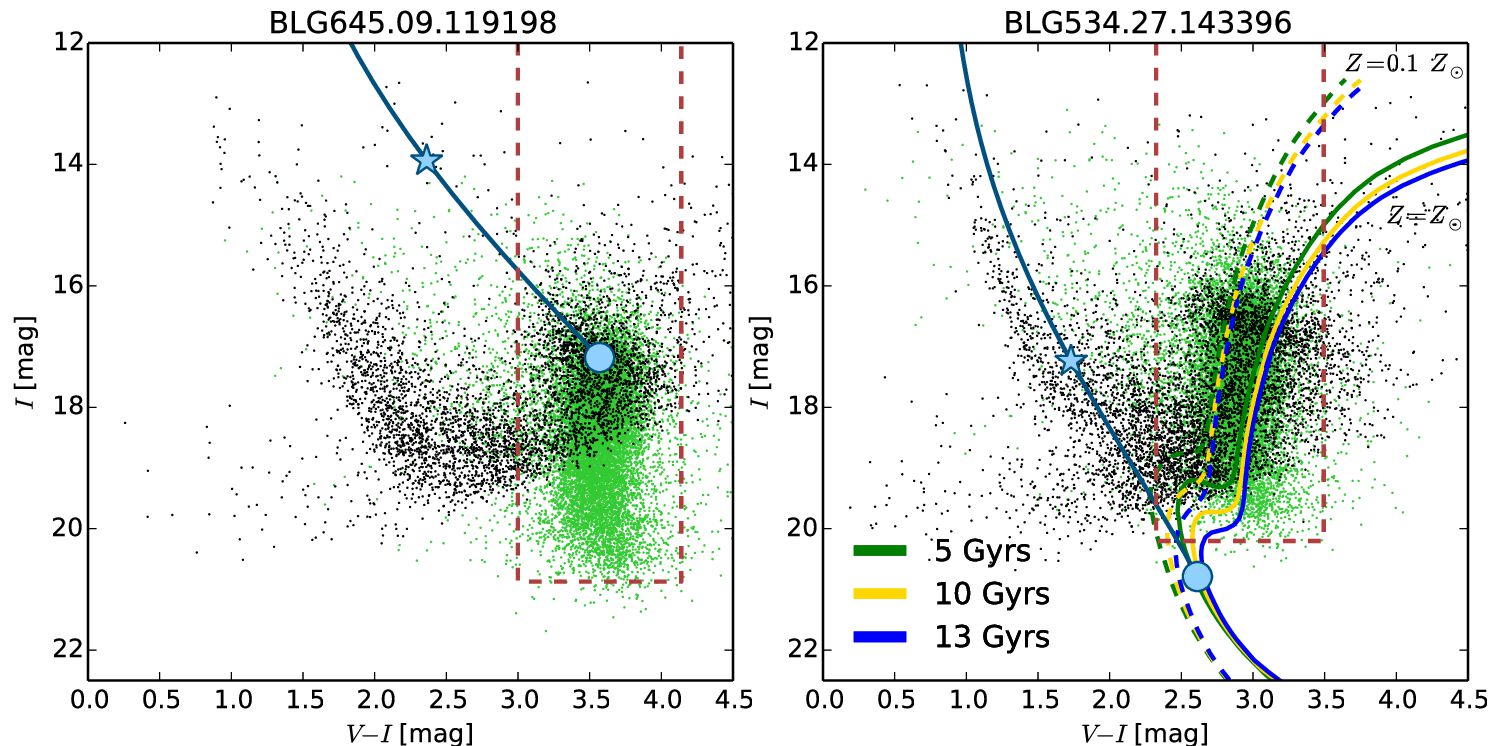}
\caption{Visualization of the dereddening procedure for two cases in the color--magnitude diagrams.
Isochrones are plotted in the same manner as in Fig. \ref{fig4}.
The blue star marks our analyzed spotted object. The solid blue line represents the reddening curve.
The blue dot marks the star location after dereddening
procedure. Green points show
all the spotted stars from our sample. Black points mark a sample of stars from the neighborhood
in the direction toward the analyzed object. Red dashed rectangle presents RGB region
defined by $(V-I)_0 = 1.10 \pm 2 \sigma \pm \sigma E(V-I)$, $M_{I} < 3.5$ mag. {\textit{Left panel:}}
A case when the spotted star is clearly a foreground giant. In this
case we set the color index
for this star at $(V-I)_0 = 1.10$ mag and the brightness $I_0$ as it appears
from the reddening curve. {\textit{Right panel:}} A case when the spotted star is a dwarf.
In this case we estimate $(V-I)_0$ and $I_0$ from isochrones.
It is worth mentioning, however, that both $I_0$ and $(V-I)_0$, 
adopted for dwarfs are very uncertain.}
\label{fig5}
\end{center}
\end{figure*}

The majority of analyzed stars fall into
the Red Giant Branch (RGB) region. We can describe this region in the
color--magnitude diagram (CMD)
by the average color index $(V-I)_0 = 1.10 \pm 0.18$ mag and the absolute magnitude
$M_{I} < 3.5$ mag. We assume that stars lying within the RGB 
region are in fact giants from the Galactic bulge, and that their reddening is as 
resulting from the interstellar extinction map presented by \citet{Nataf13}. 
Stars that appear bluer might be foreground giants or dwarf stars, and 
the amount of reddening toward them should be smaller than calculated according to \citet{Nataf13} for a given line of sight. 
We assume that all stars bluer than $2\sigma$ from the typical RGB
color $(V-I)_0 = 1.10$~mag, {\em i.e.} bluer than  $(V-I)_0 = 0.74$~mag
(vertical dashed line in both panels of Fig. \ref{fig4}), are potential foreground objects (objects located
in the top-left part of the CMDs in Fig. \ref{fig4}). In fields with large differential reddening we shift this boundary
by $\sigma E(V-I)$ as calculated by \citet{Nataf13}.

We simulate extinction between the Sun and 
the analyzed spotted star assuming that the extinction is proportional
to the dust density $\rho_{\mathrm{dust}}$, which can be modeled by a double exponential disk:

\begin{equation}
\rho_{\mathrm{dust}} \propto \exp \left(-\frac{R-R_0}{h_R}\right) \exp \left(-\frac{|z|}{h_z}\right),
\label{eqn:sharma}
\end{equation}

\noindent where $R$ is distance from the Sun to the analyzed star, $R_{0}$ is distance from the Sun to the Galactic bulge and $z$ is vertical distance of the analyzed object from the Galactic plane.
We use the length and height scales ($h_R$ and $h_z$, respectively) as calculated by \citet{Sharma11}
with values $h_R = 4200$~pc and $h_z = 88$~pc. These parameters
were found by the authors using \citet{Schlegel98} dust maps in the Milky Way. We normalize the 
dust density $\rho_{\mathrm{dust}}$ measured in the given direction to the total reddening
as measured by \citet{Nataf13}.

Using the above extinction model, we are able to spread out the entire reddening $E(V-I)$
along the line of sight toward the star with unknown reddening. 
This allows us to move the star along the reddening curve and estimate what 
brightness and color it would have, if it were located in the bulge.
With this procedure, we can assess if this is a foreground giant
or rather a dwarf star.

To decide whether the star could be a foreground giant, we take the least luminous
RGB star from the Galactic bulge and shifts this object along the reddening curve
toward us. If the observed star is more luminous than the shifted RGB bulge star
we assign to this star $(V-I)_0 =~1.10$ mag and brightness $I_0$ according to the
reddening curve.

Otherwise, the analyzed star must be a foreground dwarf. We can estimate the dwarf
color and brightness based on the isochrones limited to the Main Sequence (MS) stage,
where dwarfs spend the majority of their lives. We use PARSEC isochrones \citep{Bressan12}
for the solar metallicity $Z = Z_\odot =
0.0196$; \citep{Steiger16}, but see \citet{Serenelli16}, $10\%$ of the solar
metallicity and for three ages: 5~Gyrs, 10~Gyrs, and 13~Gyrs. We place these isochrones
in the bulge, and we shift them along the reddening curve toward the observer until
they cross the star's position on the CMD. The coordinates of this intersection are our
searched extinction-free brightness and color for the analyzed dwarf. 
However, it is worth mentioning that this approach is affected by large uncertainties
and only allows us to state if the star is indeed a dwarf.

We crossmatched our list of the spotted stars with general catalog of all sources 
published by \citeauthor{Gaia16} (\citeyear{Gaia16, Gaia18}) in the {\it Data Release 2} using angular radius equal to 0.4 arcsec. We found, that $11\;915$ out of $12\;660$ objects from our list have measurements in the Gaia database. Unfortunately, only $24$ stars have significant parallaxes with $\sigma_{\varpi}/{\varpi} < 0.05$, $108$ sources have measured parallaxes with $\sigma_{\varpi}/{\varpi} < 0.1$, and $1465$ objects have parallax measurements with $\sigma_{\varpi}/{\varpi} < 0.33$. This means that some stars from our sample are certainly
located closer than the Galactic bulge. A fair fraction of objects found in the Gaia catalog have no parallax 
measurements, or have negative parallaxes ($5296$ stars), and
the remaining objects from our list have insignificant 
parallax entries in the Gaia database.

The right panel of Figure \ref{fig4} shows CMD after applying our new
dereddening method. This approach allows us to classify $676$ stars as dwarfs, and
$11\;984$ objects as giants. Two cases of our dereddening approach discussed above
are presented in Figure \ref{fig5}. It is noteworthy, however, that while this dereddening
procedure is acceptable for statistical analysis of large sample of stars, it may overestimate
brightness and color of individual objects. This can be especially important for stars observed
toward directions where high inhomogeneities of the interstellar dust
distribution occur.

\section{General properties of spotted stars} \label{chap:properties}

Two largest, up-to-date, statistical analyzes of spotted stars were obtained
by \citet{Drake06}, who discussed the properties of $~3000$ giants and subgiants
toward the Galactic bulge and \citet{Lanzafame18b}, who analyzed $~150\;000$ BY Dra
candidates detected in the Gaia DR2 data. In the present paper, we 
conduct a similar analysis as \citet{Drake06} but based on more than four times 
larger sample. A significantly larger set of examined stars and the long-term 
OGLE photometry allow us to examine more precisely correlations presented by 
\citet{Drake06} and to find new, previously unknown dependencies.

Below we discuss statistical properties of spotted stars.
In most of the diagrams presented in this work we bin stellar parameters along 
x-axis. Usually, we calculate the mean values in each bin along the x-axis, and 
the median values along the y-axis. Diagrams with rotation periods are presented 
in a logarithmic scale. For the binning along the rotation periods, we require at 
least 50 points in each bin and a bin width of at least $\log{(P)}~=~0.1$ dex. 
For the binning along the amplitudes we also require at least 50 points, but 
the width of the bins was $0.02$ mag. We mark these averaged values by
large purple dots
with appropriate error bars. Typically, error bars are smaller than, or equal to
the size of the used symbols.

\subsection{Location of spotted stars in the sky}

\begin{figure}[h!]
\begin{center}
\includegraphics[scale=0.30]{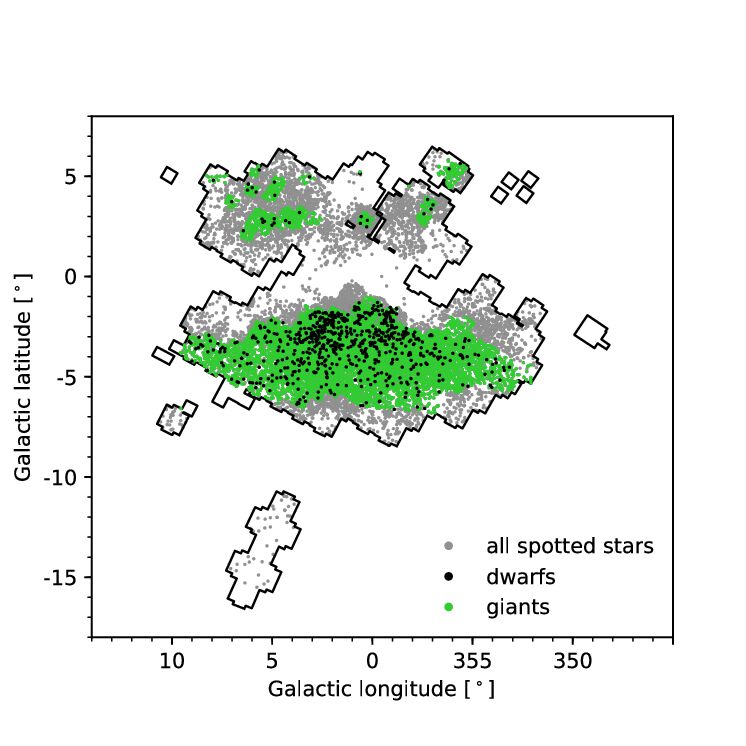}
\caption{Distribution of spotted stars in the sky toward the Galactic bulge. 
Black contours correspond to the OGLE footprint. Gray points denote all spotted 
stars discovered in the OGLE-IV data. Black and green points denote all stars analyzed
in this paper coinciding with OGLE-III footprint. With green points we marked 
stars classified as giants, while black dots present objects classified as 
dwarfs.}
\label{fig6}
\end{center}
\end{figure}

In Figure \ref{fig6}, we present the location of analyzed spotted stars in the 
sky. The black contours show the outline of combined OGLE-II, OGLE-III and OGLE-IV fields 
toward the Galactic bulge. The Galactic bulge area is dominated by giants, which fill 
almost the whole area observed by OGLE. The great majority of these
stars is located in the bulge itself. 
However some of them are, very likely, foreground Galactic disk objects.
The central region of the Galaxy is impossible to observe in the optical 
range due to huge interstellar extinction in this part of the Milky Way, hence 
there is a lack of detections near the plane.

\subsection{\textit{I}-band amplitude vs. rotation period -- the final division into dwarfs and giants} \label{chap:amplitudes}

The amplitudes of light curves of active stars vary significantly with time
as a result of spots evolution. These variations are evident in the phased 
light curves presented in Figure~\ref{fig1}. Therefore, before determining the 
average amplitudes over the time span of the OGLE project, we removed outlying 
measurements deviating more than $\pm 4 \sigma$ from the mean brightness. A 
common approach to determination of the amplitude of brightness variations assumes fitting
the Fourier series to the phased light curves. However, our 
tests show that this approach may produce spurious results for stars
with a quickly changing average brightness. Using a second-order Fourier series, we find
amplitudes of many light curves overestimated or underestimated.
Similar results are obtained for higher or lower orders of the Fourier series. For this reason, 
we propose here another method of measuring amplitudes, which is free
of the influence of average brightness variations.

We first transform each light curve into bins containing at least $200$  
measurements and covering at least $5$ cycles of rotation. \citet{Mathur14} found that the
length of the bins equal to $5$ rotation cycles is optimal for active stars. In the next step 
we calculate for each bin the difference between $95$th percentile and the $5$th 
percentile of \textit{I-}band magnitudes. This approach is similar to the activity proxy $R_{\mathrm{var}}$
defined by \citet{Basri11, Basri13} for {\it Kepler} stars.
The final amplitude is defined as a median of all measurements divided by the distance between these two percentiles, {\em i.e.} $0.9$.

\begin{figure}[h]
\begin{center}
\includegraphics[scale=0.2]{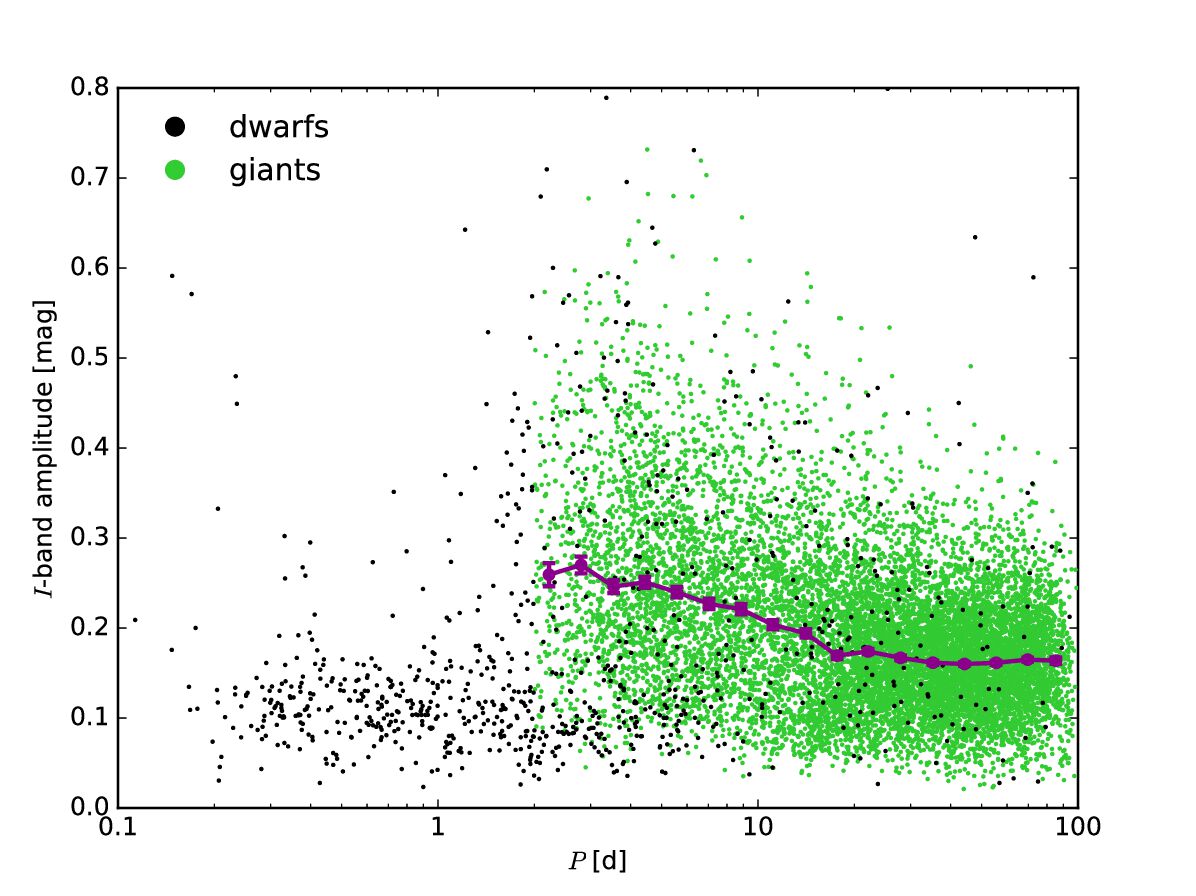}
\caption{\textit{I}-band amplitude vs. rotation period $P$. The large purple points with 
error bars indicate the medians of \textit{I}-band amplitude binned in $\log{(P)}$.}
\label{fig7}
\end{center}
\end{figure}

In Figure \ref{fig7}, we present amplitude-period plane for spotted stars from our sample.
We notice that the investigated variables can be divided into two broad
groups, depending on the rotation rate, and the amplitude of the 
brightness variations.
The first one contains fast rotators ($P \leq 2$~d) with most of the
amplitudes lower than $0.2$~mag and 
the second group contains slow rotators
with periods up to $100$~d and amplitudes up to $0.8$~mag.

Our dereddening procedure revealed that $59 \%$ stars (243 out of 415
objects) with rotation periods $P \leq 2$~d are very likely dwarfs. 
The remaining 172 stars from this group were flagged as probable
giants, yet it is commonly known that giants cannot rotate so fast
because they would be torn apart by the centrifugal force. We conclude
that all stars with rotation periods shorter than 2 days, should be
considered as dwarfs. This can be easily verified by calculating the critical
rotation period $P_{\mathrm{crit}}$ for which a star would be torn apart by the centrifugal force. Using the 
{\textit{Single Star Evolution}} code (SSE, \citealt{Hurley00}) 
we calculate the evolutionary tracks for stars with masses on the Zero Age Main 
Sequence (ZAMS) in the range $1 \leq M/M_\odot \leq 3$ with a step of $0.25$~$M_\odot$. We assume the solar metallicity ($Z = Z_\odot = 0.0196$; 
\citealt{Steiger16}) and the evolution time $T_{\mathrm{evolv}} = 12$~Gyrs. We calculate the 
radius during the first third of the time spent by the star on the RGB, when the 
radius grows steadily, weighted by the time of evolution. In addition, we neglect any mass loss.
Then we calculate the critical periods $P_{\mathrm{crit}}$ using equation 2 from \citet{Ceillier17}:

\begin{equation}
P_{\mathrm{crit}} = \sqrt{\frac{27\pi^2 R_\star^3}{2GM_\star}},
\end{equation}

\noindent where $R_\star$ and $M_\star$ are the radius and mass of the star
and $G$ is the gravitational constant. After inserting solar units ($R_\odot$ and $M_\odot$) we
obtain the following formula for the critical rotation period in days:

\begin{equation}
P_{\mathrm{crit}} = 0.213 \hspace{0.1cm} {\mathrm d} \hspace{0.1cm} \cdot 
\left(\frac{R_\star}{R_\odot}\right)^{\frac{3}{2}} 
\left(\frac{M_\star}{M_\odot}\right)^{-\frac{1}{2}}.
\label{p_crit}
\end{equation}

The resulting critical period for a one solar mass star
is equal to $P_{\mathrm{crit}} = 0.81$ d, for a two solar mass star it is equal to $P_{\mathrm{crit}} = 2.97$ d,
and for a three solar mass star it  is equal to $P_{\mathrm{crit}} = 8.92$ d. We 
see that the rotation periods around 2 days of the $172$ apparent giants
are close to, or even shorter than their critical periods. 
This is why we assume that all the objects with periods $P \leq 2$~d, 
are dwarfs. Accordingly, the final number of dwarfs increased to $848$ objects, 
while the final number of giants is $11\;812$. 

Summarizing Figure \ref{fig7}, the small-amplitude, fast-rotating group
contains dwarfs, while the group with longer rotation periods having in many cases
stars with amplitudes larger than $0.2$~mag contains dwarfs as well as giants with
the vast majority of the latter.

Apart from fast-rotating, small-amplitude dwarfs, we also detect
dwarfs with high amplitudes, but they are definitely slow rotators. 
We do not find any significant correlation between the amplitude and rotation period for dwarfs. 
However, for spotted giants we observe a clear correlation -- the slower 
the giants rotate, the smaller amplitudes they have. It is noteworthy that, on 
average, slowly rotating stars exhibit larger secular brightness changes. This effect can 
be related to the mechanism responsible for generating stellar magnetic fields
and thus starspots. It seems that slowly rotating stars have larger and slower 
evolving spots on their surfaces. However, it should be kept in mind that
large amplitudes result from an asymmetry in the spotted
area. Stars covered by spots evenly give a minimal, or just imperceptible
change of luminosity. In the case of uniformly spotted stars, the amplitude in
the optical band is insignificant, although the star can be very
active. In such stars the activity level can be measured
from the emission cores in the spectral lines CaII H and K
\citep{Baliunas95} or X-ray flux \citep{Pallavicini81}.

The extensive studies of spotted dwarfs from {\it Kepler} data
showed  that photometric variability is correlated with the rotation
period. For instance, \citet{McQuillan14} using variability index
$R_{\mathrm{var}}$ defined by \citet{Basri11, Basri13} found, that
higher amplitudes are observed in stars with shorter rotation
periods. Similar picture for fast-rotating seismic solar analogs was
obtained by \citet{Salabert16} who used
$S_{\mathrm{ph}}$ \citep{Mathur14} variability index. Recently,
\citet{Lanzafame18a, Lanzafame18b} demonstrated the existence of
bimodality in amplitude-period plane. The authors pointed out, that
the fastest rotating dwarfs concentrate in two regimes of amplitudes
-- low and high, whereas the slowly rotating dwarfs show
preferentially low photometric variability. \citet{Lanzafame18a,
Lanzafame18b} concluded, that such a multimodality suggests the
existence of different regimes of surface inhomogeneities in young,
and middle-age stars. Our data partially confirm the above results for
dwarfs. However, we do not see bimodality of amplitudes for fast
rotators but it should be mentioned that amplitudes measured in our
work are on average ten times larger. This is obviously related to the capabilities of the instrument -- much lower 
amplitudes can be detected from the outer space compared to the
ground-based instruments. Additionally, amplitudes may vary significantly depending on the 
filters in which an object is observed. On the other hand, amplitude measurements based only on the
Gaia data may not be very accurate due to the low number of epochs. As
we mentioned before, amplitudes of brightness of the spotted stars change rapidly. It is interesting, however,
to see that our giants behave similarly to the slowly rotating dwarfs
on the amplitude-period diagram, {\it i. e.} their amplitude decreases
with the increasing period. 
Separation for both groups of stars, dwarfs and giants, is also visible 
around the rotation period $P \approx 2$~d.

\subsection{Spatial distribution of the rotation periods}

\begin{figure}[h]
\includegraphics[scale=0.2]{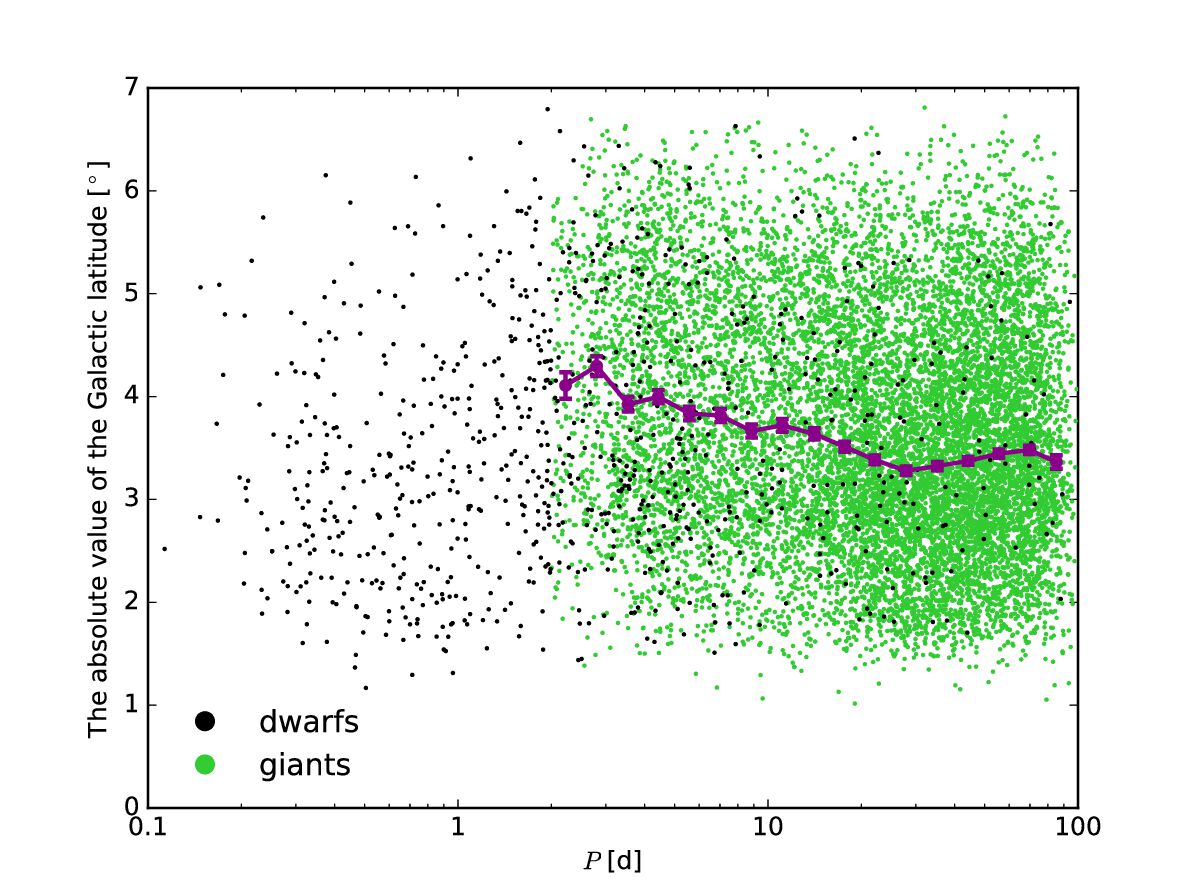}
\caption{Absolute values of the Galactic latitude $|b|$ vs. rotation period $P$ 
for the spotted variables. Large purple points with error bars indicate
the median $|b|$ values binned in 
$\log{(P)}$.}
\label{fig8}
\end{figure}

In Figure \ref{fig8}, we present a relationship between absolute values of Galactic latitudes and 
rotation periods of spotted stars from our sample. The most central region of 
the Milky Way is inaccessible for the OGLE observations, because the clouds 
of dust toward this region absorb the bulk of the light in the visual domain.
Thus, there are no points with $|b|<1^\circ$ in Figure \ref{fig8}.

\citet{Drake06} noted that chromospherically active giants located close to the 
Galactic plane, have on average longer rotation periods. This correlation is indeed visible
in Fig. \ref{fig8} where the slow rotating giants are on average closer to 
the Galactic plane, but it is absent for dwarfs.
We think, however, that the existence of this correlation for giants, and its lack for 
dwarfs is caused by the selection effect against shorter-period red
stars with typically fainter {\textit{V}}-band magnitudes which are
missing in the sample close to the Galactic plane where the extinction
is higher. In addition, stars with 
no detection in the {\textit{V}}-band are missing by definition in our sample.

As mentioned before, the dereddened luminosities and colors for dwarfs are burdened
with large uncertainties so we are left with rotation periods, variability amplitudes
and Galactic coordinates as the only certain parameters for these stars. This is why we drop dwarfs
from further analysis.

\subsection{$(V-I)_0$ vs. the rotation period}

\begin{figure}[h]
\includegraphics[scale=0.2]{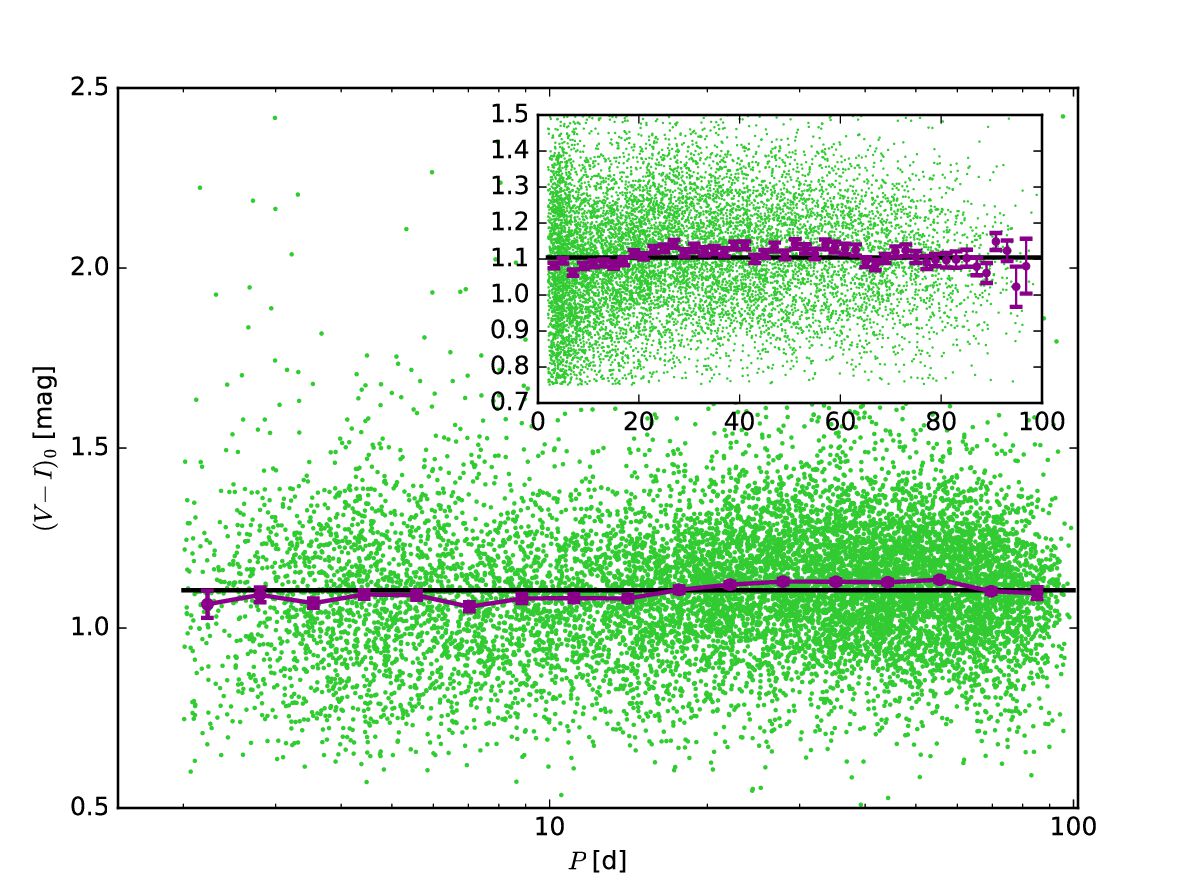}
\caption{Dereddened color index $(V-I)_0$ vs. rotation period
$P$. Large purple points with error bars indicate the median $(V-I)_0$ values binned in $\log{(P)}$. 
Inset ranges correspond to Fig. 14 of \citet{Drake06}. 
Here large purple points with error bars denote the median $(V-I)_0$ values 
binned in linear scale. Black solid lines represent the median values
of the $(V-I)_0$. over the whole range of $P$. \vspace{0.1cm}}
\label{fig9}
\end{figure}

\citet{Drake06} found that the rotation period increases with color for 
stars with periods $P \lesssim 30$ d. In Figure~\ref{fig9}, we present 
the dereddened color index $(V-I)_0$ as a function of the rotation period $P$ 
for stars from our sample. In the inset of Figure \ref{fig9}, we show
a similar color range to that presented by \citet{Drake06}. In both
plots solid black lines are plotted at the median of $(V-I)_0$.

We cannot unambiguously confirm that this correlation exists
by looking at a four times larger data set which covers substantially
wider range of colors than the range presented by \citet[][see Fig. 14 therein]{Drake06}.
However, using a similar range as in \citet{Drake06} with the linear binning, this correlation
can be marginally detected.

\subsection{$I_0$ vs. the rotation period}

\begin{figure}[h]
\includegraphics[scale=0.2]{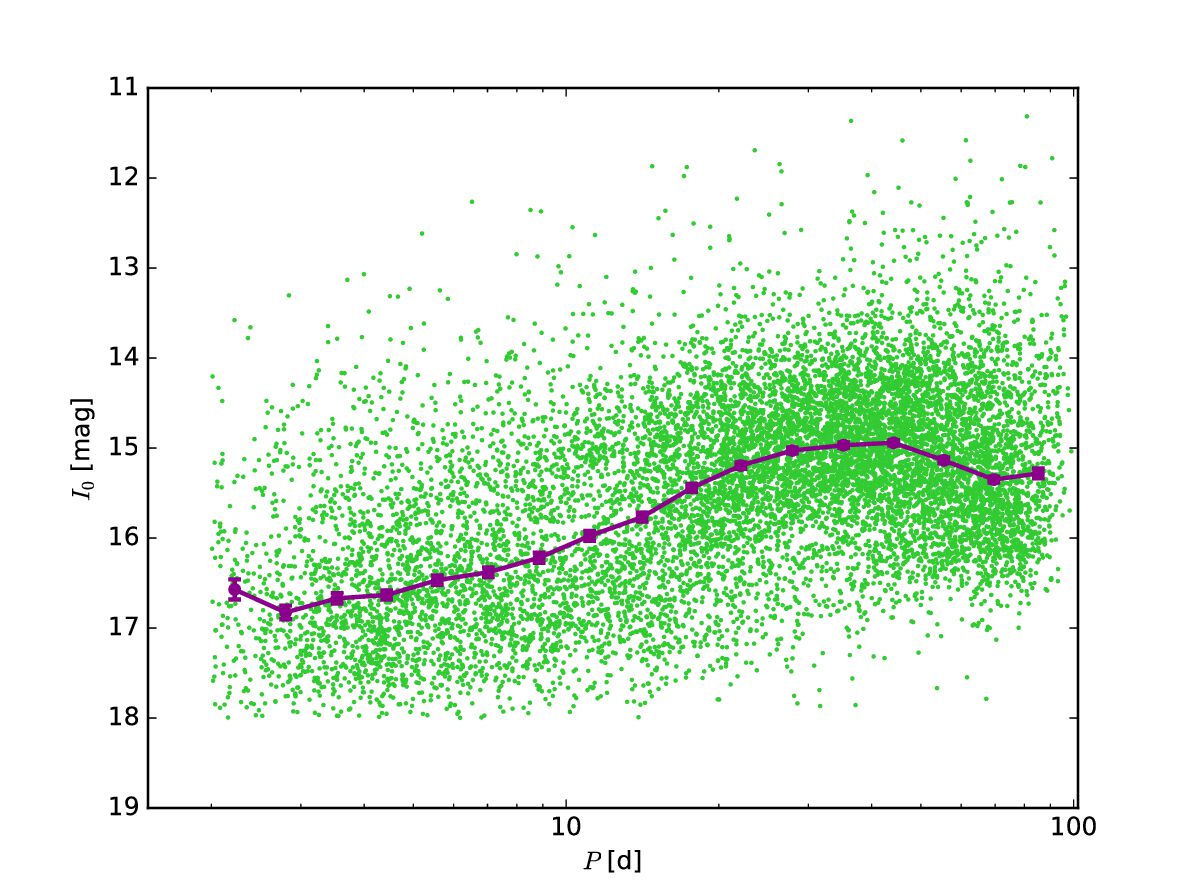}
\caption{Mean extinction-corrected \textit{I}-band magnitude ($I_0$) vs. 
rotation period $P$. Large purple points with error bars indicate the median $I_0$ binned in 
$\log{(P)}$.}
\label{fig10}
\end{figure}

Previous studies of spotted, magnetically active stars have excluded the
existence of a dependence between the brightness of the stars and the rotation 
period \citep{Drake06}. In Figure \ref{fig10}, we present a plot of 
the extinction-corrected magnitude $I_0$ versus the rotation period $P$ for our giants.

We see a clear correlation between the brightness and period, such that the 
spotted giants are brighter when they rotate slower. The correlation is visible for 
stars with rotation periods of up to $50$ days and flatten for $P > 50$ d.
A range of rotation periods $50 < P < 100$~d contains $2630$ 
stars, while a range $2 < P \leq 50$ d contains $9081$ stars.
Perhaps the future detections of very slowly rotating giants will reveal the
extension of this correlation.

\subsection{$I_0$ vs. the \textit{I}-band amplitude} \label{sec:amp}

\begin{figure}[h]
\begin{center}
\includegraphics[scale=0.2]{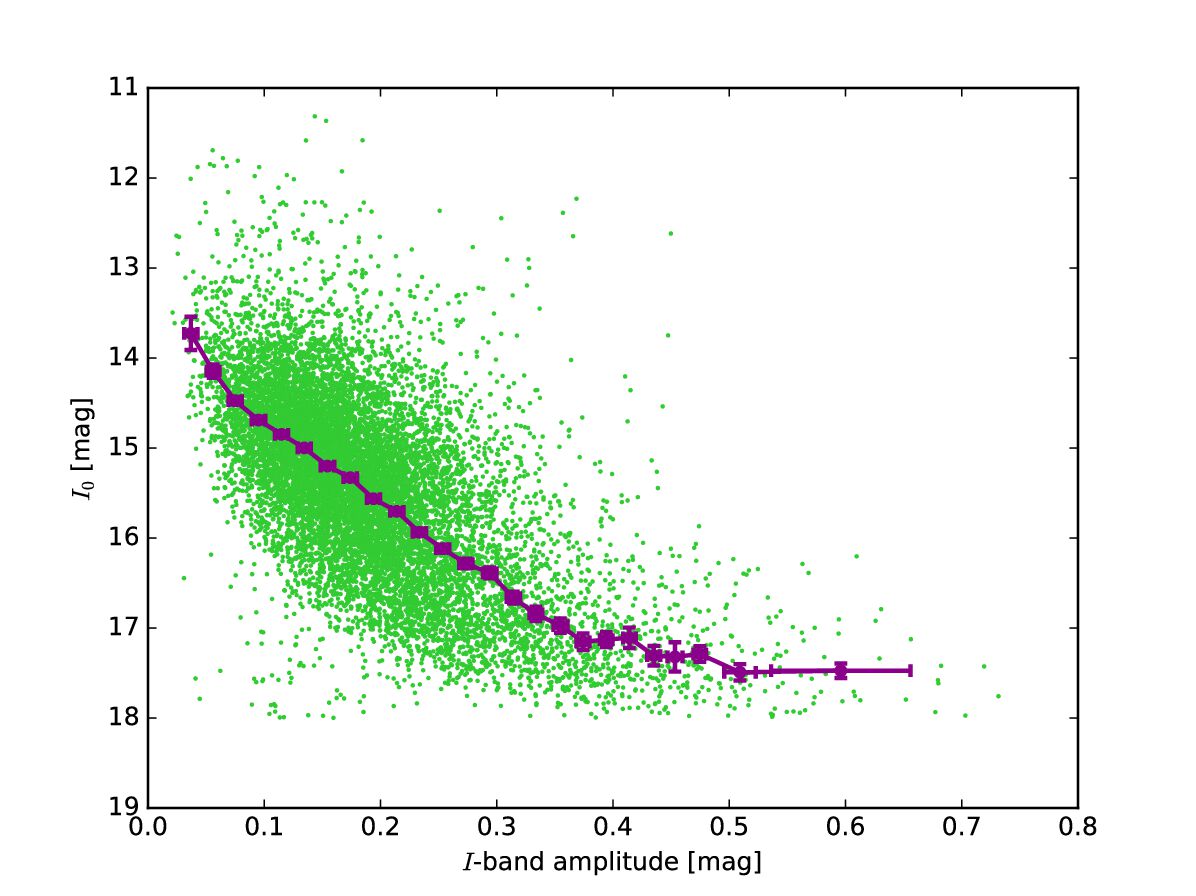}
\caption{Mean extinction-corrected \textit{I}-band magnitude ($I_0$) vs. 
\textit{I}-band amplitude. Large purple points with error bars indicate the median $I_0$ binned 
in \textit{I}-band amplitude.}
\label{fig11}
\end{center}
\end{figure}

Drake's (\citeyear{Drake06}) study of chromospherically active stars showed the existence of a weak 
correlation between stellar brightness and variability amplitude.
The author pointed out that faint stars
exhibit larger brightness variations. In Figure \ref{fig11}, we present the 
relationship between the extinction corrected \textit{I}-band magnitude $I_0$ 
and the \textit{I}-band amplitude. The amplitudes are calculated in the same way
as in Section \ref{chap:amplitudes}.

In Figure \ref{fig11}, the correlation discussed in \citet{Drake06} can be clearly
seen -- the fainter stars have a stronger manifestation of the magnetic activity 
({\em i.e.} larger spot coverage) and thus have larger amplitudes. Unfortunately, in
our opinion an accurate examination of this correlation is almost impossible due to 
a~selection effect -- faint stars with small amplitudes are less likely to be 
detected and properly classified as spotted variables, so we 
do not expect to have many faint, low-amplitude objects in our collection. 
However, we found an interesting property of bright stars with $I_0~\lesssim 
16$~mag. Save for a few exceptions, they do not exhibit amplitudes higher than $0.4$ 
mag. There is no any obvious selection effect which could produce this
feature, so we believe that this finding is related to the nature of spotted stars. More detailed
studies are needed for clarification of this phenomenon.

\section{Two types of spots} \label{chap:twospots}

\begin{figure*}
\begin{center}
\includegraphics[scale=0.17]{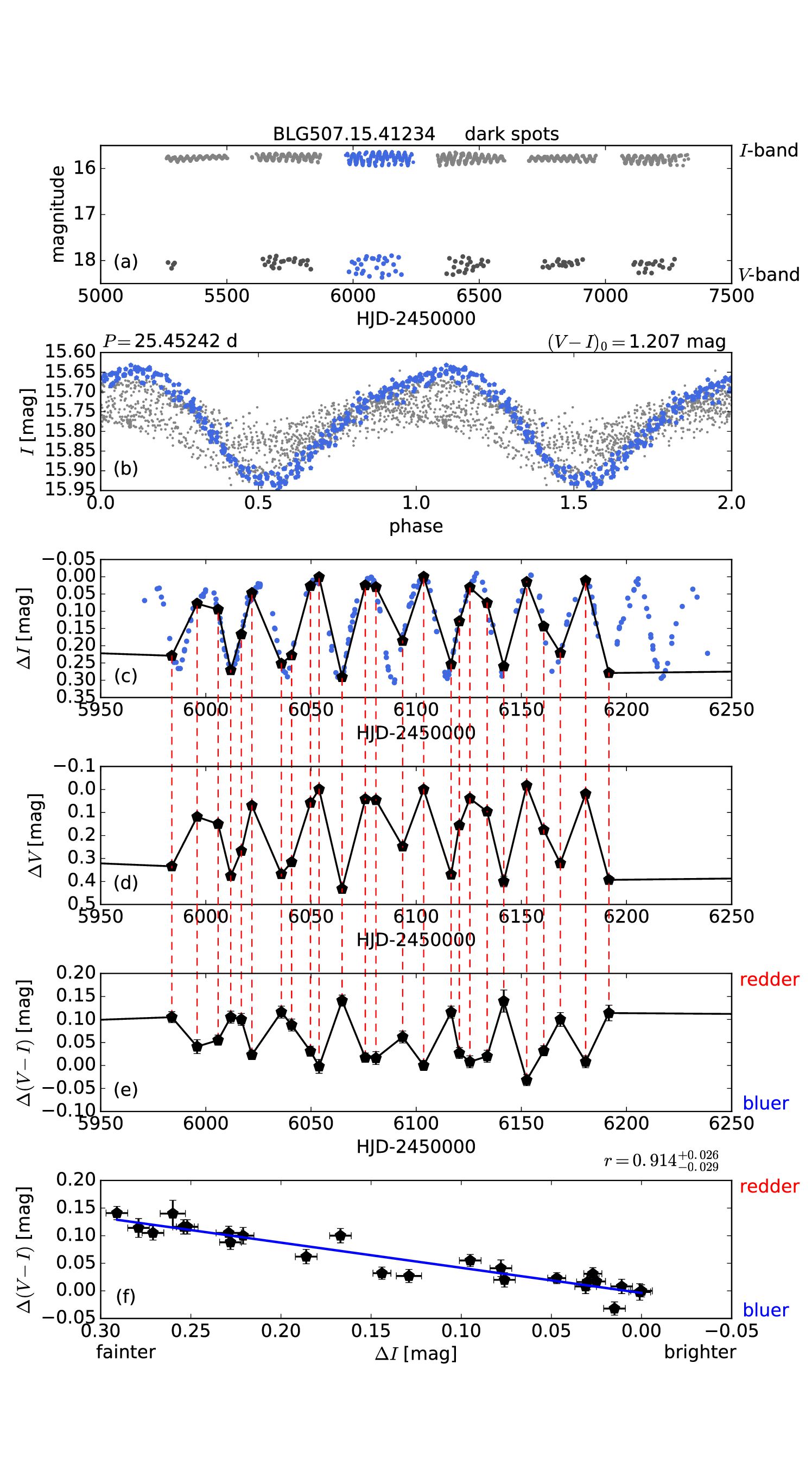}
\includegraphics[scale=0.17]{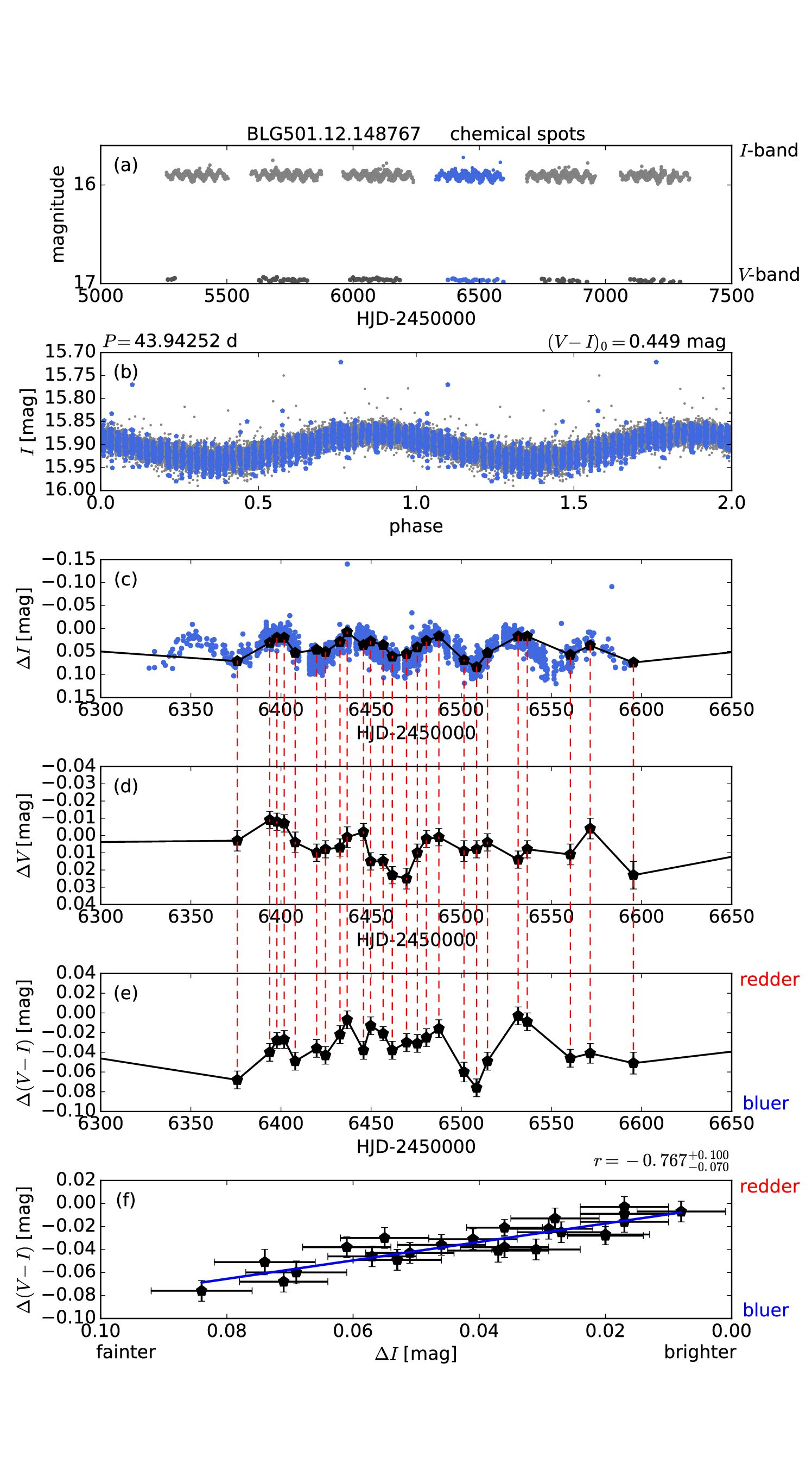}
\caption{Examples of different types of spots in two stars. The left panel
shows an example of 
a star with dark spots (BLG507.15.41234), while the right panel shows an 
example of a star with chemical spots (BLG501.12.148767). (a) Whole 
light curves in both {\it{I}}- and {\it{V}}-band. Blue points denote
observations analyzed on further plots. (b) Phase folded light curves with 
rotation periods given above each plot. Above the plot we also give
dereddened color index $(V-I)_0$. (c) Changes in the 
{\it{I}} band for the time-span marked with the blue points in the top plots. Black points show 
measurements for which we have simultaneous {\it{I}}- and  {\it{V}}-band observations. (d) 
Changes in the {\it{V}} band. (e) Color changes. (f) Correlations between 
color variations and brightness variations. The correlation coefficient 
calculated within the considered time range is given above the plot. For better 
visibility of these correlations we fitted straight lines to the data. All 
black points are plotted with the appropriate measurement uncertainties in both 
coordinates.}
\label{fig12}
\end{center}
\end{figure*}

\begin{figure*}
\begin{center}
\includegraphics[scale=0.17]{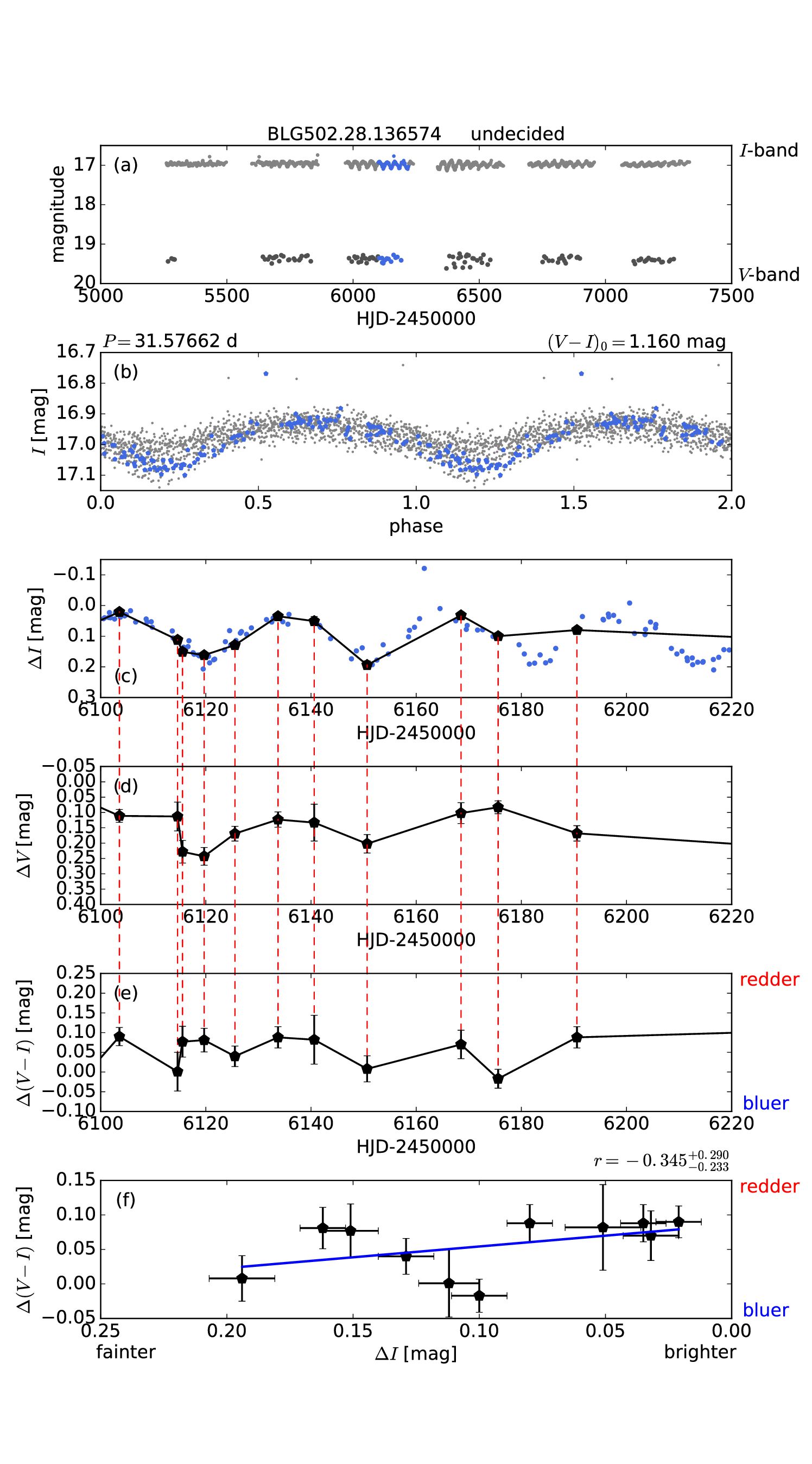}
\includegraphics[scale=0.17]{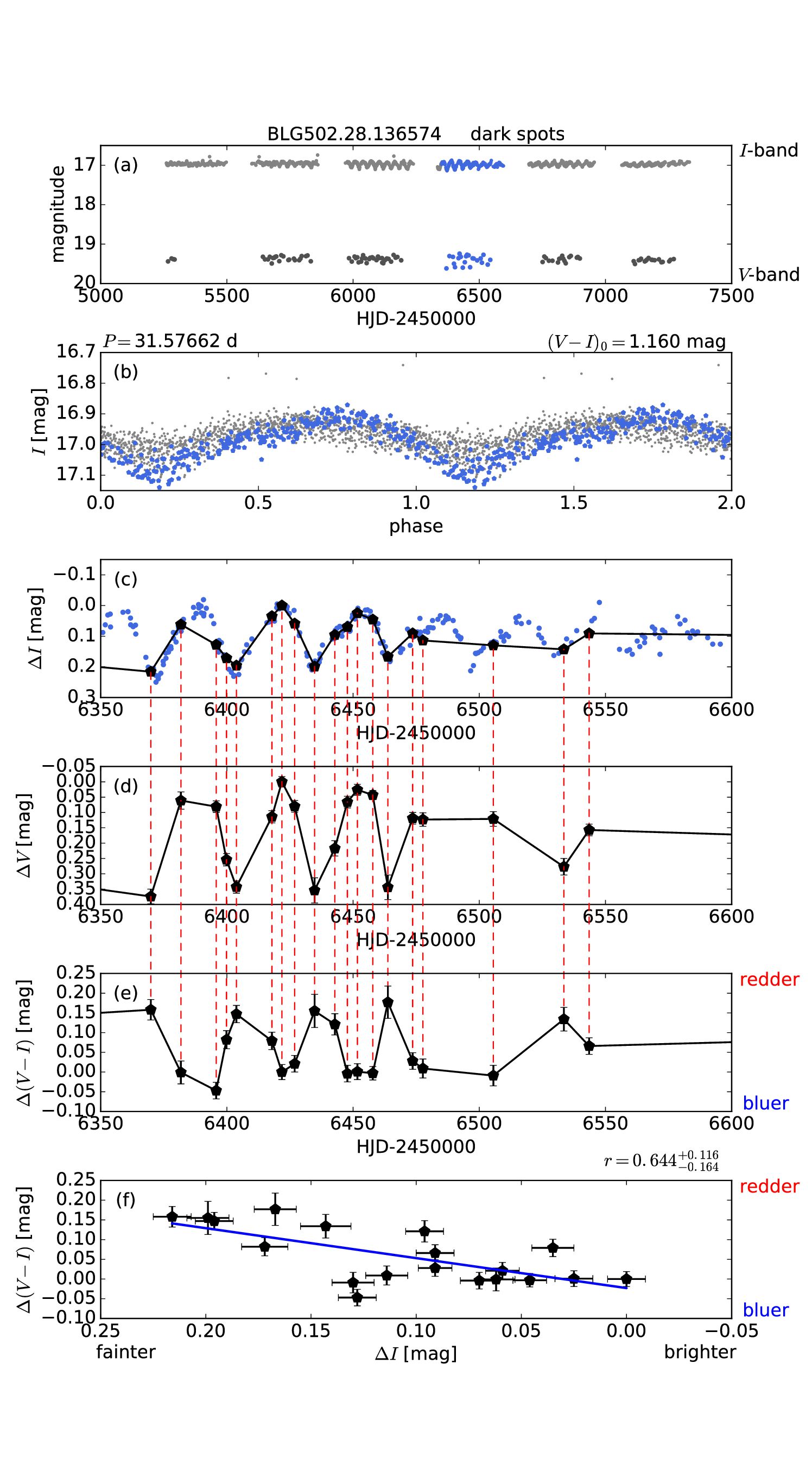}
\caption{Example of the evolution of starspots. Both panels show the same star 
(BLG502.28.136574) in two different intervals of time. This is the star 
for which the correlation coefficient is around zero 
($r = 0.163_{-0.117}^{+0.112}$) 
over the entire time-span of observations. In the time range  
HJD$-2450000 \in [6100, 6220]$ there is no, or only the slight negative correlation, which 
would suggests the apparent existence of
chemical spots on the stellar surface, or very small impact of
dark spots on the brightness and color of the star. On the other 
hand, in the time range HJD$-2450000 \in [6350, 6600]$ the correlation is 
significant and indicates the existence of dark spots. (a) All observations in 
both {\it{I}}- and {\it{V}}-bands. Blue points denote the observations 
analyzed in the further plots. (b) Phase folded light curves with the rotation period 
given above the plot together with the
dereddened color index $(V-I)_0$. (c) The {\it{I}}-band variations for the 
time-span marked in the top plot.  Black points show 
measurements for which we have simultaneous {\it{I}}- and
{\it{V}}-band observations. (d) The {\it{V}}-band variations. 
(e) Color variations. (f) Correlations between color variations and 
brightness variations. The correlation coefficient calculated in the considered time range is 
printed above the plot. For better visibility of these correlations we fitted 
a straight line to the data. All black points are plotted with the appropriate 
measurement uncertainties in both coordinates.}
\label{fig13}
\end{center}
\end{figure*}

Thanks to the long-term, well-sampled, two-band OGLE light curves, we are able 
to study the relation between brightness variability and the color 
index $(V-I)$ variability. For all stars from our collection, we choose brightness 
measurements in both filters from the same night, with the time difference between 
the frames in both colors not exceeding 0.1 day. In the next step, we calculate the $(V-I)$
indices for each pair of brightness measurements. Then we remove all outlying $(V-I)$ 
measurements, deviating more than $\pm 3\sigma$ from the median. As~a~result of 
these calculations we are left with several, up to several dozen brightness measurements 
in both filters and the color indices $(V-I)$. 
Subsequently, we take the maximum measured brightness in the \textit{I}-band 
($I_{max}$) for each object, the corresponding \textit{V}-band ($V_{max}$) magnitude
and the color index ($(V-I)_{max}$). Then we calculate $\Delta I$, $\Delta V$ and $\Delta (V-I)$ as follows:

\begin{equation}
\Delta I = I - I_{max},
\end{equation}

\begin{equation}
\Delta V = V - V_{max},
\end{equation}

\begin{equation}
\Delta (V-I) = (V-I) - (V-I)_{max}.
\end{equation}

\noindent For the uncertainty of brightness changes we use 
appropriate values as reported in the light curves. 
We calculate the uncertainty for changes in color as:

\begin{equation}
\varepsilon(V-I) = \sqrt{dV^2 + dI^2,}
\end{equation}

\noindent where $dV$ and $dI$ are brightness uncertainties as reported in the light curves. 

We expect that the brightness of a star with cool, dark spots should decrease
when highly spotted area is on the visible side of a star and the color should change to a redder one.
Indeed, we detect such a correlation for some of the stars in our sample. An example of such a behavior is shown in 
Figure~\ref{fig12} (left panel). However, we also detect a number of objects 
with the opposite dependence between the luminosity and color changes -- the stars are more luminous,
when the color is redder  (Figure~\ref{fig12}, right panel).

For all the stars from our sample we calculate the relations 
between the luminosities and colors. We use bootstrapping method to
estimate the correlation coefficient with appropriate uncertainties.
For each star we randomly draw with returning brightness in 
{\textit{I-}}band and color $(V-I)$ along the measurement uncertainties. 
We made 1000 draws, during which we calculate the correlation coefficient.
In the next step, we calculate 16th, 50th and 84th percentiles of the 
correlation coefficient distribution. We assume that we 
will use the~50th percentile (the median of the correlation
coefficient) for further analysis  with uncertainties 
calculated as follows:

\begin{equation}
\sigma_+ = 50\mathrm{th} - 16\mathrm{th},
\end{equation}
\begin{equation}
\sigma_- = 84\mathrm{th} - 50\mathrm{th},
\end{equation}
\begin{equation}
\sigma_\pm = \frac{\sigma_+ + \sigma_-}{2}.
\end{equation}

\noindent In addition, we remove from our sample the stars with $\sigma_\pm > 
0.3$ because of a statistically insignificant correlation due to
{\em e.g.} a too low
number of points. These calculations leave $11\,412$ (out of $12\,660$) stars with statistically significant correlation between 
brightness and color variations.

From $11\,412$ stars that show statistically significant correlation 
we find 5199 objects with the positive
correlation coefficient ($r - \sigma_{-} \geq 0.2$; left panel in Figure 
\ref{fig12}) and 465 stars with the negative correlation coefficient ($r + 
\sigma_{+} \leq -0.2$, right panel in Figure \ref{fig12}). Positive correlation 
coefficient means that a star is redder (and cooler) when it is less luminous, which is a 
clear confirmation of the existence of cool, dark spots on the surface.
For this reason, we successively name this group as ``dark 
spots''. On the other hand, a negative correlation coefficient means a lot more 
puzzling behavior -- a star is redder when it is more luminous. It is worth mentioning, 
however, that both types of correlations are related to rotation periods, which is a 
strong evidence of the existence of spots on the stellar surfaces.
Additionally, both spots' behaviors result from differences in brightness 
amplitudes measured in the \textit{I-} and \textit{V-}band.  
It is seen from Figure \ref{fig12} that the positive correlation
coefficient means that the
amplitude in the \textit{V-}band is larger than in \textit{I-}band. It is opposite
for stars showing the negative correlation coefficient; here the 
amplitude in the \textit{V-}band is smaller than in 
\textit{I}-band. One scenario that can explain a lower amplitude in the 
\textit{V}-band, and cyclic reddening of the star is connected with
chemical spots on the stellar surface. 
The overabundance of heavy elements inside a spot related to the surface  magnetic 
field can cause a~phenomenon called \textit{line-blanketing} which depresses 
the short-wavelength part of the electromagnetic spectrum \citep{Milne28}. The 
line-blanketing depends on the wavelength, because the number and the strength of spectral
lines of the overabundant elements increases with decreasing wavelength. The energy absorbed at 
short wavelengths is re-emitted at long wavelengths due to the backwarming effect. This causes a
star to become fainter at shorter wavelengths and
brighter at longer wavelengths hence redder  ({\em e.g.} 
\citealt{Gray09}). This effect is very well known for Ap stars 
({\em e.g.} \citealt{Molnar73, Stepien93}). Recently, \citet{Sikora19}
analyzed all known chemically peculiar stars located closer than
100 pc from the Sun and  showed that they all have spectral types
earlier than late~F. Assuming
F8 as the limiting spectral type we obtain the corresponding  color index
$(V-I)_0 = 0.5$ mag.  In addition, the observations
of many field Ap stars over several decades indicate that their average
brightnesses and amplitudes of their light curves are remarkably
stable. So only stars from our sample hotter than the above limit
(allowing, of course, for uncertainty of the derived extinction-free
indices) and showing the same properties can be considered as
candidate Ap (or, alternatively, CP) stars. Definite confirmation of
their classification should come from spectroscopic observations.
   
In another scenario that can explain the observed changes with the
negative correlation we can consider a system of two stars with a cool,
tidally distorted component and a hotter companion. Many such systems
are known as RS CVn-type binaries. Let us assume that the orbit
inclination is too low for eclipses to occur but large enough to see
light variability due to ellipticity of the cool component. 
At the maximum light we see both stars from the side. The increased
contribution from the red star results in reddening of the system. At
the minimum light we see the red star almost end-on and its
contribution is diminished. The color index becomes bluer. 
Additionally, the spots evolution on the cool component surface can also
contribute to the total light variations of the system. Note that
there is no {\it a priori} limit on  $(V-I)_0$ for stars in this category.
On the other hand, the negative correlation was noticed in Young Stellar 
Objects, and was explained as a significant change in the structure of the 
inner disk ({\em e.g.} \citealt{Gunther14, Wolk15, Rebull15, Wolk18}).

The remaining 5748 objects show the correlation coefficient around 0 ($r + \sigma_{+} > 
-0.2$ and $r - \sigma_{-} < 0.2$), so we named this group ``undecided''. The apparent
lack of correlation may result from a variable combination of dark spots and bright areas
(like faculae, plages or quiescent prominences) in magnetically active stars or a specific
combination of different elements in chemical spots of chemically peculiar stars. Activity
variation in the course of activity cycles will also blur the correlation.
One example of this situation is presented in Figure~\ref{fig13}. Looking at the 
light curve of the same star in different epochs, it is clearly seen
that at a time 
we do not see any significant spot-induces modulation of brightness and color (see
the left panel of Fig. \ref{fig13}), but somewhat later 
measurable changes of the brightness and color of the star are clearly visible (right 
panel in Fig. \ref{fig13}). Note that in the left panel of Fig. 
\ref{fig13}, the amplitudes in both filters are comparable. About 100 days 
later, when the dark spots appear on the stellar surface (right panel in Fig. 
\ref{fig13}), the amplitude in {\textit{V-}}band becomes larger.

In the remaining part of the paper we use the correlation coefficients
calculated from the entire OGLE-IV data when discussing both types of spots 
(dark and chemical), whereas the correlation coefficients presented in both panels of 
Figures \ref{fig12} and \ref{fig13} are calculated only over the time ranges marked with 
blue points.

\begin{figure*}
\begin{center}
\includegraphics[scale=0.45]{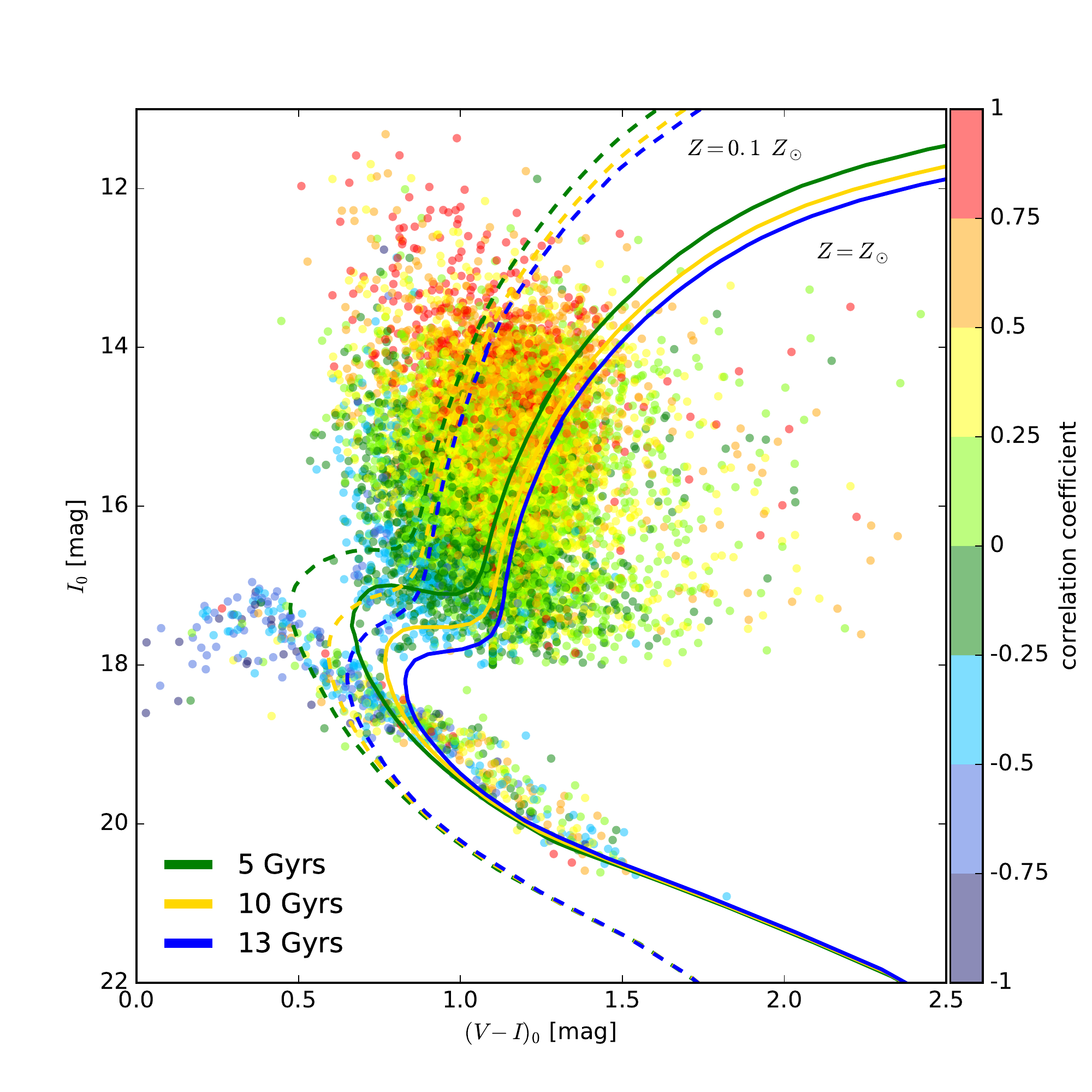}
\caption{Color--magnitude diagram (CMD) for stars, for which we found 
statistically significant dependencies between brightness variations 
and  color variations ($\sigma_\pm \leq 0.3$). Isochrones are the same 
as in Fig. \ref{fig4}. The color bar on the right describes the correlation
coefficient.}

\label{fig14}
\end{center}
\end{figure*}

\begin{figure*}
\begin{center}
\includegraphics[scale=0.28]{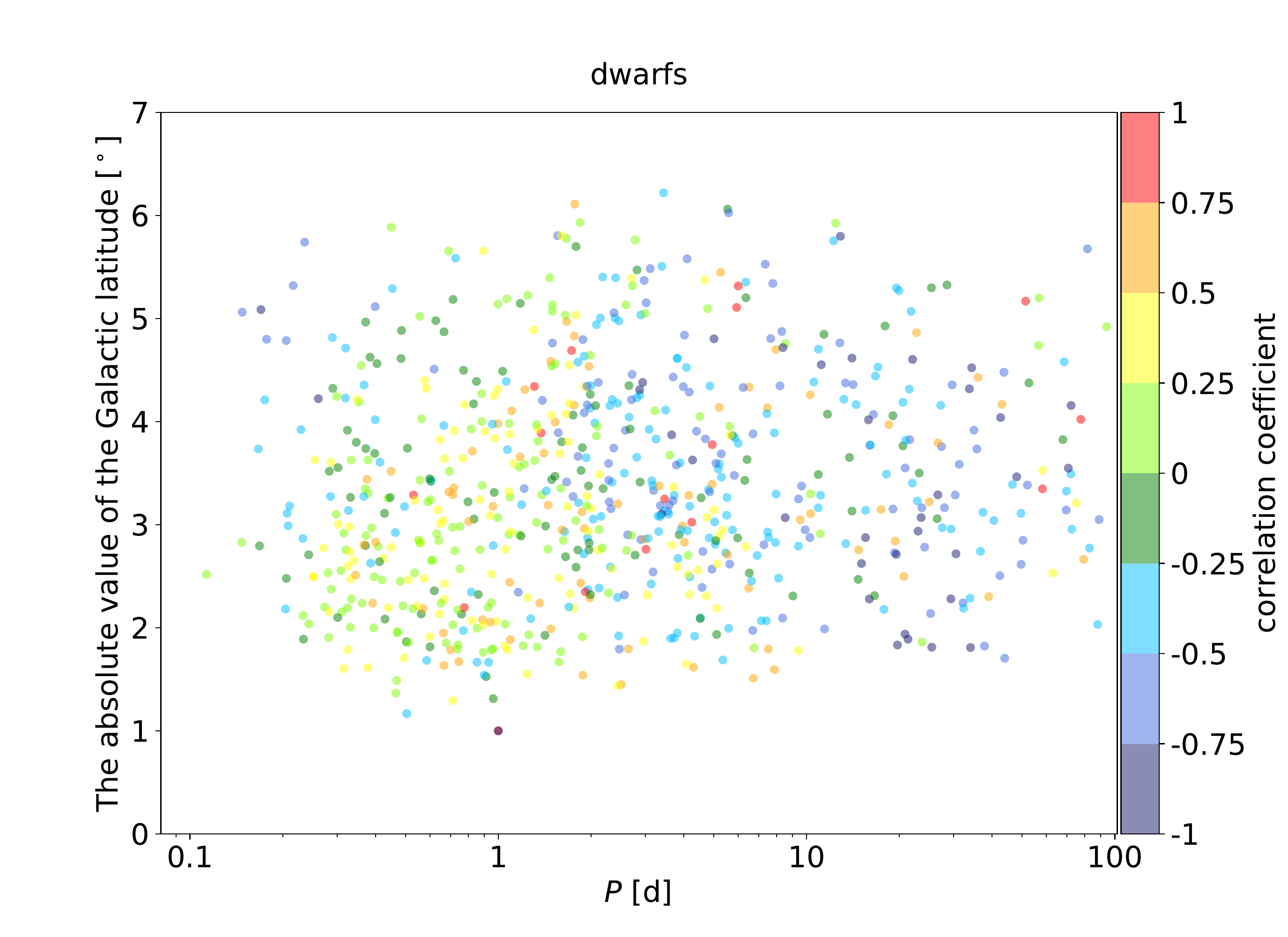}
\includegraphics[scale=0.28]{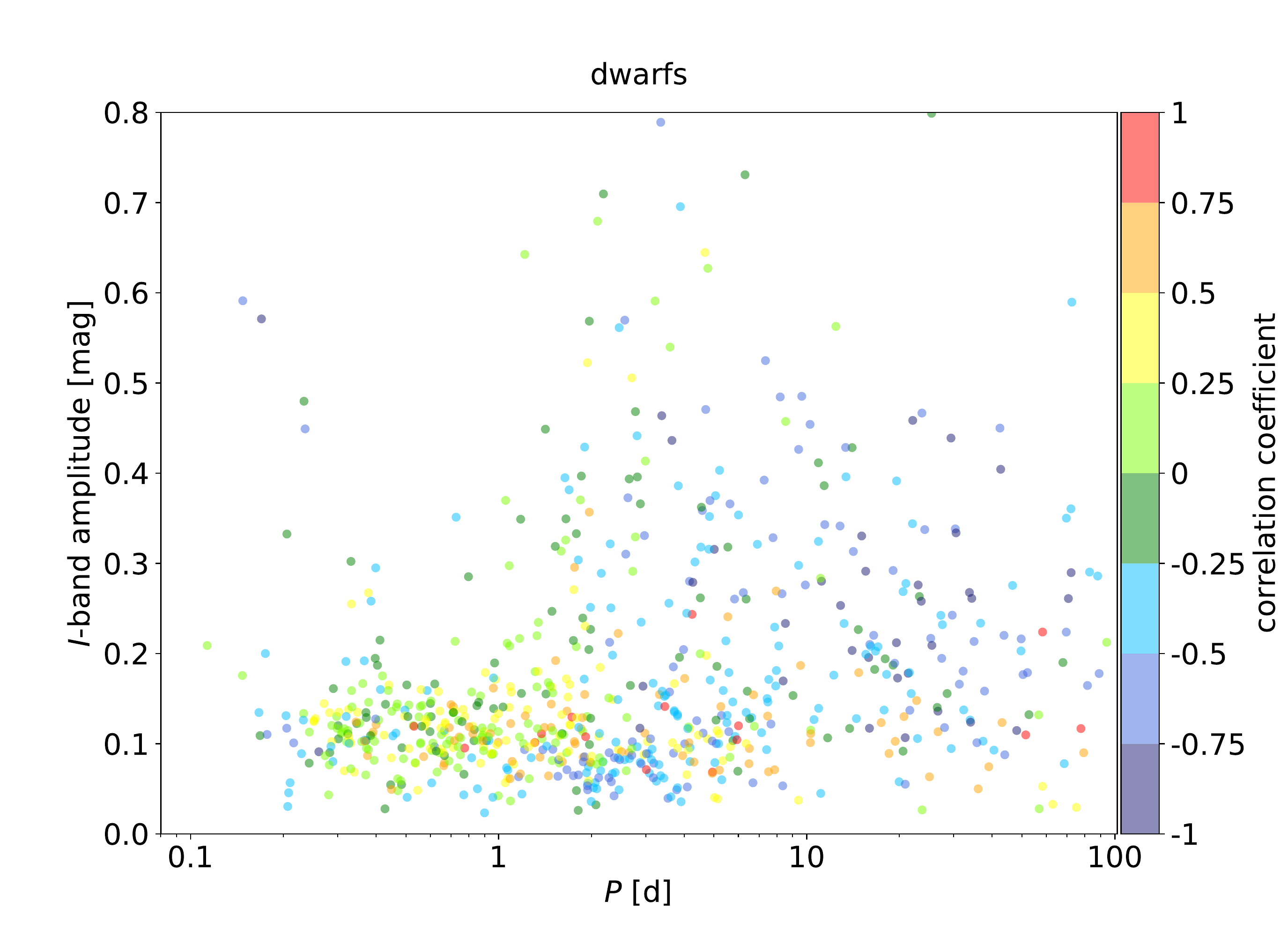} \\
\includegraphics[scale=0.28]{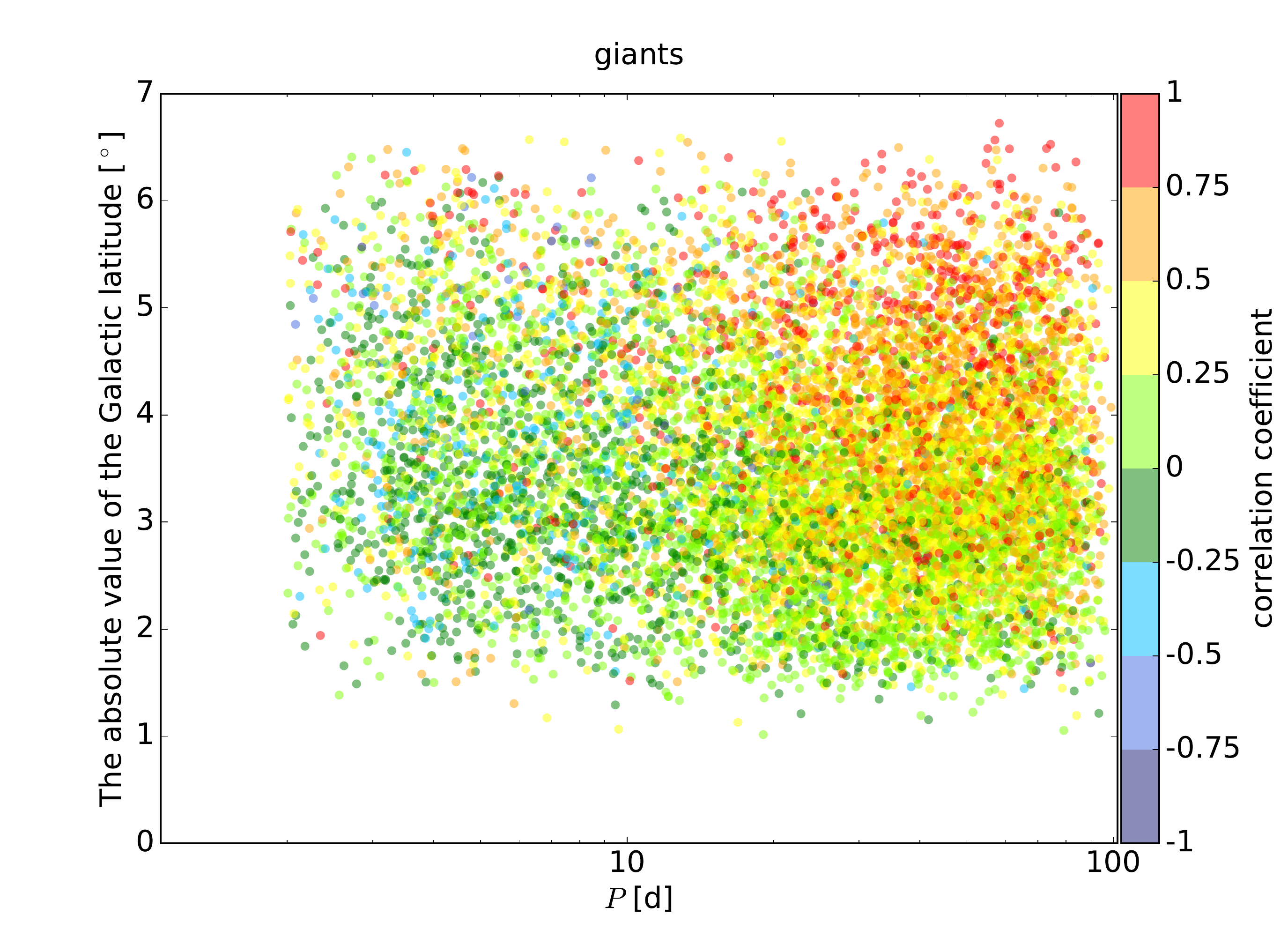}
\includegraphics[scale=0.28]{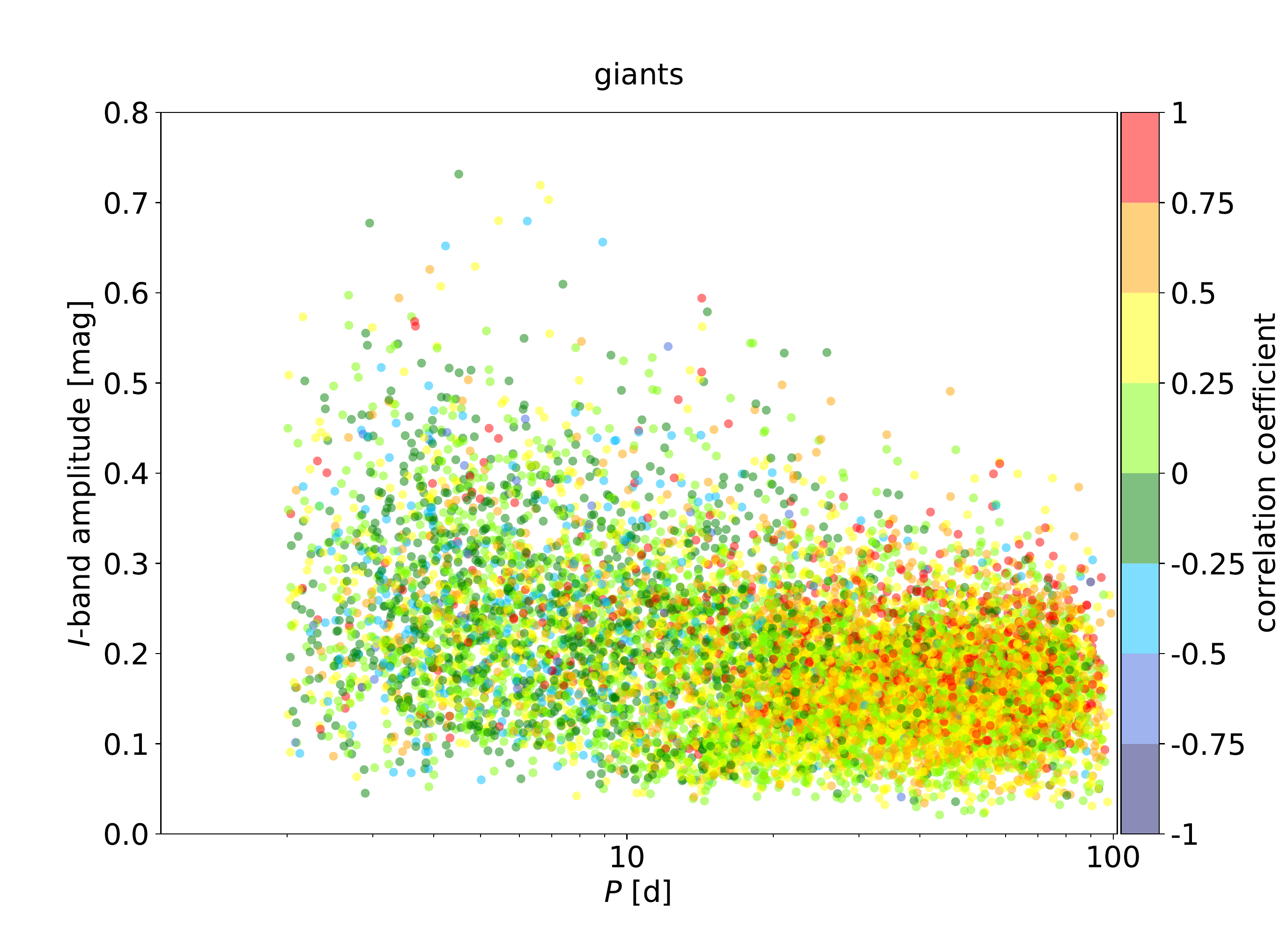}
\caption{Relations between the absolute value of the Galactic latitude $|b|$ 
and the rotation period $P$ (left panels), and between the {\it{I}}-band amplitude and the
rotation period $P$ (right panels), separately for dwarfs (top panels) and giants (bottom panels)
as functions of the color-coded correlation coefficient. 
Only stars with the statistically significant 
correlation coefficients, {\em i.e.} when $\sigma_\pm \leq 0.3$,  are plotted.}
\label{fig15}
\end{center}
\end{figure*}

\subsection{Type of spots on the color--magnitude diagram}

Using the dereddened color $(V-I)_0$, brightness $I_0$ (see Section 
\ref{chap:color_calculations}) and the statistically significant correlation 
coefficient ($\sigma_\pm \leq 0.3$), we can place each spotted star in the CMD 
and search for any possible connection between a given type of spots
and the star's location in CMD. This kind of CMD, with color-coded correlation coefficient is 
presented in Figure \ref{fig14}.

Detection  of two different, linear relations between the color and brightness changes, 
and the separation of a group of stars with the correlation
coefficient close to 0 allowed us to obtain a very interesting picture in the CMD diagram. 
Looking at Figure \ref{fig14} we distinguish four interesting points
related to the spotted  stars:

\begin{enumerate}

\item We find $161$ and $255$ dwarfs with the positive and negative
correlation coefficient, respectively. It means that the majority of
stars classified as dwarfs have negative correlation coefficients, 
which points toward chemical spots on their surfaces. By contrast,
only $210$ giants (out of several thousands) show the measurable negative correlation.

\item A systematic change of the correlation coefficient is visible along the Main 
Sequence (MS). Stars located in the upper part of the MS, 
bluer than $(V-I)_0 \approx 0.7$ mag,  have mostly negative
coefficients. The limit of 0.7 mag corresponds to the cool boundary of
chemically peculiar stars ({\em e.g.} \citealt{Sikora19}), where strong 
overabundances of heavy elements are detected ({\em e.g.}  \citealt{Paunzen15, 
Paunzen16}), blurred by the uncertainties in the color index. 
The number of dwarfs with the negative correlation
rapidly decreases in the lower part of the MS where more stars with
the positive correlation, indicative of cool spots, occur.

\item The majority of giants show positive correlation coefficients -- they have 
dark spots on their surfaces. The giants with the
negative correlation occur almost exclusively close to the bottom of
the giant branch.

\item The correlation coefficient varies smoothly along the giant
branch, from the negative value at its bottom, through zero in the middle part
up to the largest positive values at the top.
The boundaries between the successive values of the correlation 
coefficient seem to have slopes similar to the isoradius lines. 

\end{enumerate}

From the above results we conclude that the chemical spots or variations
characteristic of RS CVn-type stars with favorably inclined orbits
(producing negative correlation between light and color variations)
occur in hotter MS stars and slightly evolved giants. Mildly evolved
giants and some cool dwarfs show the simultaneous presence of dark
spots and bright areas indicative of high level of activity. Other cool MS
stars and evolved giants show predominance of dark spots.

\subsection{Correlations overview}

Now we discuss in detail two out of five relations presented in Section 
\ref{chap:properties}), taking into account the values of the
correlation coefficients between light and color variations. These are
period-absolute value of the Galactic latitude $|b|$ (left panels
of Fig. \ref{fig15}), and
period-{\it{I}}-band amplitude (right panels of
Fig. \ref{fig15}), separately for dwarfs and giants. 
The individual data points are colored according to the
value of the correlation coefficient as coded by the bar on the
right-hand side of each Figure.

It is seen that most of the fast rotating stars with $P \leq 2$~d (all assumed
to be dwarfs) have positive values of the correlation coefficient --
out of 146 stars from this range 97 show positive correlation. These
are cool, rapidly rotating dwarfs covered with dark spots. On the
other hand, negative correlation coefficients prevail in slowly
rotating dwarfs -- out of 270 stars from this range 206 show negative
correlation. A large fraction of them belong to Ap stars. As it is
well known, these objects rotate with typical periods from a fraction of a day to a few weeks,
although much longer periods, even up to centuries, are also observed
\citep{Preston74, Renson01, Bychkov16, Mathys19}. Some slowly rotating stars with negative correlation between light and color variations may belong to the RS~CVn-type variables as discussed earlier.

A reversed tendency is seen among giants where the negative correlation
occurs mostly for stars with periods shorter than 10 days whereas the largest
positive values of the correlation coefficient occur in stars with
periods longer than 20 days. This is in line with our previous
interpretation that giants with shorter periods belong to RS CVn-type
variables which, depending on the orbit inclination, may show
positive, as well as negative values of the correlation
coefficient. Giants rotating with the longest periods may belong to
wide binaries or to single, active stars.

The systematic trends of the light curve amplitude and $|b|$ with the
period, discussed in Section \ref{chap:properties} are confirmed here.

\begin{figure*}[h]
\begin{center}
\includegraphics[scale=0.13]{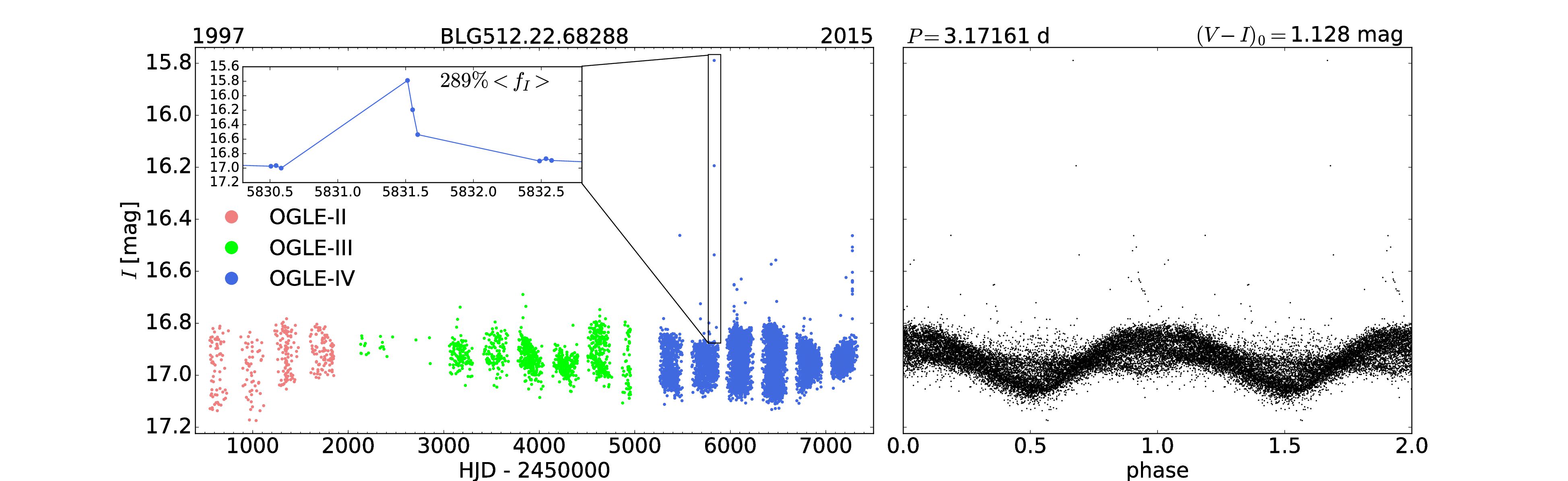}
\includegraphics[scale=0.13]{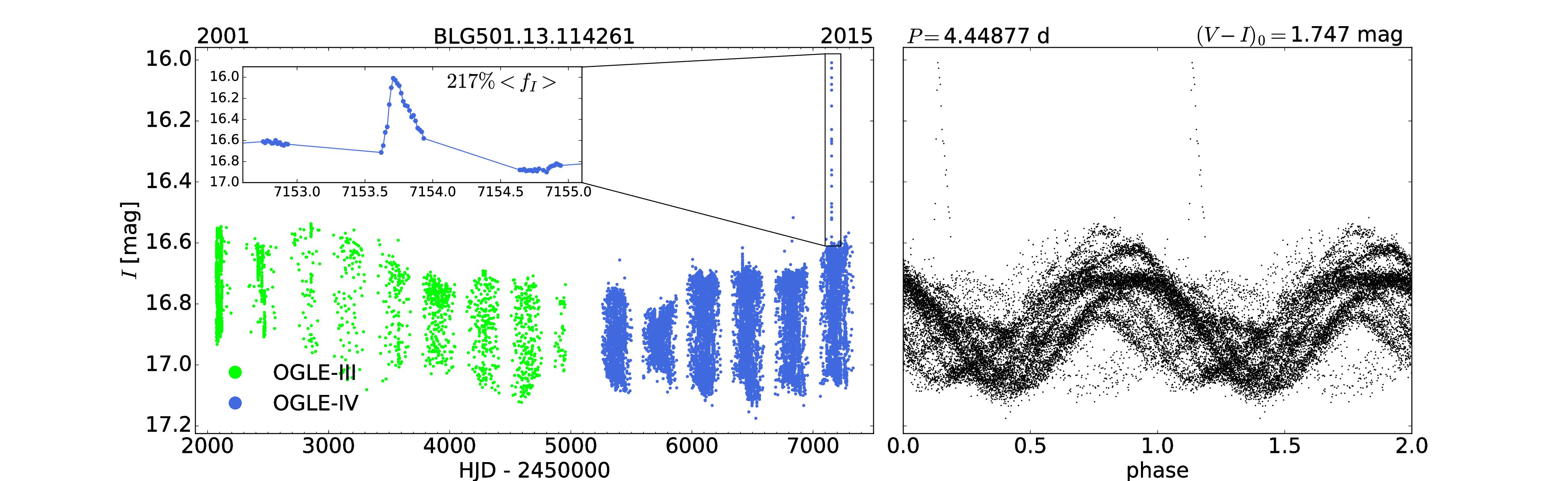}
\includegraphics[scale=0.13]{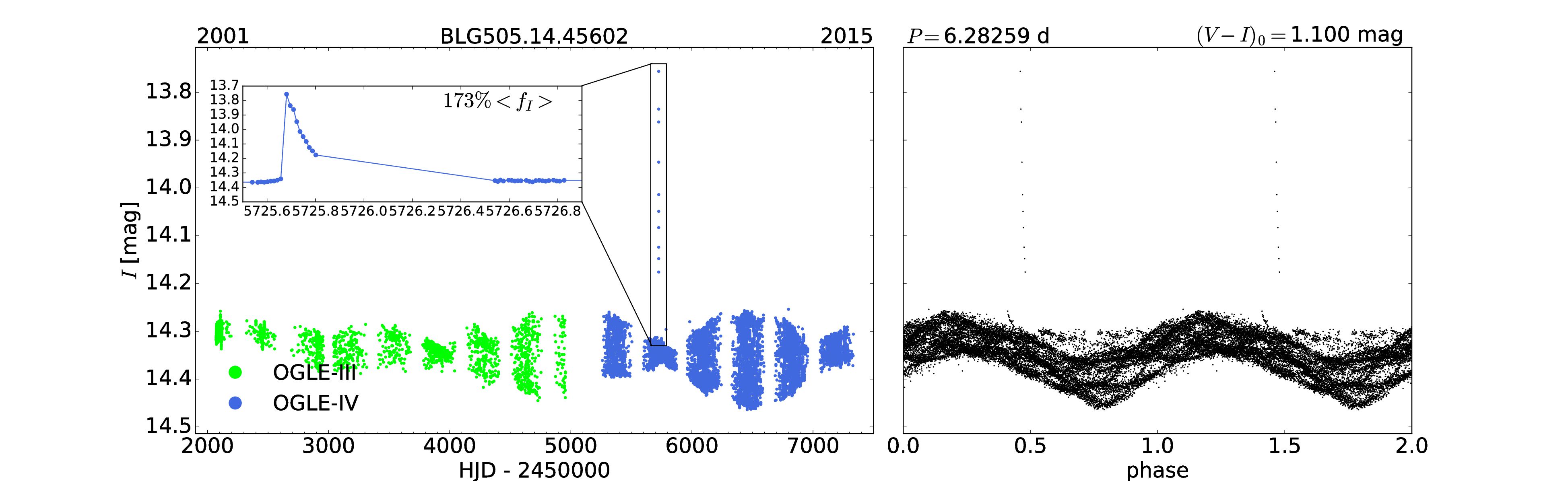}
\includegraphics[scale=0.13]{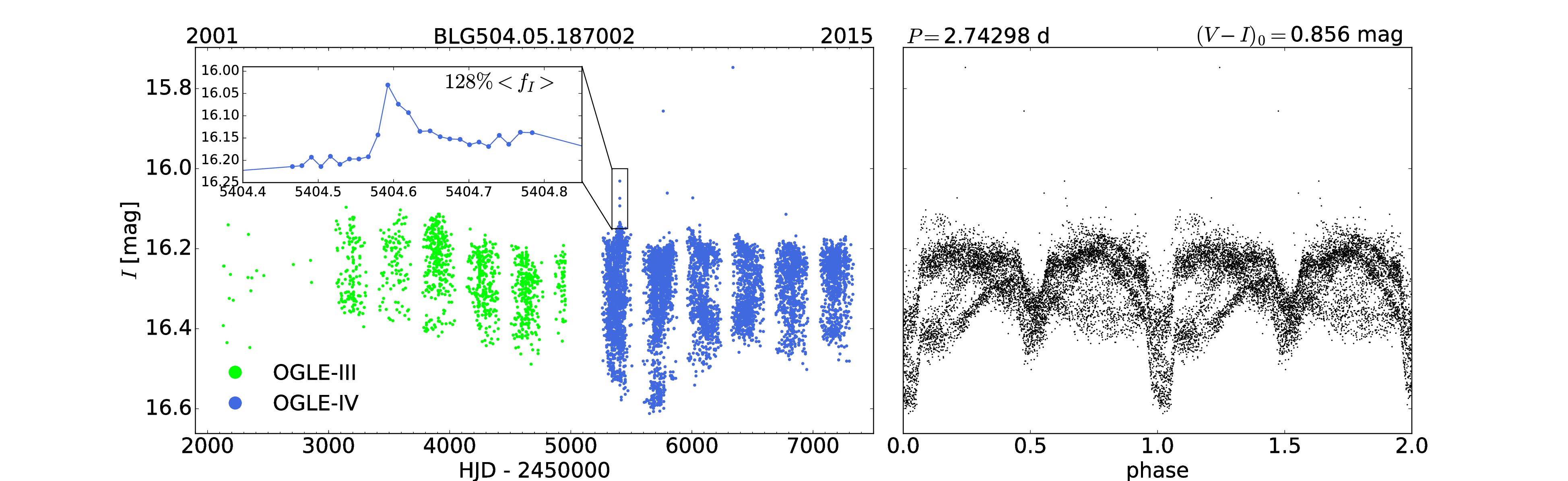}
\caption{Four examples of flaring stars found in our collection of spotted 
stars. Left panels show time-domain light curves, right panels show phased light 
curves with rotation periods $P$, given above the phased plot. Next to the periods
we also provide dereddened color index $(V-I)_0$. The dates over 
non-phased plot refer to the moment when the observations started, and the year 
from which the last observations come. The consecutive
phases of the OGLE project are marked with different colors. 
Inset graphs present the zoomed flares together with the amount of star's 
brightening during the flare relative to its mean brightness.}
\label{fig16}
\end{center}
\end{figure*}

\section{Flaring stars}

Besides spots, flares are another determinant of the stellar magnetic 
activity. Many years of research of active stars have shown that flares occur in 
stars of all spectral types -- from O to late M  ({\em e.g.} \citealt{Marchenko98, 
Balona12, Maehara12, Hawley14, Yang17}). Classical flaring events in cool stars 
are a phenomenon of a sudden release of energy accumulated by the magnetic 
field in active regions due to the magnetic reconnection, while flares in hot 
stars probably are caused by shocks and radiatively driven winds ({\em e.g.} 
\citealt{Balona12}). It is also possible that flares observed in hot stars
are not produced by these stars, but by a cooler companion. 
The sudden release of energy is manifested by a rapid 
brightening of the star, followed by a slow decline. Typical time-scales of the 
flares range from minutes to several hours or even days ({\em e.g.} 
\citealt{Henry96}). Flaring events can be observed almost along the entire spectrum 
of the electromagnetic radiation -- from X-rays to radio waves.  It is well 
known that flares on the solar surface have energies between $10^{25}$ erg and 
$10^{32}$ erg \citep{Shimizu95, Emslie12}. However, \citet{Maehara12} detected 
1000 times more energetic flares (called {\it{superflares}}) on the solar-type 
stars. \citet{Drake06} found no evidence of flares among the thousands of light 
curves of chromospherically active stars from the MACHO~data. The absence of 
flaring events in the MACHO data was the motivation to look for flaring events 
in our collection of spotted stars.

\subsection{Search for flares}

Flares can be detected as outlying points in the light curves. We 
search for flaring events in two steps. In the first step we looked for 
outliers deviating by $-4\sigma$ from the mean brightness among all of the light 
curves from our sample. This automatic scan returned 3582 stars as candidates 
for flaring objects. Further classification was based solely on the visual 
inspection of the pre-selected light curves. When at least three consecutive 
points deviate from the periodic light curve, we consider such an event as a 
flare. In each case we also check for a characteristic asymmetry between the 
rising and decaying branch. This search returned 79 stars as almost 
certain objects with flare phenomenon. It is important to note that the flare 
detection efficiency strongly depends on the cadence of the observations -- some 
of the OGLE fields have been observed more often than the others (see 
\citealt{Udalski15}). Obviously, some flares have been observed
completely, some in part, and some must have been missed. The impact of the cadence
on the flares detection efficiency has been discussed {\em e.g.} by \citet{Yang18}.  
The outliers occurring in stars located in the 
fields that were observed less frequently could look like measurement errors 
({\em e.g.} cosmic rays) and were rejected during our classification. In 
Figure \ref{fig16}, we present four examples of light curves with flares from 
our collection.

With the optical data only, it is impossible to measure the entire energy 
released during an outburst, because we lack necessary values of the
stellar parameters. Because of that, we decided to calculate the 
relative brightening during a flare for each star 
classified as a flaring object. We convert the mean brightness over
the time span of the observations and the flare maximum brightness 
into fluxes. The relative brightening can be defined as:

\begin{equation}
\frac{\Delta f_I}{\left<f_I\right>} \cdot 100\% = \frac{f_{I, \mathrm{max}} - \left<f_I\right>}{\left<f_I\right>} \cdot 
100\%,
\label{eqn:relative_brightness}
\end{equation}

\noindent where $f_{I,\mathrm{max}}$ is the flare maximum flux and
$\left<f_I\right>$ is a mean flux over the time span of the observations. 
We calculated the relative brightening for the highest data point 
during the flaring event. It is worth mentioning, however,  that the maximum 
recorded brightness during the outburst might not correspond to the
true flare brightness 
maximum -- as we discussed it earlier, it is strongly dependent on the 
cadence of the observations. The relative brightenings of the flares are provided 
in insets of Figure \ref{fig16}. The upper row of Figure \ref{fig16} shows a 
light curve with the largest brightening found in our collection 
(BLG512.22.68288) equal to $289\% \left<f_I\right>$. The value of $289\% 
\left<f_I\right>$ means that during this flare the star brightened by $189\%$ above the 
average brightness level.

\subsection{Flaring stars in the color--magnitude diagram}

\begin{figure}[h]
\begin{center}
\includegraphics[scale=0.21]{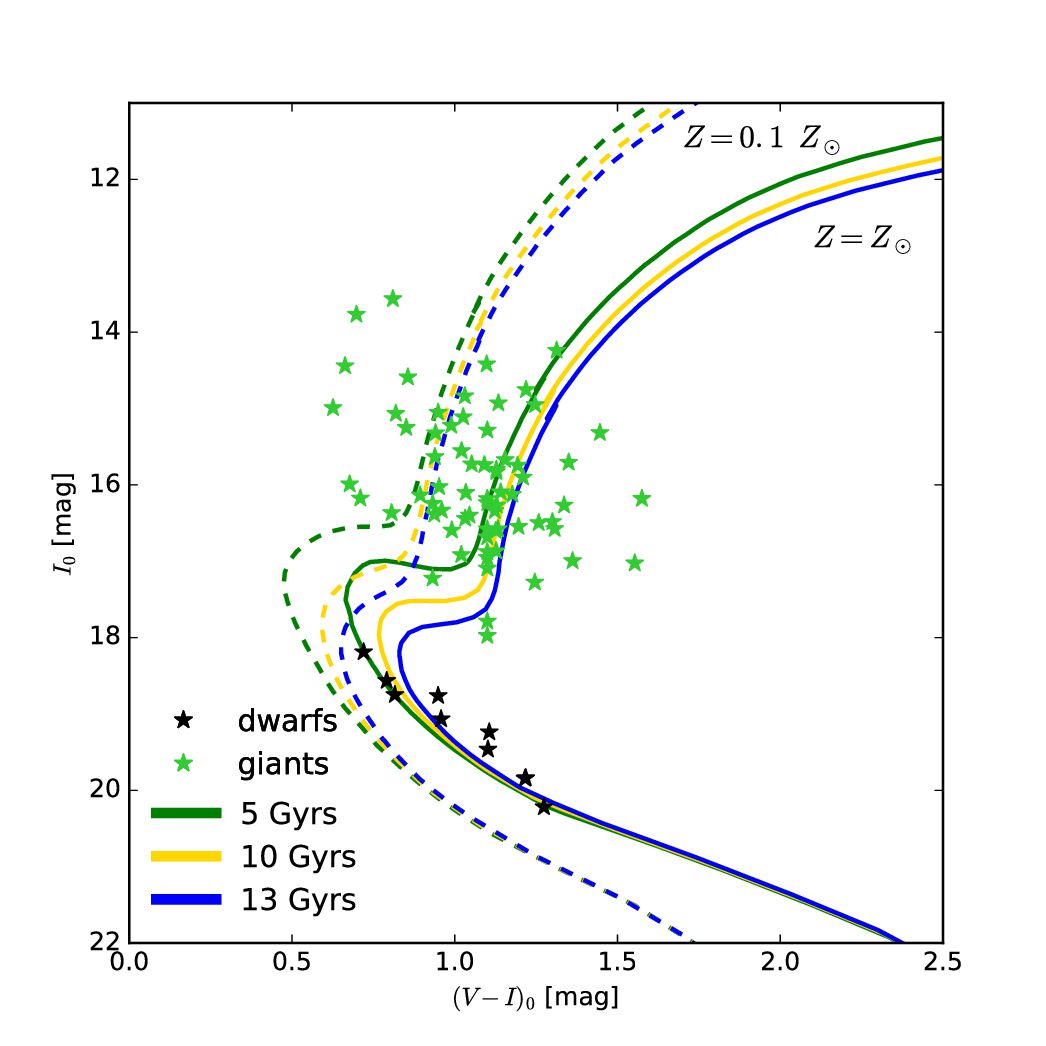}
\caption{Color--magnitude diagram (CMD) for flaring stars found in our collection 
of spotted variables. Isochrones are the same as in Fig. \ref{fig4}.}
\label{fig17}
\end{center}
\end{figure}

In Figure \ref{fig17}, we present location of flaring stars in the CMD. 
Using our dereddening procedure (described in Section 
\ref{chap:color_calculations}), we note that 11 flaring stars are dwarfs, 
and 68 are giants. Recent research shows, that flares on the giant stars are 
not unusual ({\em e.g.} \citealt{Doorsselaere17}). It is worth mentioning, however, 
that our detection efficiency for dwarfs is much smaller than for
giants because we are not 
able to detect dwarfs located in the Galactic bulge due to their low
brightness and strong interstellar extinction.

\begin{figure*}
\begin{center}
\includegraphics[scale=0.43]{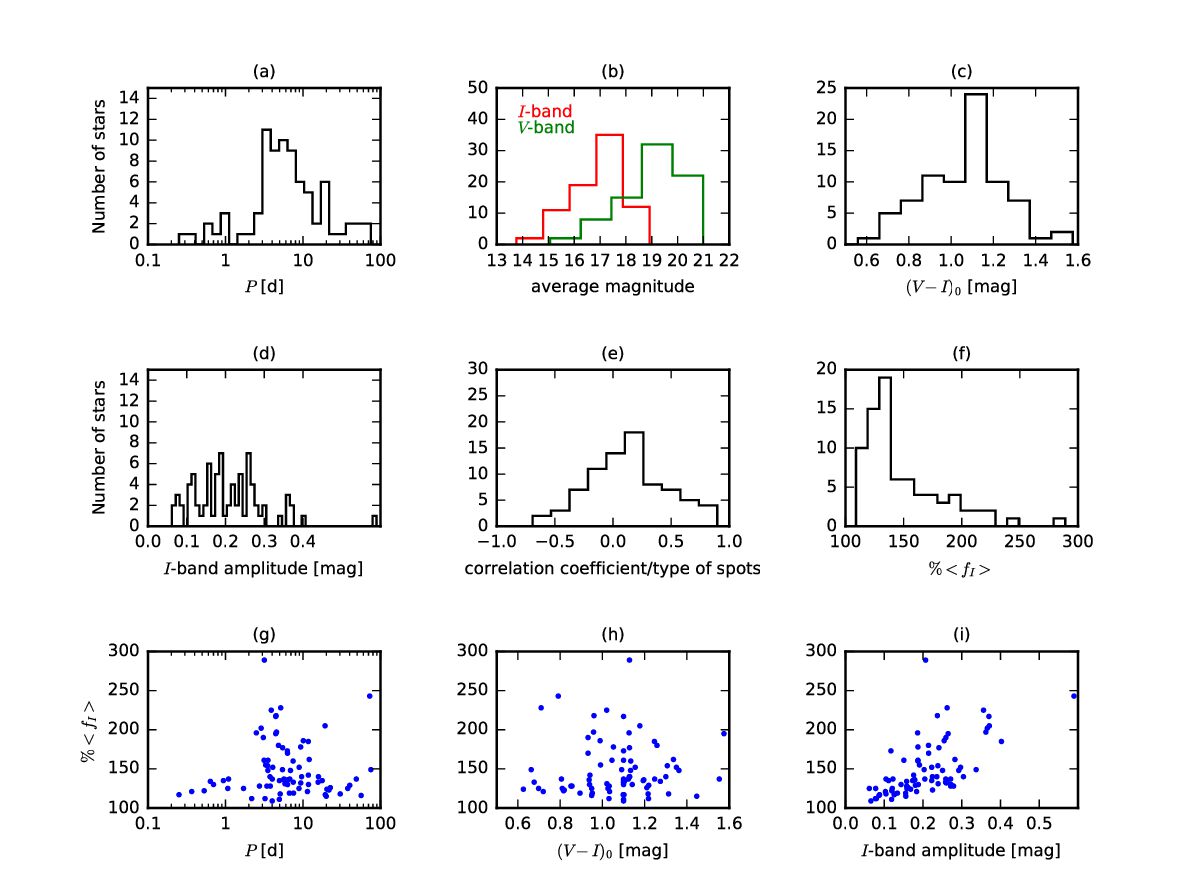}
\caption{Basic statistical properties of flaring stars. (a) The histogram of the 
rotation periods $P$. (b) The histogram of the average magnitudes in both 
filters. (c) The histogram of the dereddened color indices $(V-I)_0$. (d) The 
histogram of the {\textit{I}}-band amplitudes. (e)~The histogram of the 
correlation coefficients, which is related to the type of spots occurring on 
stellar surfaces. (f) The histogram of the observed values of the relative
flare brightenings. (g) Dependence between relative flare brightening and 
rotation period $P$. (h)~Dependence between relative flare brightening 
and dereddened color index $(V-I)_0$. (i) Dependence between relative 
flare brightening and the mean {\textit{I}}-band amplitude.}
\label{fig18}
\end{center}
\end{figure*}

\subsection{Basic statistical properties of flaring stars}

In this section we investigate the basic statistical properties of the 79 stars in which we detected 
flares. Figure \ref{fig18} shows five histograms with different
parameters of the flaring stars: the rotation periods (Figure 
\ref{fig18}a), average magnitudes (Figure \ref{fig18}b), color indices $(V-I)_0$ 
(Figure \ref{fig18}c), {\it I}-band amplitudes (Figure \ref{fig18}d) and the 
correlation coefficients between periodic light and color variations
(Figure \ref{fig18}e). In Figure \ref{fig18}f, we present the distribution 
of flares with different relative brightenings. The three bottom
diagrams (Figures \ref{fig18}g, h and i) are the plots of relative
brightening versus 
rotation period, color index $(V-I)_0$ and {\textit{I}}-band amplitude.

The rotation periods of our flare stars are within the range $ 0.251 \leq P \leq 75.132$~d
(Figure \ref{fig18}a) with the maximum number of flaring objects
occurring for $2 \leq P < 4$~d. A range of the rotation periods is wide, what
may suggests that the occurrence of flares does not depend on how 
fast a star rotates. However, \citet{Maehara17} showed that the flare occurrence rate depends
on rotation period. The small number of stars with rotation periods
shorter than about 3 days results very likely from the low number of all
investigated stars with these periods.

The most common values of the {\textit{I}}-band brightness of flaring stars occur 
in the range $16.9$-$17.9$ mag. By comparing Figure \ref{fig18}b with Figure 
\ref{fig3} we can conclude that stars with flares are on average less luminous than 
other spotted stars for which the most common values of the {\textit{I}}-band 
brightness are in the range $16.7$-$16.9$ mag. A~similar difference 
is observed in the {\textit{V}}-band.

It is well known that most of the flaring stars are red ({\em e.g.} \citealt{Davenport16, Yang17, Doorsselaere17}).
Indeed, 52 of our flaring stars ($66\%$) have colors $(V-I)_0 \geq 1.0$ mag with a peak at
$(V-I)_0 \approx 1.1$~mag corresponding to the spectral type of K4-K5 (Figure \ref{fig18}c). 
The highest brightening with the value of $289\%$ (BLG512.22.68288; upper panel in 
Figure \ref{fig16}) was also found in a red star, with a color index $(V-I)_0 = 
1.128$~mag. This is consistent with the common belief that most active
are the red stars. Flares also appear on blue 
stars, but mechanisms responsible for such events are still under discussion 
({\em e.g.} \citealt{Balona12, Pedersen17}). The bluest object from our sample
of flaring stars has $(V-I)_0 = 0.626$ mag.

Amplitudes of periodic brightness variations are related to spot sizes, their lifetimes, 
asymmetry in the surface distribution, migration and
evolution. The amplitudes of the flaring stars do not seem to differ
significantly from the bulk of the observed amplitudes although only two
of them exceeds 0.4 mag (Figures \ref{fig18}d, i). This is probably the effect of a small number of 
spotted stars with larger amplitudes. It is worth mentioning, however, that the correlation between
the relative flare brightening and the mean {\textit{I}}-band amplitude measured over the time
interval of the OGLE observations is clearly visible (Figure \ref{fig18}i). It seems that stars with larger
brightness amplitudes (and thus larger spots coverage) produce more energetic flares, what is indirect confirmation
of the work done by \citet{Yang17}, who found a positive correlation between the starspot
size and the flare activity in M~dwarfs. However, they also 
found that some of the M~dwarfs with strong flares do not exhibit any light 
variations caused by the starspots.

Using the same method as in Section \ref{chap:twospots}, we check which types of 
spots are observed in flaring stars (Figure \ref{fig18}e). All these stars show 
statistically significant correlation coefficients between the light and 
color variations ($\sigma_{\pm} \leq 0.3$)but the extreme values close
to  $-1$ or $1$ are rarely present. The majority of the flare stars is
grouped around the null correlation coefficient. It seems that
frequently flaring stars 
show the less stable brightness-color correlation, and thus the faster 
evolving spots and more strongly variable magnetic fields.

Most of the flare brightenings does not exceed 
$150\%$ (52 stars out of 79; Figure \ref{fig18}f). This means that strong 
outbursts in which the star's brightness increases twice or more are 
infrequent. It is an indirect confirmation of Maehara's (\citeyear{Maehara12}) research,
who found only 148 stars with {\textit{superflares}} among 
$83\;000$ stars from the {\textit{Kepler}} data.

\section{Conclusions}

In this paper, we presented the detection and statistical analysis of $12\;660$ 
spotted stars lying toward, and in the Galactic bulge. We introduced 
a new dereddening procedure, as the bulge is one of the most challenging regions of the sky to 
explore due to the large, irregular interstellar extinction.  As an result of our analysis we classified $848$ spotted stars as dwarfs and $11\;812$ stars as giants. In the first 
part of this work, we conducted a similar analysis to that done by \citet{Drake06}, what allowed 
us to confirm accurately some of the correlations reported by them, and to analyze these correlations in greater details. Moreover, we found previously unknown interesting features
of the spotted variables. The second part of our paper contains the analysis
based on the brightness-color correlation we have disclosed. We found 79 
flaring objects among all spotted variables and we examined their statistical 
properties.

Our analysis shows that the correlation between the Galactic latitude
and rotation period of giants reported by \citet{Drake06} is
probably caused by the selection effect. We think this is an artifact
of the period magnitude relation existing for giants which means that
it is harder to detect faster rotating hence fainter giants due to the
increasing extinction close to the Galactic plane. We also did not confirm
the period-color relation suggested by \citet{Drake06}. Four times
larger sample of stars does not show any systematic trend of color
with period. On the other hand, our data show a clear
correlation between luminosity and rotation period which is not visible in the data analyzed by \citet{Drake06}.
We confirm that fainter stars from our sample show larger brightness 
variations, and thus larger and/or more nonuniform spots coverage. However, the existence of this 
correlation is most likely a selection effect -- for fainter stars we were able 
to detect only large amplitudes. It is noteworthy, however, that 
bright stars ($I_0 \lesssim 16$ mag) do not show amplitudes larger than 0.4 mag. 
This is an unexpected finding and it seems to be real. We notice
that the spotted stars can be divided into two groups: fast-rotating ($P 
\leq 2$ d) dwarfs with small amplitudes ({\it{I-}}band amplitude $< 0.2$~mag), and 
slow-rotating dwarfs and giants with amplitudes of up to 0.8 mag. The existence of 
these two distinct, well-resolved groups suggests that the mechanisms responsible 
for spot formation, thus the magnetic activity, may differ between 
these two classes. On the other hand, large amplitudes are recorded on stars with
large spots coverage, but with clear longitudinal asymmetry in spotted
areas. Uniformly spotted stars or with dominant polar spots will
produce very small brightness variations despite the heavy spot coverage on their surfaces.
In such cases another measure of the magnetic activity should be used
{\em e.g.} core emission in the calcium lines CaII~H~and~K
\citep{Baliunas95} or X-ray flux \citep{Pallavicini81}.
Hereafter, we indicate both groups as interesting for a more 
detailed study. The correlation between the {\it{I-}}band amplitude and rotation 
period for giants is clearly visible -- the slower a giant rotates, the smaller 
amplitude it shows. By contrast, slow rotating dwarfs seem to show larger 
amplitudes than the rapidly rotating ones.

Using accurate, long-term, two-band OGLE photometry we revealed the
existence of the correlation 
between brightness variations and color variations of the spotted variables. We find 
that based on the value of the correlation coefficient stars from our
collection can be divided into three groups according to the 
strength of this correlation. This dependence is an indicator of which type of spots 
prevails on the stellar surface. First group contains stars which show positive 
correlation coefficient ($r-\sigma_{-} \geq 0.2$) what suggests that
variability is due to the existence of 
dark, cool spots on the stellar surface. Second group consists of stars with the
negative correlation coefficient ($r+\sigma_{+} \leq -0.2$). 
Negative values of the coefficient mean that 
the stars are redder when more luminous because their brightness variations 
in the {\textit{V}}-band are smaller than in {\textit{I}}-band. This
kind of variations can be produced by nonuniform distribution of
elements over the stellar surface {\em i.e.} chemical spots associated with
the stable magnetic field. The increased abundance of heavy elements 
in a chemical spot rotating with the star results in a variable 
{\textit{line-blanketing}} effect. Field chemically peculiar stars are
observed only among early spectral type stars. It would be valuable to
observe spectroscopically our Ap star candidates to confirm their
status. The negative correlation observed in cool stars may result from the rotation of a
tidally distorted cool component in a binary of the RS CVn-type
variable with a favorable orbit inclination.  Observations show 
that the correlation coefficients are remarkably stable throughout the OGLE-IV phase.
This suggests that spots are slowly evolving and their lifetimes can be very long. Going 
further, we can expect that the magnetic fields on these stars should be
quite stable on the time scale of at least a decade. The 
last group contains stars for which the correlation coefficient is around 0 
($r+\sigma_{+} > -0.2$ and $r-\sigma_{-} < 0.2$). It seems that stars
from the last group have rapidly evolving spots together with variable bright
regions hence their surface magnetic field rapidly vary in time, which makes the correlation between brightness and 
color blurred. A very interesting result is obtained when all the
spotted stars are plotted in the CMD with the correlation coefficient
color coded. All three groups are clearly separated with the
coefficient varying along the MS from the negative values for the
hottest stars, through null values up to the positive values for the
coolest dwarfs. In the case of giants the coefficient varies along the
giant branch: the negative correlation prevails among least massive
giants and its value increases with the giant mass. The lines of
constant coefficient lie approximately along the lines of constant
stellar radius. This would suggest that the type of
this correlation depends on the stellar radius, and thus on the strength and configuration of the 
surface magnetic fields. Among our spotted stars we found 79 objects with flare 
footprints. Our results show that the correlation coefficient of the
flaring stars is concentrated around null value. This indicates
that these stars have unstable, fast-evolving magnetic 
fields producing frequent outbursts. Moreover, correlation between the flare brightening
and the {\textit{I}}-band amplitude is clearly visible. We also notice that slow 
rotators show the highest brightenings during flares, however  
stars with very strong flares, at least doubling the stellar brightness are very rare.

\section{Acknowledgments}

We are grateful to the anonymous Referee for the careful reading of our manuscript and for his/her supportive and constructive comments which greatly improved this publication. We also thank the anonymous Statistics Consultant for important comments related to the Fourier time-series analysis. Thanks to these remarks the Section about period searching has been significantly expanded. We would like to thank Profs. M. Kubiak, G.~Pietrzy\' nski and Dr. M. Pawlak for their contribution to the collection of the OGLE photometric 
data over the past years. We are grateful to late Z. Ko\l aczkowski for discussion 
and ideas that improved this paper. 

This work has been supported by the National
Science Centre, Poland, grant MAESTRO no. 2016/22/A/ST9/00009 to I. Soszy\' nski.
P. Iwanek is partially supported by the 
``Kartezjusz'' programme no. POWR.03.02.00-00-I001/16-00 founded by the National 
Centre for Research and Development, Poland. The OGLE project has received 
funding from the National Science Center, Poland, grant MAESTRO 
no. 2014/14/A/ST9/00121 to A.~Udalski.


\newpage
\begin{thebibliography}{}
\bibitem[Alard \& Lupton(1998)]{Alard98} Alard, C., Lupton, R. H., 1998, \apj, 503, 325
\bibitem[Alcock et al.(1995)]{Alcock95} Alcock, C., Allsman, R. A., Axelrod, T. S., Bennett, D. P., Cook, K. H., Freeman, K. C., Griest, K., Marshall, S. L., Peterson, B. A., Pratt, M. R., Quinn, P. J., Reimann, J., Rodgers, A. W., Stubbs, C. W., Sutherland, W., Welch, D. L., 1995, \aj, 109, 1653
\bibitem[Baliunas et al.(1995)]{Baliunas95} Baliunas, S. L., Donahue, R. A., Soon, W. H., Horne, J. H., Frazer, J., Woodard-Eklund, L., Bradford, M., Rao, L. M., Wilson, O. C., Zhang, Q., Bennett, W., Briggs, J., Carroll, S. M., Duncan, D. K., Figueroa, D., Lanning, H. H., Misch, T., Mueller, J., Noyes, R. W., Poppe, D., Porter, A. C., Robinson, C. R., Russell, J., Shelton, J. C., Soyumer, T., Vaughan, A. H., Whitney, J. H., 1995, \apj, 438, 269
\bibitem[Balona(2012)]{Balona12} Balona, L. A., 2012, \mnras, 423, 3420
\bibitem[Balona et al.(2015)]{Balona15} Balona, L. A., Baran, A. S., Daszy\' nska-Daszkiewicz, J., De Cat, P., 2015, \mnras, 451, 1445
\bibitem[Balona(2017)]{Balona17} Balona, L. A., 2017, \mnras, 467, 1830
\bibitem[Basri et al.(2011)]{Basri11} Basri, G., Walkowicz, L. M., Batalha, N., Gilliland, R. L., Jenkins, J., Borucki, W. J., Koch, D., Caldwell, D., Dupree, A. K., Latham, D. W., Marcy, G. W., Meibom, S., Brown, T., 2011, \aj, 141, 20
\bibitem[Basri et al.(2013)]{Basri13} Basri, G., Walkowicz, L. M., Reiners, A., 2013, \apj, 769, 37
\bibitem[Bressan et al.(2012)]{Bressan12} Bressan, A., Marigo, P., Girardi, L., Salasnich, B., Dal Cero, C., Rubele, S., Nanni, A., 2012, \mnras, 427, 127
\bibitem[Bychkov et al.(2016)]{Bychkov16} Bychkov, V. D., Bychkova, L. V., Madej, J., 2016, \mnras, 455, 2567
\bibitem[Ceillier et al.(2017)]{Ceillier17} Ceillier, T., Tayar, J., Mathur, S., Salabert, D., García, R. A., Stello, D., Pinsonneault, M. H., van Saders, J., Beck, P. G., Bloemen, S., 2017, \aap, 605, 111
\bibitem[Davenport(2016)]{Davenport16} Davenport, R. A. J., 2016, \apj, 829, 23
\bibitem[Drake(2006)]{Drake06} Drake, A. J., 2006, \aj, 131, 1044
\bibitem[Drake et al.(1989)]{Drake89} Drake, S. A., Simon, T., Linsky, J. L., 1989, \apjs, 71, 905
\bibitem[Durney and Latour(1978)]{Durney78} Durney, B. R., Latour, J., 1978, GApFD, 9, 241
\bibitem[Emslie et al.(2012)]{Emslie12} Emslie, A., Dennis, B., Shih, A., Chamberlin, P., Mewaldt, R., Moore, C., Share, G., Vourlidas, A., Welsch, B.,  2012, \apj, 759, 71
\bibitem[Gaia Collaboration(2018)]{Gaia18} Gaia Collaboration; Brown, A. G. A., et al., 2018, \aap, 616, 1
\bibitem[Gaia Collaboration(2016)]{Gaia16} Gaia Collaboration; Prusti, T., et al., 2016, \aap, 595, 1
\bibitem[Gray and Corbally(2009)]{Gray09} Gray, R.~O., Corbally, J. C., 2009, Princeton University Press, ~ISBN: {\^A} 978-0-691-12511-4
\bibitem[Guerrero et al.(2013)]{Guerrero13} Guerrero, G., Smolarkiewicz, P. K., Kosovichev, A. G., Mansour, N. N., 2013, \apj, 779, 176
\bibitem[Guerrero et al.(2016a)]{Guerrero16a} Guerrero, G., Smolarkiewicz, P. K., de Gouveia Dal Pino, E. M., Kosovichev, A. G., Mansour, N. N., 2016 \apj, 819, 104
\bibitem[Guerrero et al.(2016b)]{Guerrero16b} Guerrero, G., Smolarkiewicz, P. K., de Gouveia Dal Pino, E. M., Kosovichev, A. G., Mansour, N. N., 2016, \apjl, 828, 3
\bibitem[Guerrero et al.(2018)]{Guerrero18} Guerrero, G., Zaire, B., Smolarkiewicz, P. K., de Gouveia Dal Pino, E. M., Kosovichev, A. G., Mansour, N. N., 2018, eprint arXiv:1810.07978
\bibitem[G\"unther et al.(2014)]{Gunther14} G\"unther, H. M., Cody, A. M., Covey, K. R., Hillenbrand, L. A., Plavchan, P., Poppenhaeger, K., Rebull, L. M., Stauffer, J. R., Wolk, S. J., Allen, L., Bayo, A., Gutermuth, R. A., Hora, J. L., Meng, H. Y. A., Morales-Calder\'on, M., Parks, J. R., Song, I., 2014, \aj, 148, 122
\bibitem[Guthnick and Prager(1914)]{Guthnick14} Guthnick, P., Prager, R., 1914, Ver\"off. Berlin-Babelsberg, 1, 44
\bibitem[Hawley et al.(2014)]{Hawley14} 	Hawley, S. L., Davenport, J. R. A., Kowalski, A. F., Wisniewski, J. P., Hebb, L., Deitrick, R., Hilton, E. J., 2014, \apj, 797, 121
\bibitem[Henry \& Newsom(1996)]{Henry96} Henry, G. W., Newsom, M. S., 1996, \pasp, 108, 242
\bibitem[Hoffmeister(1915)]{Hoffmeister15} Hoffmeister, C., 1915, \an, 200, 177
\bibitem[Hubrig et al.(2019)]{Hubrig19} Hubrig, S., K\" uker, M., J\" arvinen, S. P., Kholtygin, A. F., Sch\" oller, M., Ryspaeva, E. B., Sokoloff, D. D., 2019, \mnras, 484, 4495
\bibitem[Hurley et al.(2000)]{Hurley00} Hurley, J. R., Pols, O. R., Tout, C. A., 2000, \mnras, 315, 543
\bibitem[Irwin et al.(2009)]{Irwin09} Irwin, J., Aigrain, S., Bouvier, J., Hebb, L., Hodgkin, S., Irwin, M., Moraux, E., 2009, \mnras, 392, 1456
\bibitem[Kiraga(2012)]{Kiraga12} Kiraga, M., 2012, \actaa, 62, 67
\bibitem[Kiraga and St\k epie\'n(2013)]{Kiraga13} Kiraga, M., St\k epie\'n, K., 2013, \actaa, 63, 53
\bibitem[Kochukhov et al.(2019)]{Kochukhov19a} Kochukhov, O., Shultz, M., Neiner, C., 2019, \aap, 621, 47
\bibitem[Kochukhov and Shulyak(2019)]{Kochukhov19b} Kochukhov, O., Shulyak, D., 2019, eprint arXiv:1902.04157
\bibitem[Kraft(1967)]{Kraft67} Kraft, R. P., 1967, \apj, 150, 551
\bibitem[Kron(1947)]{Kron47} Kron, G. E., 1947, \pasp, 59, 261
\bibitem[Kron(1950)]{Kron50} Kron, G. E., 1950, ASPL, 6, 52
\bibitem[Kurtz(1985)]{Kurtz85} Kurtz, D. W., 1985, \mnras, 213, 773
\bibitem[Lanzafame et al.(2018a)]{Lanzafame18a} Lanzafame, A. C., Distefano, E., Barnes, S. A., Spada, F., 2018, eprint arXiv:1805.11332
\bibitem[Lanzafame et al.(2018b)]{Lanzafame18b} Lanzafame, A. C., Distefano, E., Messina, S., Pagano, I., Lanza, A. F., Eyer, L., Guy, L. P., Rimoldini, L., Lecoeur-Taibi, I., Holl, B., Audard, M., de Fombelle, G. J., Nienartowicz, K., Marchal, O., Mowlavi, N., 2018, \aap, 616, 16
\bibitem[Lomb(1976)]{Lomb76} Lomb, N. R., 1976, \apss, 39, 447
\bibitem[Maehara et al.(2012)]{Maehara12} Maehara, H., Shibayama, T., Notsu, S., Notsu, Y., Nagao, T., Kusaba, S., Honda, S., Nogami, D., Shibata, K., 2012, \nat, 485, 478
\bibitem[Maehara et al.(2017)]{Maehara17} Maehara, H., Notsu, Y., Notsu, S., Namekata, K., Honda, S., Ishii, T. T., Nogami, D., Shibata, K., 2017, \pasj, 69, 41
\bibitem[Marchenko et al.(1998)]{Marchenko98} Marchenko, S., Moffat, A., van der Hucht, K., Seggewiss, W., Schrijver, H., Stenholm, B., Lundstrom, I., Setia Gunawan, D., Sutantyo, W., van den Heuvel, E., de Cuyper, J.-P., Gomez, A., 1998, \aap, 331, 1022
\bibitem[Mathur et al.(2014)]{Mathur14} Mathur, S., Garc\' ia, R. A., Ballot, J., Ceillier, T., Salabert, D., Metcalfe, T. S., R\' egulo, C., Jim\' enez, A., Bloemen, S., 2014, \aap, 562, 124
\bibitem[Mathys et al.(2019)]{Mathys19} Mathys, G., Romanyuk, I. I., Hubrig, S., Kudryavtsev, D. O., Landstreet, J. D., Sch\"oller, M., Semenko, E. A., Yakunin, I. A., 2019, eprint arXiv:1902.05869
\bibitem[Maury and Pickering(1897)]{Maury97} Maury, A. C., Pickering, E. C., 1897, AnHar, 28, 1
\bibitem[McQuillan et al.(2014)]{McQuillan14} McQuillan, A., Mazeh, T., Aigrain, S., 2014, \apjs, 211, 24
\bibitem[Mekkaden(1985)]{Mekkaden85} Mekkaden, M. V., 1985, \apss, 117, 381
\bibitem[Messina and Guinan(2002)]{Messina02} Messina, S., Guinan, E. F., 2002, \aap, 393, 225
\bibitem[Milne(1928)]{Milne28} Milne, E. A., 1928, \theobservatory, 51, 88
\bibitem[Molnar(1973)]{Molnar73} Molnar, M. R., 1973, \apj, 179, 527
\bibitem[Montet et al.(2017)]{Montet17} Montet, B. T., Tovar, G., Foreman-Mackey, D., 2017 \apj, 851, 116
\bibitem[Nataf et al.(2013)]{Nataf13} Nataf, D. M., Gould, A., Fouqu\' e, P., Gonzalez, O. A., Johnson, J. A.,  Skowron, J., Udalski, A., Szyma\' nski, M. K., Kubiak, M., Pietrzy\' nski, G., Soszy\' nski, I., Ulaczyk, K., Wyrzykowski, \L., Poleski, R., 2013, \apj, 769, 88
\bibitem[Ol\' ah \& Strassmeier(2002)]{Olah02} Ol\' ah, K., Strassmeier, K. G., 2002, \an, 323, 361
\bibitem[Ol\' ah et al.(2009)]{Olah09} Ol\' ah, K., Koll\' ath, Z., Granzer, T., Strassmeier, K. G., Lanza, A. F., J\" arvinen, S., Korhonen, H., Baliunas, S. L., Soon, W., Messina, S., Cutispoto, G., 2009, \aap, 501, 703
\bibitem[Ol\' ah et al.(2016)]{Olah16} Ol\' ah, K., K\" ov\' ari, Zs., Petrovay, K., Soon, W., Baliunas, S., Koll\' ath, Z., Vida, K., 2016, \aap, 590, 133
\bibitem[Pallavicini et al.(1981)]{Pallavicini81} Pallavicini, R., Golub, L., Rosner, R., Vaiana, G.S., Ayres, T., Linsky, J. L., 1981, \apj, 248, 279
\bibitem[Parker(1955)]{Parker55} Parker, E. N.,  1955, \apj, 122, 293
\bibitem[Pedersen et al.(2017)]{Pedersen17} Pedersen, M. G., Antoci, V., Korhonen, H., White, T. R., Jessen-Hansen, J., Lehtinen, J., Nikbakhsh, S., Viuho, J., 2017, \mnras, 466, 3060
\bibitem[Paunzen et al.(2015)]{Paunzen15} Paunzen, E., Fr\"ohlich, H.-E., Netopil, M., Weiss, W. W., L\"uftinger, T., 2015, \aap, 574, 57
\bibitem[Paunzen et al.(2016)]{Paunzen16} Paunzen, E., Netopil, M., Bernhard, K., H\"ummerich, S., 2016, BlgAJ, 24, 97
\bibitem[Pojma\' nski(1997)]{Pojmanski94} Pojma\' nski, G., 1997, \actaa, 47, 467
\bibitem[Pojma\' nski(2002)]{Pojmanski02} Pojma\' nski, G., 2002, \actaa, 52, 397
\bibitem[Preston(1974)]{Preston74} Preston, G. W., 1974, \araa, 12, 257
\bibitem[Rebull et al.(2015)]{Rebull15} Rebull, L. M., Stauffer, J. R., Cody, A. M., G\"unther, H. M., Hillenbrand, L. A., Poppenhaeger, K., Wolk, S. J., Hora, J., Hernandez, J., Bayo, A., Covey, K., Forbrich, J., Gutermuth, R., Morales-Calder\'on, M., Plavchan, P., Song, I., Bouy, H., Terebey, S., Cuillandre, J. C., Allen, L. E., 2015, \aj, 150, 175
\bibitem[Reinhold et al.(2017)]{Reinhold17} Reinhold, T., Cameron, R. H., Gizon, L., 2017, \aap, 603, 52
\bibitem[Renson and Catalano(2001)]{Renson01} Renson, P., Catalano, F. A., 2001, \aap, 378, 113
\bibitem[Roettenbacher et al.(2016)]{Roettenbacher16} Roettenbacher, R. M., Monnier, J. D., Korhonen, H., Aarnio, A. N., Baron, F., Che, X., Harmon, R. O., K\" ov\` ari, Zs., Kraus, S., Schaefer, G. H., Torres, G., Zhao, M., Ten Brummelaar, T. A., Sturmann, J., Sturmann, L., 2016, \nat, 533, 217
\bibitem[Ross(1926)]{Ross26} Ross, F. E., 1926, \aj, 36, 124 
\bibitem[Salabert et al.(2016)]{Salabert16} Salabert, D., Garc\' ia, R. A., Beck, P. G., Egeland, R., Pall\'e, P. L., Mathur, S., Metcalfe, T. S., do Nascimento, J.-D., Jr., Ceillier, T., Andersen, M. F., Trivi\~n o Hage, A., 2016, \aap, 596, 31
\bibitem[Scargle(1982)]{Scargle82} Scargle, J. D., 1982, \apj, 263, 835
\bibitem[Schlegel et al.(1998)]{Schlegel98} Schlegel, D. J., Finkbeiner, D. P., Davis, M., 1998, \apj, 500, 525
\bibitem[Schwabe(1844)]{Schwabe44} Schwabe, H.,  1844, \an, 21, 233
\bibitem[Schwabe(1845)]{Schwabe45} Schwabe, H.,  1845, \an, 22, 365
\bibitem[Serenelli et al.(2016)]{Serenelli16} Serenelli, A., Scott, P., Villante, F.L., Vincent, A.C., Asplund, M., Basu, S., Grevesse, N., Pen\~ a-Garay, C., 2016, \mnras, 463, 2
\bibitem[Sharma et al.(2011)]{Sharma11} Sharma, S., Bland-Hawthorn, J., Johnston, K. V., Binney, J., 2011, \apj, 730, 3
\bibitem[Shimizu(1995)]{Shimizu95} Shimizu, T.,  1995, \pasj, 47, 251
\bibitem[Sikora et al.(2019)]{Sikora19} Sikora, J., Wade, G. A., Power, J., Neiner, C., 2019, \mnras, 483, 2300
\bibitem[Simon \& Fekel(1987)]{Simon87} Simon, T., Fekel, F.C. Jr., 1987, \apj, 316, 434
\bibitem[Skumanich(1972)]{Skumanich72} Skumanich A.,  1972, \apj, 171, 565
\bibitem[Soszy\' nski et al.(2013)]{Soszynski13} Soszy\' nski, I., Udalski, A., Szyma\' nski, M. K., Kubiak, M., Pietrzy\' nski, G., Wyrzykowski, \L., Ulaczyk, K., Poleski, R., Koz\l owski, S., Pietrukowicz, P., Skowron, J., 2013, \actaa, 63, 21
\bibitem[Soszy\' nski et al.(2016)]{Soszynski16} Soszy\' nski, I., Pawlak, M., Pietrukowicz, P., Udalski, A., Szyma\' nski, M. K., Wyrzykowski, \L., Ulaczyk, K., Poleski, R., Koz\l owski, S., Skowron, D. M., Skowron, J., Mr\' oz, P., Hamanowicz, A., 2016, \actaa, 66, 405
\bibitem[St\k epie\'n(1968)]{Stepien68} St\k epie\'n, K., 1968, \apj, 154, 945
\bibitem[St\k epie\'n and Czechowski(1993)]{Stepien93} St\k epie\'n, K., Czechowski, W., 1993, \aap, 268, 187
\bibitem[Strassmeier et al.(2019)]{Strassmeier19} Strassmeier, K. G., Carroll, T. A., Ilyin, I. V., 2019, eprint arXiv:1902.11201
\bibitem[Szyma\' nski et al.(2011)]{Szymanski11} Szyma\' nski, M. K., Udalski, A., Soszy\' nski, I., Kubiak, M., Pietrzy\' nski, G., Poleski, R., Wyrzykowski, \L., Ulaczyk, K., 2011, \actaa, 61, 83
\bibitem[Udalski et al.(2002)]{Udalski02} Udalski, A., Szyma\' nski, M., Kubiak, M., Pietrzy\' nski, G., Soszy\' nski, I., Wo\' zniak, P., Zebru\'n, K., Szewczyk, O., Wyrzykowski, \L., 2002, \actaa, 52, 217
\bibitem[Udalski et al.(2015)]{Udalski15} Udalski, A., Szyma\' nski, M. K., Szyma\' nski, G., 2015, \actaa, 65, 1
\bibitem[Udalski et al.(2018)]{Udalski18} Udalski, A., Soszy\' nski, I., Pietrukowicz, P., Szyma\' nski, M. K., Skowron, D. M., Skowron, J., Mr\' oz, P., Poleski, R., Koz\l owski, S., Ulaczyk, K., Rybicki, K.,  Iwanek, P., Wrona, M., 2018 \actaa, 68, 315
\bibitem[VanderPlas(2018)]{VanderPlas18} VanderPlas, J. T., 2019, \apjs, 236, 16
\bibitem[Van Doorsselaere et al.(2017)]{Doorsselaere17} Van Doorsselaere, T., Shariati, H., Debosscher, J., 2017, \apjs, 232, 26
\bibitem[von Steiger \& Zurbuchen(2016)]{Steiger16} von Steiger, R., Zurbuchen, T. H., 2016, \apj, 816, 13
\bibitem[Wade et al.(2016)]{Wade16} Wade, G. A., Neiner, C., Alecian, E., Grunhut, J. H., Petit, V., Batz, B. de, Bohlender, D. A., Cohen, D. H., Henrichs, H. F., Kochukhov, O., Landstreet, J. D., Manset, N., Martins, F., Mathis, S., Oksala, M. E., Owocki, S. P., Rivinius, Th., Shultz, M. E., Sundqvist, J. O., Townsend, R. H. D., ud-Doula, A., Bouret, J.-C., Braithwaite, J., Briquet, M., Carciofi, A. C., David-Uraz, A., Folsom, C. P., Fullerton, A. W., Leroy, B., Marcolino, W. L. F., Moffat, A. F. J., Naz\'e, Y., Louis, N. St., Auri\' ere, M., Bagnulo, S., Bailey, J. D., Barb\'a, R. H., Blaz\' ere, A., B\" ohm, T., Catala, C., Donati, J.-F., Ferrario, L., Harrington, D., Howarth, I. D., Ignace, R., Kaper, L., L\" uftinger, T., Prinja, R., Vink, J. S., Weiss, W. W., Yakunin, I., 2016, \mnras, 456, 2
\bibitem[Wolk et al.(2015)]{Wolk15} Wolk, S. J., G\"unther, H. M., Poppenhaeger, K., Cody, A. M., Rebull, L. M., Forbrich, J., Gutermuth, R. A., Hillenbrand, L. A., Plavchan, P., Stauffer, J. R., Covey, K. R., Song, I., 2015, \aj, 150, 145
\bibitem[Wolk et al.(2018)]{Wolk18} Wolk, S. J., G\"unther, H. M., Poppenhaeger, K., Winston, E., Rebull, L. M., Stauffer, J. R., Gutermuth, R. A., Cody, A. M., Hillenbrand, L. A., Plavchan, P., Covey, K. R., Song, I., 2018, \aj, 155, 99
\bibitem[Wo\' zniak(2000)]{Wozniak00} Wo\' zniak, P. R., 2000, \actaa, 50, 421
\bibitem[Yang et al.(2017)]{Yang17} Yang, H., Liu, J., Gao, Q., Fang, X., Guo, J., Zhang, Y., Hou, Y., Wang, Y., Cao, Z., 2017, \apj, 849, 36
\bibitem [Yang et al.(2018)]{Yang18} Yang, H., Liu, J., Qiao, E., Zhang, H., Gao, Q., Cui, K., Han, H., 2018, \apj, 859, 87

\end{thebibliography}
\end{document}